\documentclass{aa} 
\usepackage{lscape}
\usepackage{graphicx}
\usepackage{txfonts}
\usepackage{rotating}

\begin{document}
   \title{Stellar and circumstellar properties of visual binaries in the Orion Nebula Cluster\thanks{Based on observations obtained at the Gemini Observatory, 
   which is operated by the Association of Universities for Research in Astronomy, Inc., under a cooperative agreement
   with the NSF on behalf of the Gemini partnership: the National Science Foundation (United
   States), the Science and Technology Facilities Council (United Kingdom), the
   National Research Council (Canada), CONICYT (Chile), the Australian Research Council
   (Australia), MinistŽrio da Cincia e Tecnologia (Brazil), and SECYT (Argentina)
  }}

   \subtitle{}

   \author{S. Correia\inst{1,2}
          \and
         G. Duch\^ene\inst{3,4}
          \and 
          B. Reipurth\inst{5,6}
         	\and 
	H. Zinnecker\inst{1,7,8}
	\and 
	S. Daemgen\inst{9,10}
	\and 
	M. G. Petr-Gotzens\inst{9}
          \and 
         R. K\"ohler\inst{11,12}
          \and 
         Th. Ratzka\inst{13}
         	\and 	
	C. Aspin\inst{5}
	\and
	Q. M. Konopacky\inst{14}
	\and 
	A. M. Ghez\inst{15,16}
            	}

   \institute{Leibniz-Institut f\"ur Astrophysik Potsdam, An der Sternwarte 16, 14482 Potsdam, Germany
         \and
              	Institute for Astronomy, University of Hawaii, 34 Ohia Ku Street, Pukalani, HI 96768, USA
         \and
             	Astronomy Department, University of California, Berkeley, CA 94720-3411, USA
          \and
             	UJF-Grenoble\,1/CNRS-INSU, Institut de Plan\'etologie et d'Astrophysique de Grenoble\,UMR 5274, 38041 Grenoble, France
          \and   
                	Institute for Astronomy, University of Hawaii, 640 N. Aohoku Place, Hilo, HI 96720, USA
	\and
         		NASA Astrobiology Institute
       	\and
             	Deutsches SOFIA Institut, Universit\"at Stuttgart, Pfaffenwaldring 29, 70569 Stuttgart, Germany
         \and
	    	SOFIA Science Center, NASA-Ames Research Center, MS 232-12, Moffett Field, CA 94035, USA
          \and
             	European Southern Observatory, Karl Schwartzschild Str. 2, 85748 Garching bei M\"unchen, Germany	
	\and 
		Department of Astronomy \& Astrophysics, University of Toronto, 50 St. George Street, Toronto, ON M5S 3H4, Canada
	\and
		Max-Planck-Institut f\"ur Astronomie, K\"onigstuhl 17, 69117 Heidelberg, Germany
	\and
 	    	Landessternwarte, Zentrum f\"ur Astronomie der Universit\"at Heidelberg, K\"onigstuhl, 69117 Heidelberg, Germany
	\and
             	Universit\"ats-Sternwarte M\"unchen, Ludwig-Maximilians-Universit\"at, Scheinerstr. 1, 81679 M\"unchen, Germany
         \and
             	Dunlap Institute for Astronomy and Astrophysics, University of Toronto, 50 St. George Street, Toronto M5S 3H4, Ontario, Canada
         \and    
         	    	Department of Physics and Astronomy, UCLA, Los Angeles, CA 90095-1547, USA
	\and
	    	Institute of Geophysics and Planetary Physics, UCLA, Los Angeles, CA 90095-1565, USA        
             }

   \date{Received 31 October 2012 / Accepted 7 May 2013}

\abstract
{
Our general understanding of multiple star and planet formation is primarily based on observations of young multiple systems in low density regions like Tau-Aur and Oph. 
Since many, if not most, of the stars are born in clusters, observational constraints from young binaries in those environments are fundamental for understanding both the formation 
of multiple systems and planets in multiple systems throughout the Galaxy. 
}
{
We build upon the largest survey for young binaries in the Orion Nebula Cluster (ONC) which is based on Hubble Space Telescope observations 
to derive both stellar and circumstellar properties of newborn binary systems in this cluster environment. 
} 
{
We present Adaptive Optics spatially-resolved JHKL'-band photometry and K-band R$\sim$\,5000 spectra for a sample of 8 ONC binary systems from this database. 
We characterize the stellar properties of binary components and obtain a census of protoplanetary disks through K-L' color excess. 
For a combined sample of ONC binaries including 7 additional systems with NIR spectroscopy from the literature, we derive mass ratio and relative age distributions.
We compare the stellar and circumstellar properties of binaries in ONC with those in Tau-Aur and Oph from samples of binaries with stellar properties 
derived for each component from spectra and/or visual photometry and with a disk census obtained through K-L color excess.
}
{
The mass ratio distribution of ONC binaries is found to be indistinguishable from that of Tau-Aur and, to some extent, to that of Oph in the separation range 85-560\,AU and 
for primary mass in the range 0.15 to 0.8\,M$_{\sun}$. 
A trend toward a lower mass ratio with larger separation is suggested in ONC binaries which is not seen in Tau-Aur binaries. 
The components of ONC binaries are found to be significantly more coeval than the overall ONC population and as coeval as components of binaries in Tau-Aur and Oph. 
There is a hint of a larger fraction of mixed pairs, i.e. systems with a disk around only one component, in wide ONC binaries in comparison to wide binaries in Tau-Aur and Oph 
within the same primary mass range that could be caused by hierarchical triples. The mass ratio distributions of mixed and unmixed pairs in the overall population of T Tauri 
binaries are shown to be different. Some of these trends require confirmation with observations of a larger sample of binary systems.
}
{}

\titlerunning{}
\authorrunning{S. Correia et al.}

   \maketitle


%
\begin{table*}
\scriptsize
\caption{Observed sample.}
\begin{center}
\renewcommand{\arraystretch}{0.7}
\setlength\tabcolsep{11pt}
\begin{tabular}{l@{\hspace{4mm}}
			l@{\hspace{3mm}}
			r@{}
			r@{\hspace{4mm}}
			r@{\hspace{4mm}}
			r@{\hspace{4mm}}
			c@{\hspace{4mm}}
			c@{\hspace{2mm}}
			c@{\hspace{2mm}}
			c@{\hspace{2mm}}
			c@{\hspace{2mm}}
			c@{\hspace{2mm}}
			c@{\hspace{2mm}}
			c@{\hspace{4mm}}
			c@{\hspace{2mm}}
			c}

\hline\noalign{\smallskip}

Name 						& 
\multicolumn{1}{c}{R.A.} 			& 
\multicolumn{2}{c}{Decl.} 			& 
V 							& 
K 							& 
SpT$^a$ 						& 
Av$^a$ 						& 
Sep.$^b$ 						& 
P.A.$^b$ 						& 
$\Delta$m$_{H\alpha}$$^b$		& 
R$_{proj}$$^{b,c}$				&
Class$^d$					&
Ref.							&
\multicolumn{2}{c}{Observation date} \\

 	  						&   	    
\multicolumn{3}{c}{(J2000.0)}		&    
							&     
							&   
							&   
							& 
(arcsec) 						& 
(deg.) 						& 
							& 
(arcmin)						&
							&
							&
NIFS 						& 
NIRI 							\\
	  
\noalign{\smallskip}
\hline

\noalign{\smallskip}
\object{JW 63}       &  05 34 43.0 & -  & 05 20 07 	& 14.81 & 10.1 & K6   		& 1.58  	&  0.27 	& 0 		& 1.5		&  8.99  	& III	&  1		&  2008 Feb 16 	& 2008 Feb 16,19 \\
\noalign{\smallskip}
\object{JW 81}       &  05 34 46.4 & -  & 05 24 32 	& 14.95 & 10.5 & M0   		& 0.63 	&  1.34  	& 272 	& 1.9 	& 7.61 	& III	& 1 		&  2010 Nov 23 	& 2010 Sept 23 \\
\noalign{\smallskip}
\object{JW 128}     &  05 34 52.2 & -  & 05 22 32 	& 15.46 &  9.6  & M2.5 		& 1.13 	& 0.39  	& 215 	& $>$0.6  	&  6.14	& II	&  1  		& 2008 Feb 17 		& 2008 Feb 19 \\
\noalign{\smallskip}
\object{JW 176}     &  05 34 57.8 & -  & 05 23 53 	& 13.47 &  9.1  & K8 			& 0.45 	& 1.28  	& 336  	& $>$1.1  	& 4.71 	& II	& 1,2  	& 2008 Mar 07 		& 2008 Feb 24 \\
\noalign{\smallskip}
\object{JW 248}     & 05 35 04.4 & - & 05 23 14 		& 14.90 &  9.8  & M0.5-M2e 	& 0.75 	& 0.90 	& 320 	& $>$3.7 	& 3.04 	& II	& 1,2 	&  2010 Nov 23 	& 2010 Sept 21 \\
\noalign{\smallskip}
\object{JW 391}     & 05 35 12.8 & - & 05 20 44 		& 14.63 &  9.9  & M1e 		& 0.00 	& 0.29 	& 94  	& $>$3.3  	& 2.81  	& II	&  1  		& 2008 Feb 19 		& 2008 Feb 19 \\
\noalign{\smallskip}
\object{JW 709}     & 05 35 22.2 & - & 05 26 37 		& 15.13 & 10.2 & M0.5 		& 0.85 	& 0.22 	& 309 	&  0.5  	& 3.54 	& III	&  1  		& 2008 Feb 18 		& 2008 Feb 20 \\
\noalign{\smallskip}
\object{JW 867}     & 05 35 31.3 & - & 05 18 56 		& 14.12 & 9.5   & K8e 		& 0.56 	& 0.29 	& 212 	& $>$0.3 	& 5.78  	& II	&  1  		& 2008 Feb 23 		& 2008 Feb 23 \\
\noalign{\smallskip}

\hline
\end{tabular}
\end{center}
\label{Table : sample}
\begin{minipage}[position]{18cm}
$^a$\,: Unresolved spectral types and visual extinctions from Hillenbrand (\cite{Hillenbrand_1997}). \\   
$^b$\,: From Reipurth et al. (\cite{Reipurth_etal_2007}). \\   
$^c$\,: projected distance from $\theta^1$\,Ori\,C. \\
$^d$\,: YSO classification from Spitzer IRAC photometry made public in the first Orion data release of the 
YSOvar project\footnote{\tiny{http://ysovar.ipac.caltech.edu/first\_data\_release.html}} (Morales-Calder\'on et al. \cite{Morales-Calderon_etal_2011}). \\
References\,: (1) Reipurth et al. (\cite{Reipurth_etal_2007}), (2) Getman et al. (\cite{Getman_etal_2005}).
\end{minipage}
\end{table*}

\section{Introduction}
The prevalence of double or higher order multiple systems among low-mass stars is now a well established fact for main sequence field stars 
(e.g., Duquennoy \& Mayor \cite{Duquennoy_Mayor_1991}, Raghavan et al. \cite{Raghavan_etal_2010}), and there is growing evidence that 
young stellar populations generally harbor as much as twice as many physical companions (Reipurth \& Zinnecker \cite{Reipurth_Zinnecker1993}, Duch\^ene et al. 
\cite{Duchene_etal_2007}, Duch\^ene \& Kraus \cite{Duchene_Kraus_2013}). Significant differences between star-forming regions have been documented (Duch\^ene \cite{Duchene_1999}), 
and it remains unclear whether they stem from a different, more intense, dynamical evolution in dense clusters (like the Orion Nebula Cluster, ONC) 
or trace intrinsic variations in the output of the star formation process (K\"ohler et al. \cite{Koehler_etal_2006}, Reipurth et al. \cite{Reipurth_etal_2007}). 

In addition to the frequency of multiple systems (and its dependence on environment and primary mass) and to orbital characteristics like the 
separation distribution, stellar properties such as mass ratio and relative age distributions are able to give insight into the multiple star formation process 
(e.g. Hartigan et al. \cite{Hartigan_etal1994}, Brandner \& Zinnecker \cite{Brandner_Zinnecker_1997}, Duch\^{e}ne et al. \cite{Duchene_etal_1999}, 
Woitas et al. \cite{Woitas_etal_2001}, White \& Ghez \cite{White_Ghez_2001}, Hartigan \& Kenyon \cite{Hartigan_Kenyon2003}, Prato et al. \cite{Prato_etal2003}, 
Kraus \& Hillenbrand \cite{Kraus_Hillenbrand_2009a}, Kraus et al. \cite{Kraus_etal_2011}). The usefulness of such distributions to constrain formation models 
requires, however, determining the mass and age of young binary components in an unbiased way and with some accuracy, which implies the use of 
spectroscopy and multiwavelength photometry. 

While our understanding of the physical processes responsible for the high degree of stellar multiplicity among young stars remains partial (e.g. 
Tohline \cite{Tohline_2002}), this empirical fact has major consequences for theories of planet formation, which have been primarily developed 
in the context of isolated stars. Nevertheless, over 40 planet-hosting stars are members of multiple systems (Eggenberger \& Udry \cite{Eggenberger_Udry_2010}, 
and references therein), a substantial number when considering that binaries are usually excluded from planet surveys because of technical considerations. 
In fact, while there is marginal evidence that binaries tighter than $\sim$\,200\,AU are less likely to form planets (Eggenberger \& Udry \cite{Eggenberger_Udry_2010}), 
planet formation in such environments is not as rare as once thought. Instead, Duch\^ene (\cite{Duchene_2010}) has argued that planet formation follows a 
different path in tight binaries compared to wider binaries and single stars. 

To better understand the processes of star and planet formation in binary systems, it is necessary to draw a complete picture of the stellar properties 
and of circumstellar disks in pre-main sequence (T~Tauri) binary stars. Indeed, these disks are a ubiquitous outcome of star formation and serve as 
birthsite for planetary systems. 
In young binary systems, there is circumstantial evidence that the more massive primaries possess more massive and longer lived disks 
(e.g., Patience et al. \cite{Patience_etal_2008}). Although it is tempting to relate this to the finding that extrasolar planets in multiple systems are almost 
exclusively found around primaries, strong selection effects prevent us from reaching a definitive conclusion.

While spatially unresolved studies, such as recent Spitzer surveys, have revealed intriguing trends (most notably the shorter disk lifetime for tighter 
binaries, Cieza et al. \cite{Cieza_etal_2009}, see also Kraus et al. \cite{Kraus_etal_2012}), it is critical to analyze the disk properties of individual components 
of multiple systems. A major reason is that all the studies using spatially unresolved disk indicators yield intrinsically biased comparisons between the frequency of 
disks in binaries and in singles, since they can only attest to the presence of {\it at least} one disk in a binary system. In addition, of particular interest are 
the so-called "mixed" systems, i.e. pairs of two T Tauri stars with only one of the two components hosting a circumstellar disk. The occurrence of these systems 
and any correlation with binary properties such as mass ratio and separation distributions may provide critical clues to the early phases of star 
formation, as well as the dissipation of disks and the prospects for planet formation in binary systems. 
Early studies found that such mixed systems are quite rare (e.g. Prato \& Simon \cite{Prato_Simon_1997}, Duch\^{e}ne et al. \cite{Duchene_etal_1999}, 
White \& Ghez \cite{White_Ghez_2001}, Hartigan \& Kenyon \cite{Hartigan_Kenyon2003}). However, recent analysis of a broader compilation of multiple 
systems revealed a much higher fraction of mixed systems (Monin et al. \cite{Monin_etal_2007}). Most importantly, these authors concluded that there are 
substantial variations from one star-forming region to another in a way that cannot simply be explained by the age of the populations studied, again raising 
the possibility that star (and possibly planet) formation differs substantially from one cloud to the next. 

Virtually all the studies about the physical properties of binaries have focused on nearby clouds, mostly Taurus-Auriga and Ophiuchus. However, it has become 
clear that clusters like the ONC are at least as likely a representative mode of star and planet formation (e.g. Lada \& Lada \cite{Lada_Lada_2003}, Bressert et al. 
\cite{Bressert_etal_2010}). It is therefore important to revisit some of the "well-established" trends found in earlier studies in the light of a young dense cluster. 
The purpose of the present paper is to extend the previous studies about the stellar and circumstellar properties of pre-main sequence binaries to the ONC, 
the closest and prototypical young cluster. Specifically, we build on the hitherto largest ONC multiplicity survey by Reipurth et al. (\cite{Reipurth_etal_2007}, hereafter R07) 
using HST/ACS that has yielded the most complete census of binaries and higher-order multiples to date in the region.

%
\begin{figure*}
\begin{center}
\includegraphics[width=18cm, angle=0]{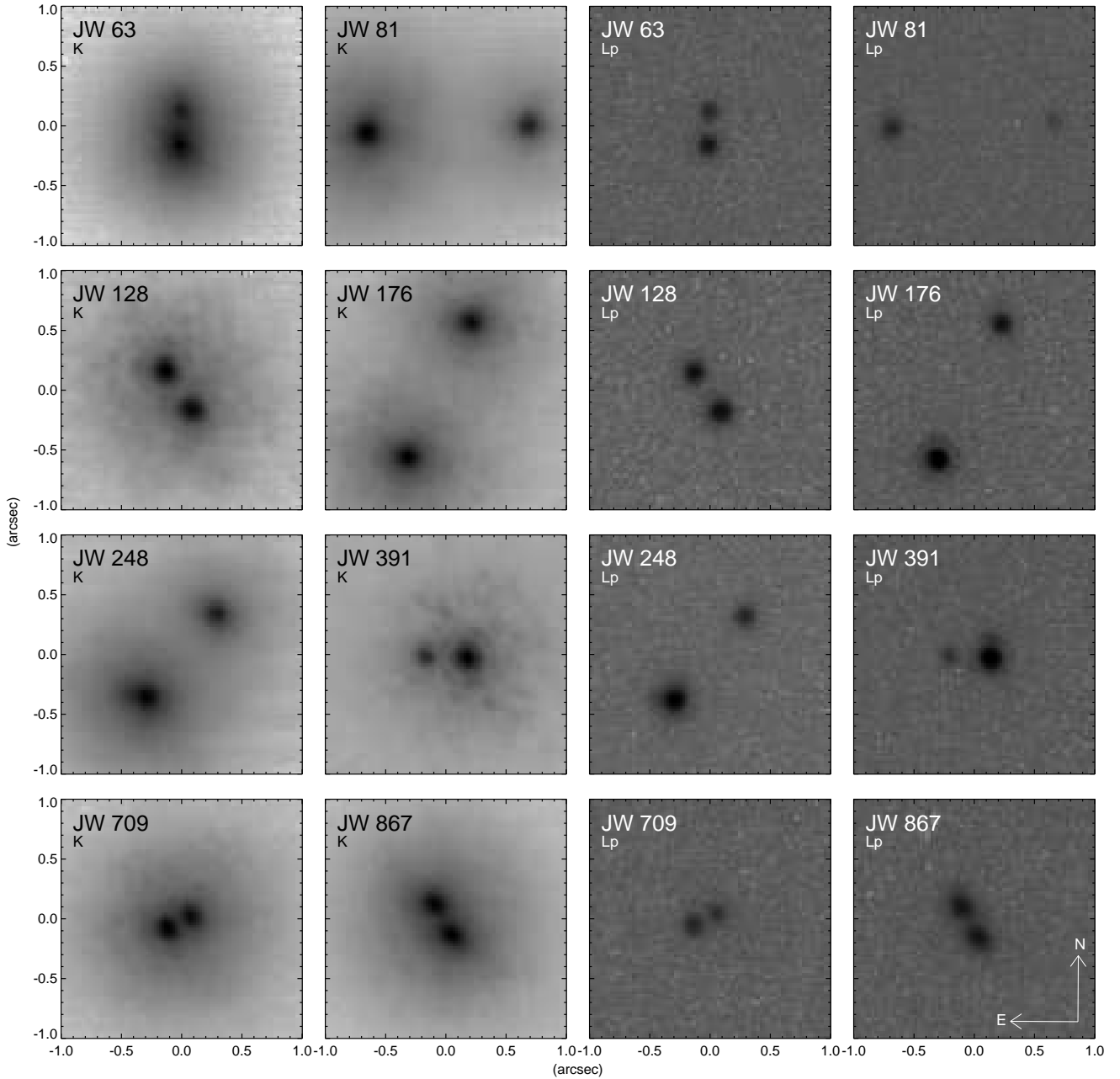}
\end{center}
\caption{K-band (two leftmost columns) and L'-band (two rightmost columns) NIRI imaging of the ONC binaries. 
The field of view is 2$\arcsec$\,$\times$\,2$\arcsec$ and images are displayed with a logarithmic stretch.}  
\label{fig: KL' images}
\end{figure*}

This paper is organized as follows\,: in Sect.\,\ref{sect:sample,obs_data_red} we present the sample of ONC binaries and describe the spectroscopic and 
photometric observations, while in Sect.\,\ref{sect:Astrometry and Photometry} we derive the spatially-resolved photometry and check the relative astrometry. 
The stellar properties are derived in Sect.\,\ref{sect:stellar properties} and the circumstellar properties in Sect.\,\ref{sect:circumstellar disks}. 
The results are used to investigate the distributions of mass ratio and of relative age in ONC binary components, as well as the occurrence of circumstellar disks. 
In Sect.\,\ref{sect:Discussion} we discuss the properties of the mass ratio distribution and the presence of disks around wide binary components in the ONC 
and compare them to those in Tau-Aur and Oph.

%
\begin{figure}
\begin{center}
\includegraphics[width=8.5cm, angle=0]{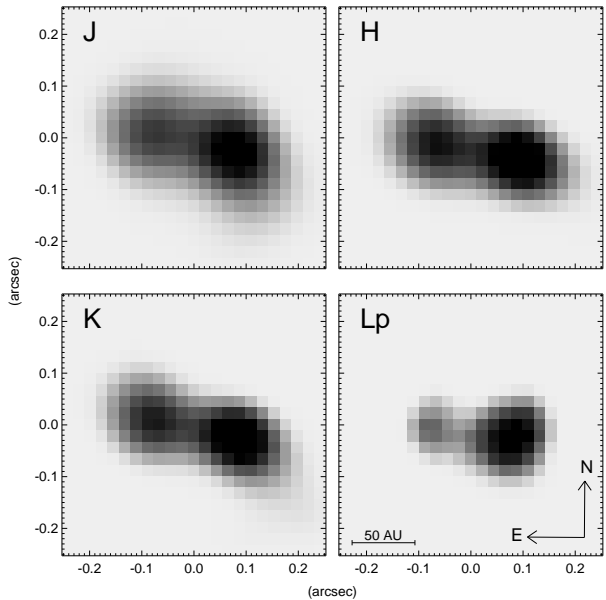}
\end{center}
\caption{Sharper images of JW\,248\,Aa-Ab obtained after 100 iterations of the multiple LR deconvolution algorithm using JW\,248\,B as the PSF.}
\label{fig: JW248 Aa-Ab}
\end{figure}

%
\begin{figure}
\begin{center}
\begin{tabular}{c}
\includegraphics[width=9cm, angle=0]{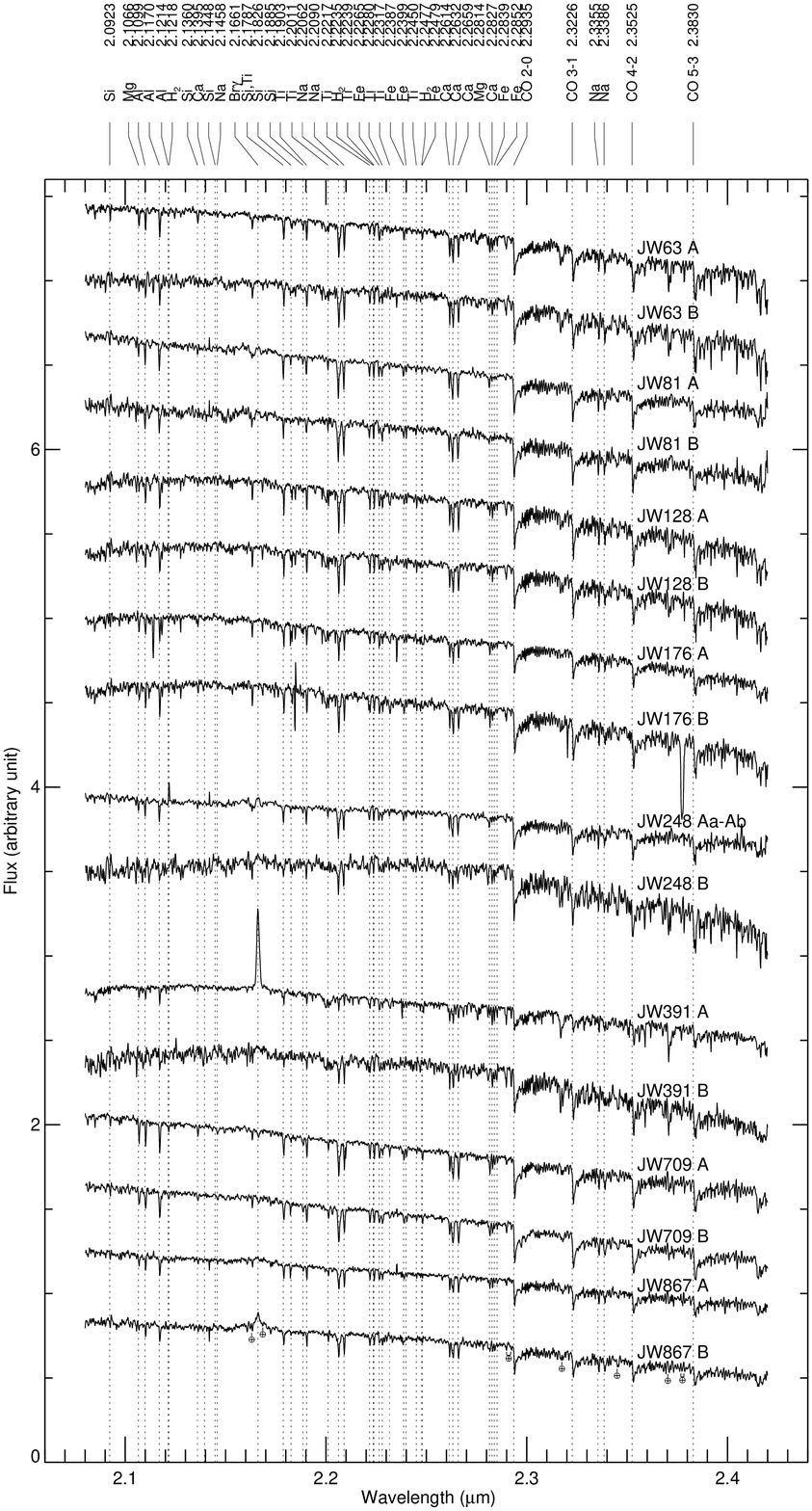}
\end{tabular}
\end{center}
\caption{K-band R\,$\sim$\,5000 spatially-resolved NIFS spectra of the ONC binaries. Atomic and molecular features are labeled, 
as well as the main telluric correction residuals.}
\label{fig: K-band spectra}
\end{figure}

%
\begin{table*}
\scriptsize
\caption{Astrometric and photometric measurements.}
\begin{center}
\renewcommand{\arraystretch}{0.7}
\setlength\tabcolsep{7pt}
\begin{tabular}{l@{\hspace{4mm}}
			r@{\,$\pm$\,}l@{\hspace{4mm}}
			r@{\,$\pm$\,}l@{\hspace{4mm}}
			c
			r@{\,$\pm$\,}l@{\hspace{4mm}}
			r@{\,$\pm$\,}l@{\hspace{4mm}}
			r@{\,$\pm$\,}l@{\hspace{4mm}}
			r@{\,$\pm$\,}l@{\hspace{4mm}}
			}

\hline\noalign{\smallskip}

Name 						&  
\multicolumn{2}{c}{Separation} 	& 
\multicolumn{2}{c}{P.A.}  			&   
Comp.						&
\multicolumn{2}{c}{J}  			&   
\multicolumn{2}{c}{H}  			&   
\multicolumn{2}{c}{K}  			&   
\multicolumn{2}{c}{L'}  			\\   

 	   						&   
\multicolumn{2}{c}{(arcsec)} 		& 
\multicolumn{2}{c}{(deg.)}  		& 
							&  
\multicolumn{2}{c}{(mag)}	  		&  
\multicolumn{2}{c}{(mag)}	  		&  
\multicolumn{2}{c}{(mag)}	  		&  
\multicolumn{2}{c}{(mag)}			\\

\noalign{\smallskip}
\hline

\noalign{\smallskip}
\object{JW 63}		&	0.2693   &   0.0023   &   358.1   &   0.5	&   A   &  11.52  &  0.03    &  10.74  &  0.04    &  10.53  &  0.03    &  10.43  &  0.02	\\
\noalign{\smallskip}
				&\multicolumn{4}{c}{}&   B   &  12.11  &  0.03    &  11.33  &  0.04    &  11.28  &  0.03    &  11.20  &  0.03 	\\
\noalign{\smallskip}
\noalign{\smallskip}

\noalign{\smallskip}
\object{JW 81}		&	1.3281   &   0.0023   &   272.6   &   0.1	&   A   &  11.71  &  0.03    &  10.96  &  0.04    &  10.73  &  0.04    &  10.63  &  0.07  	\\
\noalign{\smallskip}
				&\multicolumn{4}{c}{}&   B   &  13.15  &  0.03    &  12.42  &  0.04    &  12.13  &  0.04    &  12.02  &  0.08  	\\
\noalign{\smallskip}
\noalign{\smallskip}

\object{JW 128}	&	0.3853   &   0.0030   &   213.5   &   0.5	&   A   &  11.34  &  0.03    &  10.49  &  0.03    &  10.20  &  0.02    &   9.93  &  0.02	\\
\noalign{\smallskip}
				&\multicolumn{4}{c}{}&   B   &  11.58  &  0.03    &  10.78  &  0.03    &  10.39  &  0.02    &   9.65  &  0.02	\\
\noalign{\smallskip}
\noalign{\smallskip}

\object{JW 176}	&	1.2334   &   0.0029   &   335.0   &   0.1	&   A   &  10.77  &  0.03    &   9.83  &  0.03    &   9.52  &  0.02    &   8.67  &  0.02		\\
\noalign{\smallskip}
				&\multicolumn{4}{c}{}&   B   &  11.19  &  0.03    &  10.26  &  0.03    &  10.06  &  0.02    &   9.80  &  0.03	\\
\noalign{\smallskip}
\noalign{\smallskip}

\noalign{\smallskip}
\object{JW 248}	&	0.8971   &   0.0031   &   320.0   &   0.2	&   A   &  11.53  &  0.03    &  10.55  &  0.03    &   9.99  &  0.02    &   8.98  &  0.04  	\\
\noalign{\smallskip}
									&\multicolumn{4}{c}{}&   B   &  12.44  &  0.03    &  11.73  &  0.03    &  11.29  &  0.03    &  10.64  &  0.04  	\\
\noalign{\smallskip}
				&	0.1432   &   0.0016   &   78.9   &   1.1		&   Aa   &  11.59  &  0.04    &  10.60  &  0.04    &  10.03  &  0.03    &  9.03  &  0.05  	\\
\noalign{\smallskip}
									&\multicolumn{4}{c}{}&   Ab   &  14.57  &  0.18    &  13.82  &  0.12    &  13.58  &  0.13    &  12.51  &  0.33  	\\
\noalign{\smallskip}
\noalign{\smallskip}

\object{JW 391}	&	0.3378   &   0.0023   &    88.4   &   0.4	&   A   &  11.45  &  0.03    &  10.47  &  0.03    &   9.83  &  0.02    &   8.93  &  0.02		\\
\noalign{\smallskip}
				&\multicolumn{4}{c}{} &   B   &  13.12  &  0.03    &  12.36  &  0.03    &  11.95  &  0.02    &  11.63  &  0.03	\\
\noalign{\smallskip}
\noalign{\smallskip}

\object{JW 709}	&	0.2089   &   0.0029   &   296.8   &   0.8	&   A   &  11.80  &  0.03    &  11.00  &  0.03    &  10.72  &  0.03    &  10.53  &  0.09	\\
\noalign{\smallskip}
				&\multicolumn{4}{c}{}&   B   &  12.12  &  0.03    &  11.34  &  0.03    &  11.08  &  0.03    &  10.93  &  0.09	\\
\noalign{\smallskip}
\noalign{\smallskip}

\object{JW 867}	&	0.2798   &   0.0030   &    28.1   &   0.6	&   A   &  11.35  &  0.04    &  10.56  &  0.03    &  10.09  &  0.03    &   9.49  &  0.03	\\
\noalign{\smallskip}
				&\multicolumn{4}{c}{}&   B   &  11.40  &  0.04    &  10.67  &  0.03    &  10.22  &  0.03    &   9.53  &  0.03	\\
\noalign{\smallskip}

\hline
\end{tabular}
\end{center}
\label{Table : Astrometric and Photometric results}
\end{table*}

\section{Sample, observations and data reduction}
\label{sect:sample,obs_data_red}
\subsection{Sample}
\label{sect:sample}
Our sample of T~Tauri binaries in the ONC were selected from the list of Reipurth et al. (\cite{Reipurth_etal_2007}). 
The 8 binaries of the sample presented in Table\,\ref{Table : sample} are the optically brightest sources of that list 
which can be observed with the Adaptive Optics (AO) system Altair on Gemini North in natural guide star mode. 
The unresolved spectral types range from late-K to early-M and the separations from 0$\farcs$2 to 1$\farcs$3, 
i.e. 85-560\,AU at the distance of the ONC of 415\,pc (Menten et al. \cite{Menten_etal_2007}). 
As a first step towards building a relatively large sample of ONC binaries in order to obtain enough statistics 
to study their stellar and disk properties, we complement our analysis with a sample of 7 ONC binaries 
from a similar recent spatially-resolved spectro-photometric study (Daemgen, Correia \& Petr-Gotzens \cite{Daemgen_etal_2012}). 
These 7 additional systems are among the brightest systems found in the small and localized NIR surveys of Petr et al. (\cite{Petr_etal_1998}) 
and K\"ohler et al. (\cite{Koehler_etal_2006}), some of them also detected in the optical R07 survey. All together our sample amounts to 15 T~Tauri binaries, 
which represent a 10--15\,\% fraction of the known population of ONC binaries with separations between a few tens of AU and $\sim$\,500 AU.

\subsection{Observations and data reduction}
\label{sect:obs_data_red}
We obtained spatially-resolved J,H,K and L'-band photometry and R$\sim$\,5000 K-band spectra for each individual 
component of the 8 close binaries listed in Table\,\ref{Table : sample} using the Near IR Integral Field Spectrograph (NIFS) 
and the Near IR Imager (NIRI) at the Gemini North Telescope with both instruments fed by the AO system Altair 
(McGregor et al. \cite{McGregor_2002}).

\subsubsection{NIRI J,H,K and L'-band imaging}
\label{sect:NIRI}
NIRI is a near-IR imager using a 1024\,$\times$\,1024 Aladdin InSb array (Hodapp et al. \cite{Hodapp_etal_2003}). 
The f/32 mode was used in all four filters with a plate scale of 0$\farcs$0218 pixel$^{-1}$ and an orientation of 
P.A. = 0\,$\pm$\,0$\fdg$05 (Beck et al. \cite{Beck_etal_2004}) providing a field of view (FOV) of $\sim$\,22$\arcsec$\,$\times$\,22$\arcsec$.
In L' the faster read-out mode required the use of a subarray reducing the FOV by half. Except for JW\,63, the observations 
in all filters were done consecutively in a single night. For JW\,63, the JHK observations were executed three days before the 
L' observations (see Table\,\ref{Table : sample}). Observing conditions were for the most part photometric and with good seeing. 
The on-source integration time ranged from 50 to 200\,s in J, 43.8 to 200\,s in H, 15 to 200\,s in K, and 333 to 804.8\,s in L'. 
The JHK photometric standards GSPC S840-F (J=11.365\,$\pm$\,0.016, H=11.099\,$\pm$\,0.013, K=11.019\,$\pm$\,0.009, 
MKO photometric system, Leggett et al. \cite{Leggett_etal_2006}) and HD\,289907 (J=9.857\,$\pm$\,0.007, H=9.818\,$\pm$\,0.010, 
K=9.810\,$\pm$\,0.010, MKO photometric system, Leggett et al. \cite{Leggett_etal_2006}) and the L' photometric standards 
HD\,287736 (L'=8.559\,$\pm$\,0.010, MKO photometric system, Leggett et al. \cite{Leggett_etal_2003}) and HD\,40335 (L'=6.441\,$\pm$\,0.025, 
MKO photometric system, Leggett et al. \cite{Leggett_etal_2003}) 
were observed close in time (typically within 30\,min) and airmass (within $\lesssim$\,0.2 airmass) to the targets. The difference in 
airmass between targets and photometric standards is distributed with a mean and standard deviation of 0.06 and 0.05, respectively, 
which is small enough to neglect extinction correction. Due to an error in the preparation of the observations, GSPC S840-F was used 
as L' photometric standard for JW\,709. In this case we estimated the L' magnitude of GSPC S840-F from its K magnitude. Since the K-L' 
color of this star should be no more than 0.05 mag (from Bessell \& Brett \cite{Bessel_Brett_1988}) as it is a G-type, around G5 (B-V = 0.69, 
Lasker et al. \cite{Lasker_etal_1988}), we set L'= 10.969\,$\pm$\,0.050. Lamp flat and dark fields were also obtained. 

The data were reduced using the NIRI tasks in the Gemini IRAF package version 1.10. A sky frame derived from the on-source raw frames 
was subtracted from the raw frames. The frames were subsequently divided by the lamp flat field taken from the calibration unit in JHK or 
by a sky flat field in L' which was derived from the on-source frames.  The resulting images were then average-combined excluding the bad pixels 
identified in the dark frames. The images of the binary systems in K and L'-band are shown in Fig.\,\ref{fig: KL' images}. JW\,248\,A turned out to 
have a close ($\sim$\,0$\farcs$15 separation) faint ($\sim$\,3mag difference) visual companion. This companion is a discovery since it is 
not seen on the HST-ACS images probably due to a large flux ratio in the visible between the primary and the close companion candidate. 
Because of the relative faintness and closeness of the companion in our AO images, we sharpened the images at each wavelength using multiple 
Lucy-Richardson (LR) deconvolution of each single frame (e.g. Correia \& Richichi \cite{Correia_Richichi_2000})  with JW\,248\,B taken as the 
Point-Spread Function -- PSF (Fig.\,\ref{fig: JW248 Aa-Ab}).

\subsubsection{NIFS R$\sim$\,5000 K-band spectroscopy}
\label{sect:NIFS}
NIFS is an image-slicing integral field unit (IFU). 
The NIFS field is 3$\arcsec$\,$\times$\,3$\arcsec$ and the individual IFU pixels are 0$\farcs$1\,$\times$\,0$\farcs$04 on the sky. 
Seeing conditions were in the range 0$\farcs$3-1$\farcs$0. The signal-to-noise was built up using single exposure times 
of 60\,s to 230\,s per frame, 4 to 24 frames, and 50\% of time on-source, i.e. 10 to 20\,min effective integration time per source. 
The instrument was rotated along the PA of the binary in order to have the smallest pixel scale along the binary separation.  
 
The data were reduced using the NIFS tasks in the Gemini IRAF package version 1.10. For each observation, a standard set of 
calibrations were acquired using the Gemini facility calibration unit, GCAL (see details in Beck et al. \cite{Beck_etal_2008}). The raw 
IFU frames were sky-subtracted using images of a nearby blank sky field, flat-fielded, and cleaned out from residual bad pixels. 
A spatial rectification was applied using a Ronchi calibration mask and the data were wavelength calibrated with an arc exposure 
using Ar and Xe emission lamps taken shortly after the observations. Telluric standards were observed and their data reduced following 
the same procedure. Each target spectra were extracted through the data cubes with the IRAF task {\tt nfextract} using 0$\farcs$1 
and 0$\farcs$25 radius circular apertures depending on the binary separation. JW\,248\,Aa-Ab was not resolved and therefore considered 
as a composite spectrum. This composite spectrum is assumed to be that of JW\,248\,Aa given the faintness of JW\,248\,Ab. Telluric spectra 
were extracted using 0$\farcs$25 radius apertures. The individual spectra were median combined, the resulting spectrum of each target 
divided by the corresponding telluric spectrum and multiplied by a blackbody spectrum of appropriate temperature. 

HIP\,22028 (A1V) and HIP\,30594 (A0V) were used as telluric standard for JW\,63, 128, 176, 391, 709 and JW\,81, 248, 867, respectively. 
The spectra of HIP\,22028 appeared to be time-variable, showing significant changes in continuum shape and occasionally emission-lines 
which cannot be attributed to atmospheric transmission variability. While the continuum of HIP\,22028 for the observations corresponding 
to JW\,63 and 709 were similar and well matched by a T=9500\,K blackbody, the continuum of HIP\,22028 during the observations of JW\,128, 
176, and 391 changed locally dramatically by up to 20\,\%. Moreover, emission-lines appeared which were also time-variable from one individual spectrum to another. 
The continuum of HIP\,22028 was therefore fitted for JW\,128, 176, and 391 with a 10th order Legendre polynomial with the IRAF task {\tt continuum} 
and used in place of the blackbody spectrum to obtain the atmospheric transmission. We could also interpolate over the few strongest emission-lines. 
The Brackett gamma line of the telluric standards was removed using a Gaussian fit, except for JW\,709 and 867 for which a linear interpolation 
gave better results (i.e. less residuals because the line profile is more peaked than a Gaussian). The resulting spectra are shown in 
Fig.\,\ref{fig: K-band spectra} with the strongest atomic and molecular features labeled (compiled and cross-checked from several sources, 
notably Wallace \& Hinkle \cite{Wallace_Hinkle_1996}, Wallace \& Hinkle \cite{Wallace_Hinkle_1997} and Hinkle, Wallace \& Livingston \cite{Hinkle_Wallace_Livingston_1995}).


\section{Photometry and Relative astrometry}
\label{sect:Astrometry and Photometry}

Photometry and relative astrometry of the binary systems were performed using the tasks {\tt phot} and {\tt mkapfile} 
from the IRAF package APPHOT. The position of each system's component is defined as the centroid of the brightness distribution 
within a box of 5 pixels width. At each wavelength, the error in the position measurement is the rms variation in our astrometric 
analysis with a minimum of 0.1\,pixel. The final values of separation and position angle are the quadratic averages over the wavelengths, 
and the errors are the quadratic averages of the measurements errors which are added quadratically to the uncertainty in the plate scale 
and detector orientation. Since the plate scale of NIRI has not yet been characterized, we assumed a rather conservative plate scale 
uncertainty of 1\,\%. Table\,\ref{Table : Astrometric and Photometric results} presents our resulting measurements.

%
\begin{figure}
\begin{center}
\includegraphics[width=8.5cm, angle=0]{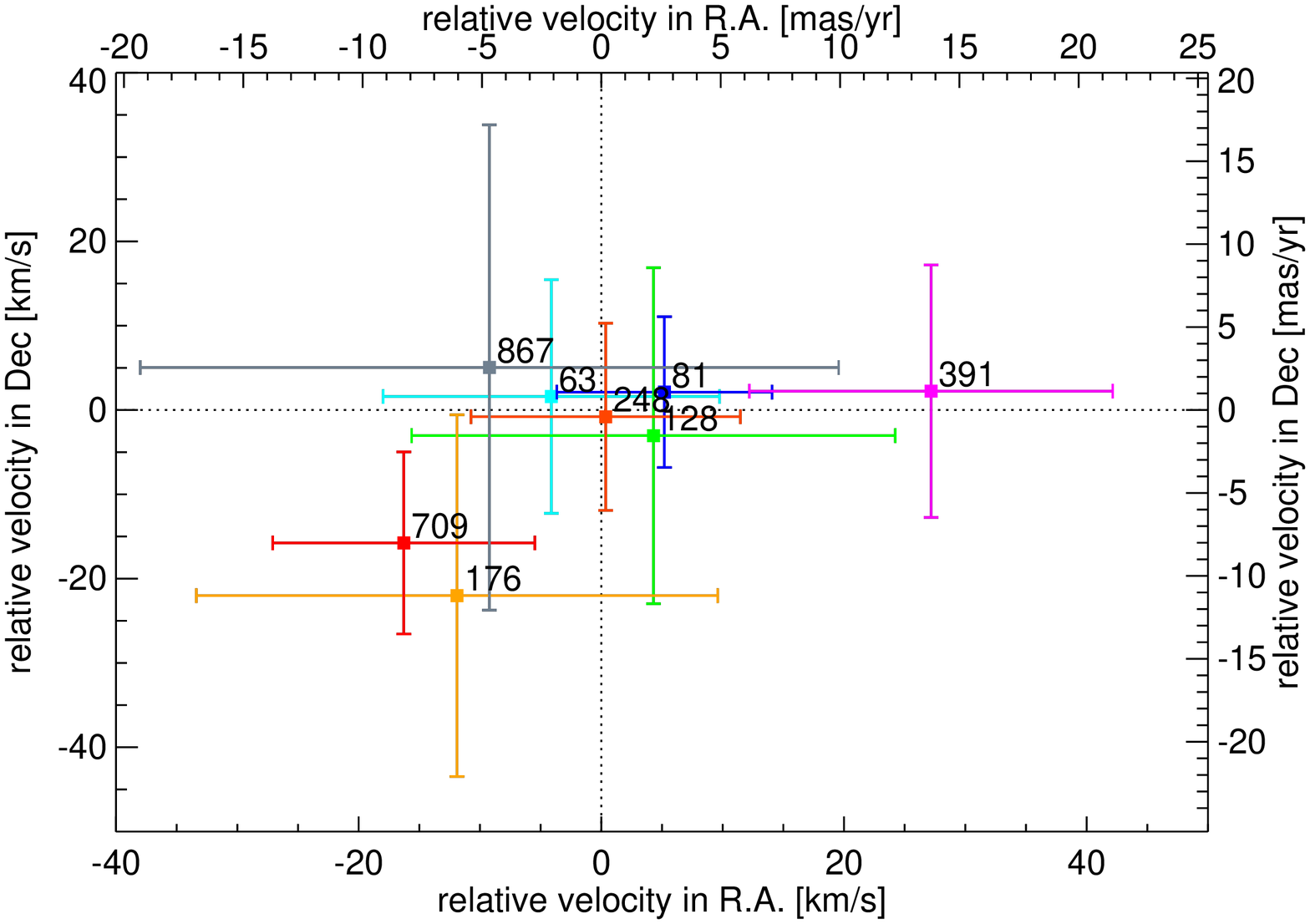} \\
\includegraphics[width=8.5cm, angle=0]{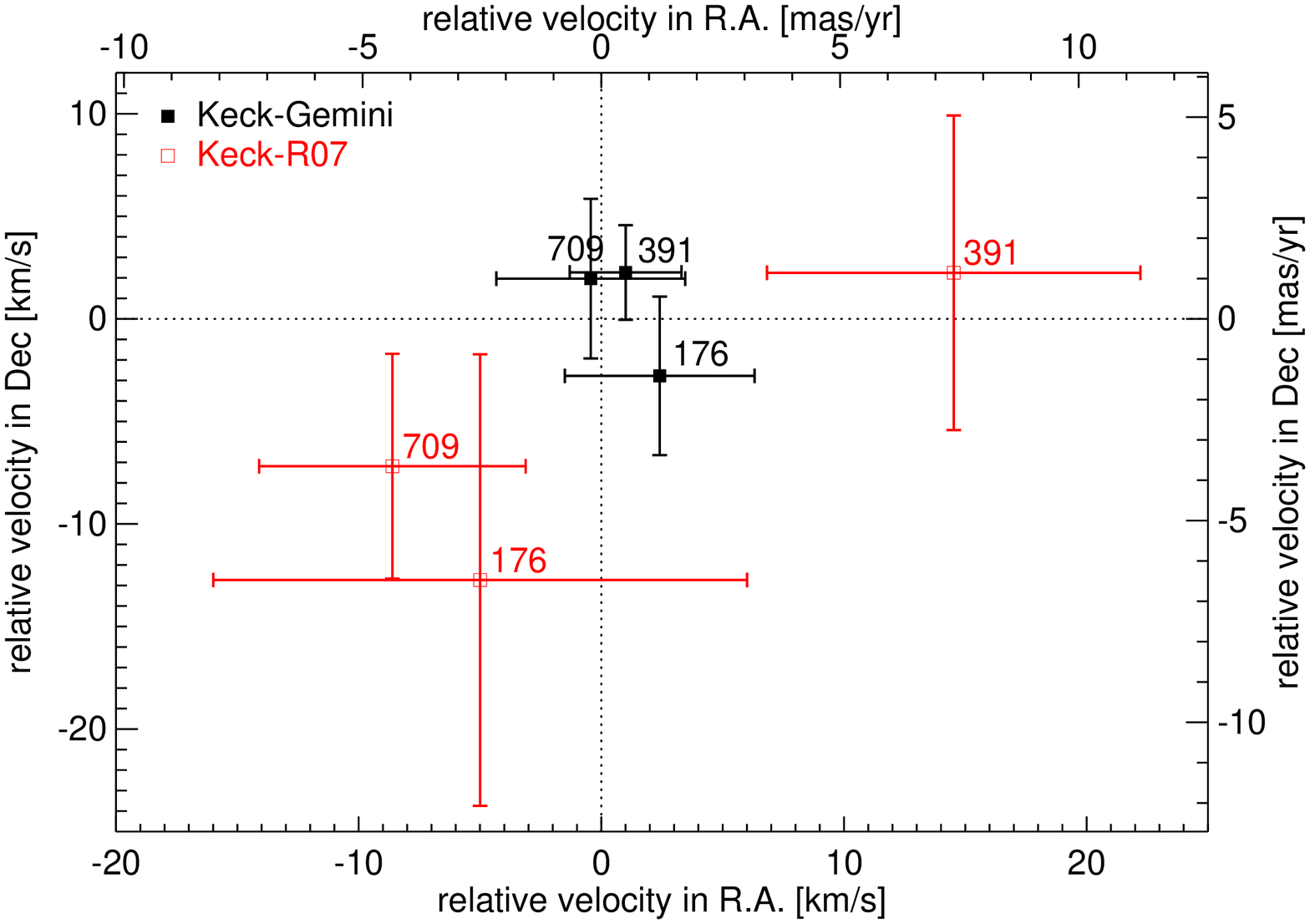}
\end{center}
\caption{ Top\,: Relative velocity of the binary components in RA and Dec between the positions from Reipurth et al. (\cite{Reipurth_etal_2007}-R07) 
reported in Table\,\ref{Table : sample} and those measured in this study (our measured positions minus R07 positions). 
Bottom\,: Relative velocity between Keck and Gemini measurements (Keck-Gemini), as well as between Keck and R07 measurements 
(Keck-R07) for JW\,176, 391, and 709. The small relative motions between Keck and Gemini epochs are consistent with bound systems within the errors.}
\label{fig: Variation_linear_vel_RA_DEC_with_R07}
\end{figure}

%
\begin{table}
\scriptsize
\caption{Additional astrometric measurements from Keck performed on 2011 Dec 16.}
\begin{center}
\renewcommand{\arraystretch}{0.9}
\setlength\tabcolsep{9pt}
\begin{tabular}{l@{\hspace{4mm}}
			r@{\,$\pm$\,}l@{\hspace{4mm}}
			r@{\,$\pm$\,}l@{\hspace{4mm}}
			}

\hline\noalign{\smallskip}

Name 						&  
\multicolumn{2}{c}{Separation} 	& 
\multicolumn{2}{c}{P.A.}  			\\   

 	   						&   
\multicolumn{2}{c}{(arcsec)} 		& 
\multicolumn{2}{c}{(deg.)}  		\\

\noalign{\smallskip}
\hline

\noalign{\smallskip}
\object{JW\,176}	&	1.2281   &   0.0026   &   335.2   &   0.1	\\
\noalign{\smallskip}
\noalign{\smallskip}

\object{JW\,391}	&	0.3377   &   0.0021   &    87.6   &   0.4	\\
\noalign{\smallskip}
\noalign{\smallskip}

\object{JW\,709}	&	0.2125   &   0.0027   &   297.2   &   0.7	\\
\noalign{\smallskip}

\hline
\end{tabular}
\end{center}
\label{Table : Keck Astrometric and Photometric results}
\end{table}

We compared the separations and position angles with those measured in Reipurth et al. (\cite{Reipurth_etal_2007}) 
from HST-ACS images and reported in Table\,\ref{Table : sample}. The upper plot of Fig.\,\ref{fig: Variation_linear_vel_RA_DEC_with_R07} 
shows the relative velocity between the two measurements in R.A. and Declination. 
Uncertainties of the HST-ACS astrometry were considered quite conservatively and include a position measurement error of 5\,mas (0.1\,pixel) 
for each component and 15\,mas in case the component is saturated. The detector scale and orientation uncertainties were assumed to be 
2.45\,mas (5\,\% of plate scale) and 0.05$^{\circ}$, respectively. 
The majority of the systems are consistent with no detectable relative motion between the two epochs, as expected for physical systems. 
As a comparison, the Keplerian velocity of a solar mass binary with a circular orbit in the plane of the sky would be in range of 1 to 3\,km\,s$^{-1}$ 
for this range of system separations (90 to 560\,AU). JW\,176, 391, and 709 are outliers at 2\,$\sigma$. 
The large relative velocities found for these systems could however be the result of the uncertainties from the measurements performed on the HST-ACS images 
due to small separations and large flux ratios and/or saturated components. In order to further explore this possibility, we obtained an additional 
epoch for these systems using AO-imaging with NIRC2 on Keck. The Keck observations were performed on 2011 Dec 16 therefore providing 
a $\sim$\,3.5\,yr time baseline with respect to the Gemini observations, as well as a similar astrometric accuracy. 
Relative astrometry was extracted in the same way as for the Gemini measurements, considering plate scale and orientation uncertainties of 
0.06\% and 0.02$^{\circ}$, respectively (Ghez et al. \cite{Ghez_etal_2008}). The derived astrometry is reported in 
Table\,\ref{Table : Keck Astrometric and Photometric results} and the lower plot of Fig.\,\ref{fig: Variation_linear_vel_RA_DEC_with_R07} shows 
the relative velocity between these measurements and those from Gemini and HST-ACS data. 
The lack of significant relative motions between the Keck and the Gemini epochs indicates that the above outliers were indeed the result of biases 
in the HST images. While chance projection with a cluster member cannot in principle be excluded given the accuracy of these relative 
velocities and the velocity dispersion of the ONC ($\sim$\,2.5\,mas\,yr$^{-1}$ i.e. $\sim$\,5\,km\,s$^{-1}$), 
we conclude that all systems are consistent with being bound systems within the uncertainties. 

Relative photometry of the individual binary components were performed with aperture radii of 0$\farcs$11 (5 pixels), except for JW\,248\,Aa and Ab 
where they were 0$\farcs$055 (2.5 pixels). The derived binary flux ratios were combined with the aperture photometry of the entire systems to 
determine the individual instrumental magnitudes. Aperture photometry was also employed for the photometric standards. In both cases, small 
apertures were used and distinct aperture corrections applied. The size of the apertures for the targets were the minimum radii to encompass both 
components.  For the standards, they were smaller and generally close to the optimum aperture size (4-5 times FWHM). The convergence of the 
curves of growth determined the values of the outer radius considered in the aperture corrections. The latter ranged from $\sim$\,1$\farcs$5 to 
$\sim$\,5$\farcs$0, depending on the wavelength of observation, the brightness of the source, the AO-correction, the presence of neighboring 
sources, and the binary separation in the case of the targets. The final errors for the component magnitudes take into account the errors in the 
instrumental magnitudes, the flux ratios, the zero points, the aperture corrections, and the standard star magnitudes.

%
\begin{figure}
\begin{center}
\includegraphics[width=8.5cm, angle=0]{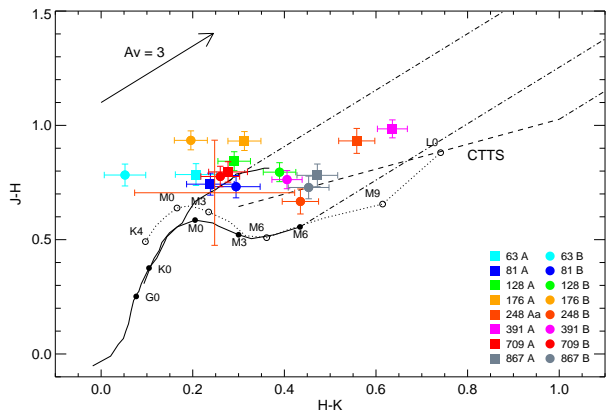}
\end{center}
\caption{The locus of the components of the ONC binaries in a J-H/H-K color-color diagram. Solid lines are the locus of photospheric dwarf 
and giant colors from Bessell \& Brett (\cite{Bessel_Brett_1988}), while in dotted lines are shown the colors of young stellar photospheres of 
given spectral type in the range K4-L0 (Luhman et al. \cite{Luhman_etal_2010}). The dashed line is the CTTS locus from Meyer et al. 
(\cite{Meyer_etal_1997}). The data point with large uncertainties is JW\,248\,Ab. All data are in the MKO photometric system.}
\label{fig: JH-HK color-color diagram}
\end{figure}

We compared our JHK magnitudes with those derived using a photometric calibration against the Two Micron All Sky Survey (2MASS) 
and converting our photometry from the MKO to the 2MASS photometric system with the color transformations  from Carpenter et al. 
(2001\footnote{and update for the final 2MASS data release.}). A 1\,$\sigma$ scatter of 0.08, 0.05, and 0.06\,mag was found in J, H, and K-band, 
respectively, which is consistent with the combined (2MASS and our) photometric errors. The largest deviations of $\sim$\,0.2\,mag may also 
be explained by photometric variability. JW\,391 exhibits the maximum deviation in all JHK bands followed by JW\,176 and 867 with 
$\sim$\,0.15-0.10\,mag in H and K. For all these systems active disk accretion is detected through the presence of Br$\gamma$ emission 
and/or K-L color excess (Sect.\,\ref{sect:circumstellar disks}) which could explain such photometric variability. It is interesting to note that 
the 2MASS magnitudes are found to be systematically fainter than our magnitudes by 0.07, 0.12, and 0.07\,mag in J, H, and K-band, respectively. 
These differences cannot be explained by photometric variations but are most probably a result of photometric measurement methods. 
There are essentially no systematics in colors between our photometry and that derived from 2MASS except for JW\,63 and JW\,176 
which are found to be clear outliers with HK$_{\mathrm{2MASS}}$-HK\,$\sim$\,0.15 and JH$_{\mathrm{2MASS}}$-JH\,$\sim$\,$-$0.2, 
respectively. The JH-HK color-color diagram of the binary components (Fig.\,\ref{fig: JH-HK color-color diagram}) shows the unusual location 
of these two systems. While a genuine color variation and/or contribution from scattered-light emission cannot be excluded, 
it is likely that this is due to a rapid variation in the sky transparency in H-band during the observation of these two systems.  

Astrometry of JW\,248\,Aa-Ab was performed on the deconvolved images. The values of separation and position angle were measured at 
each iteration of the LR algorithm in order to ensure convergence of these parameters which was observed to take place after typically 3000 
iterations. The values and errors for each wavelength are the mean and rms of the values of the individual frames, and the final astrometry 
is the weigthed mean and standard deviation of the weighted mean of the astrometry over the JHK filters. In L'-band, only the co-added image 
was used. Because of possible biases introduced by the deconvolution process on the relative photometry, the flux ratio was estimated 
employing PSF-subtraction using the relative positions determined by the deconvolution. As before, the individual frames in JHK were 
used to derive the mean values and uncertainties. The JW\,248\,Aa-Ab pair is likely to be physically bound given the small separation and 
the low density of sources brighter than the brightness of the companion in the field. Using the 2MASS catalog in a field of 1.5~arcmin radius, 
and given the separation and magnitude of JW\,248\,Ab, the probability of a chance projection is $\sim$\,2\,$\times$\,10$^{-4}$ in JHK. 
Although the H-K color is too blue but possibly contaminated by scattered-light emission (Fig.\,\ref{fig: JH-HK color-color diagram}), 
the JHKL' magnitudes and K-L' color (1.07\,$\pm$\,0.35) are suggestive of a late-M substellar source if bound.


%
\begin{figure*}
\begin{center}
\begin{tabular}{c}
\includegraphics[width=16cm, angle=0]{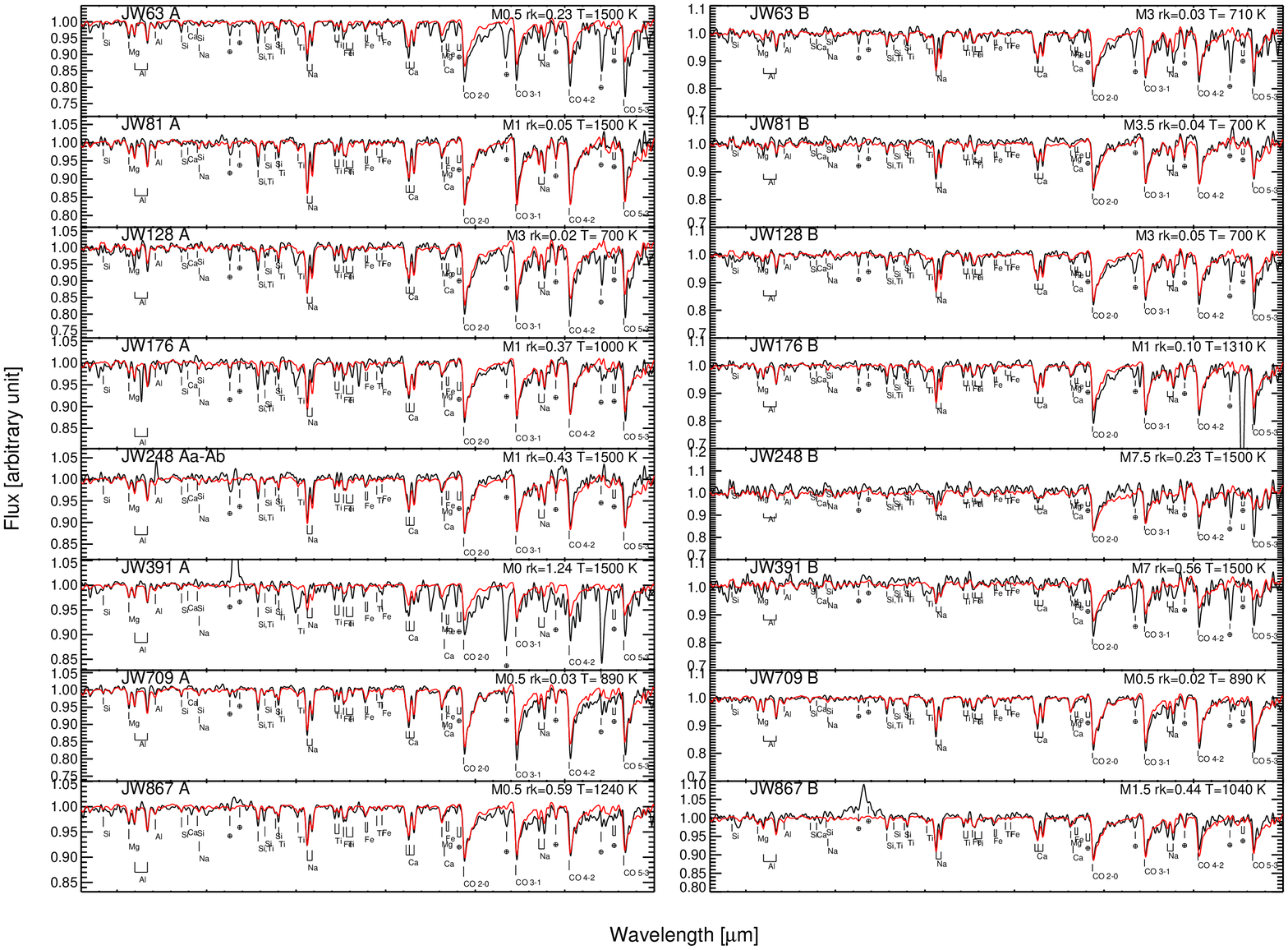} \\
\includegraphics[width=16cm, angle=0]{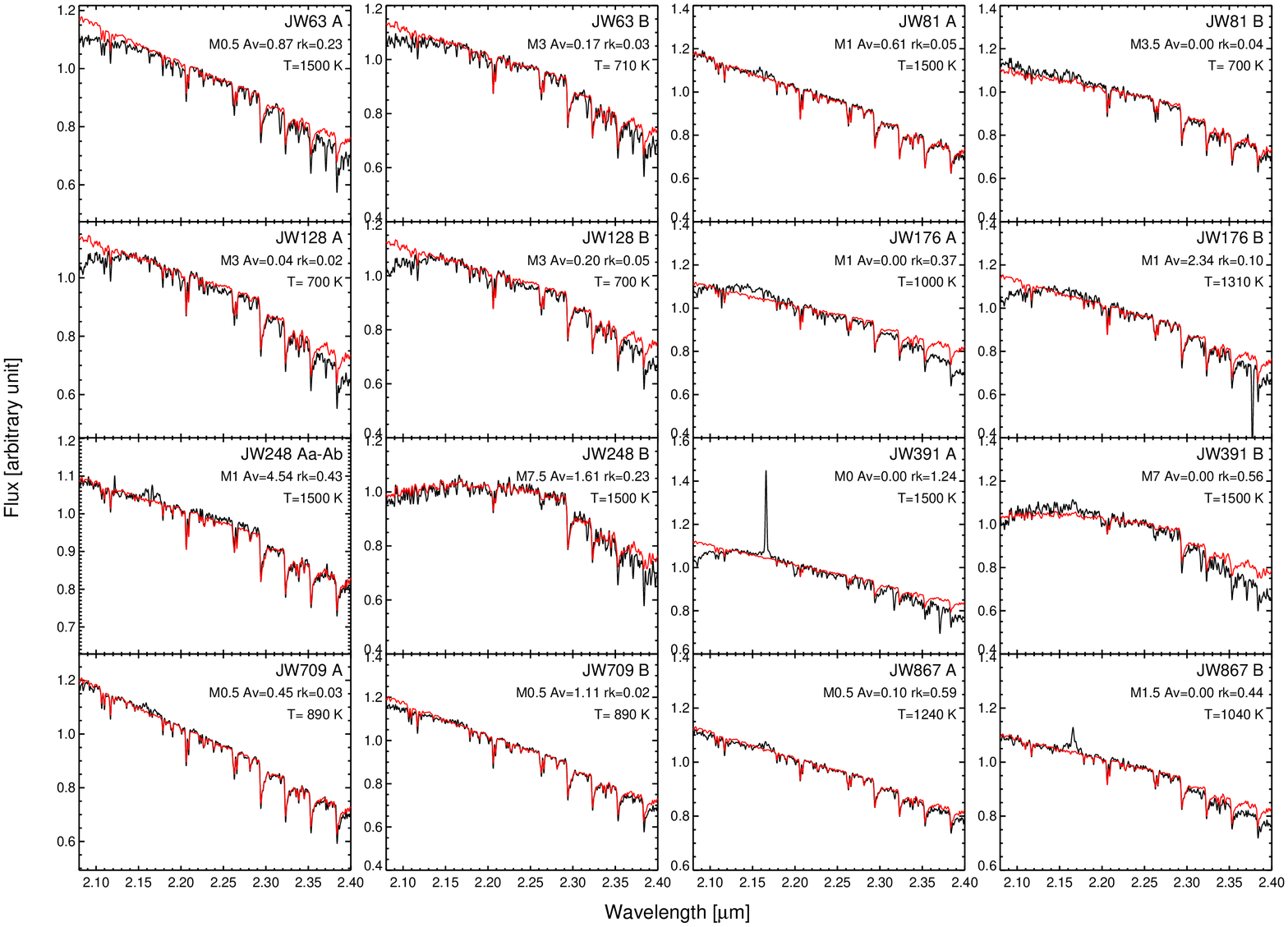}
\end{tabular}
\end{center}
\caption{Continuum-normalized spectra (upper panel) and non-continuum-normalized spectra (lower panel). The best-fit models (red) 
are also plotted and the derived spectral types, visual extinction A$_{\mathrm{V}}$, K-band IR-excess r$_{\mathrm{K}}$, and its blackbody 
temperature T reported. Atomic and molecular features are labeled, as well as the main uncorrected telluric features.}
\label{fig: spectra with best-fit models - cont-norm and non-cont-norm}
\end{figure*}

\section{Stellar properties}
\label{sect:stellar properties}
For each binary component we derived spectral type, visual extinction A$_{\mathrm{V}}$, infrared-excess in K-band r$_K$, 
bolometric luminosity, mass, and age. Our method to derive spectral type, visual extinction, and infrared-excess is a composite 
of several methods and the results for each of them are shown in Table\,\ref{Table : stellar parameter - intermediate results}. 
The adopted values are reported in Table\,\ref{Table : stellar parameter}.

\subsection{Spectral Type, Extinction, and Infrared Excess}
\label{sect:spectral type, extinction, and infrared excess emission}

\begin{figure}
\begin{center}
\begin{tabular}{c}
\includegraphics[width=8.5cm, angle=0]{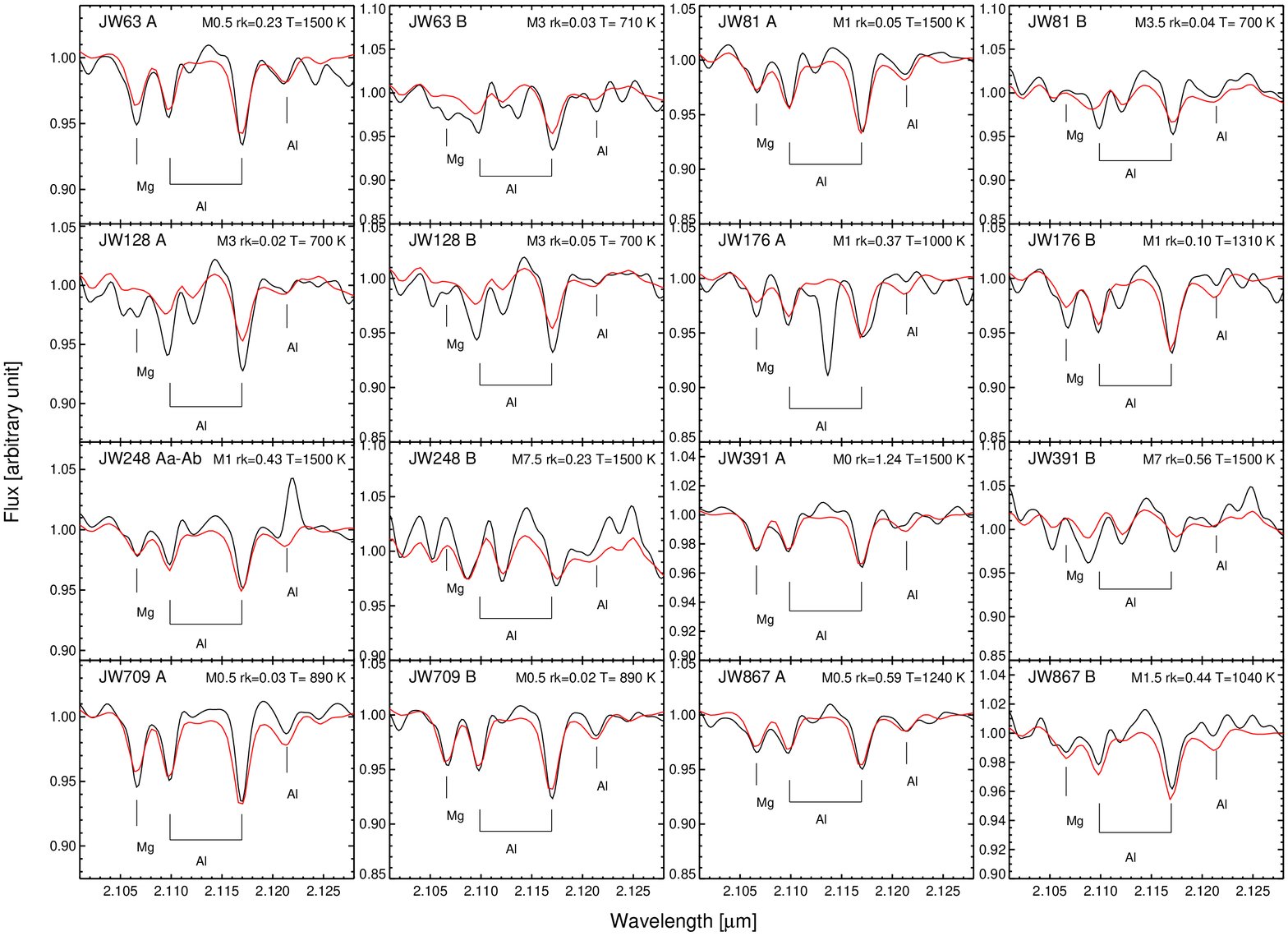} \\
\includegraphics[width=8.5cm, angle=0]{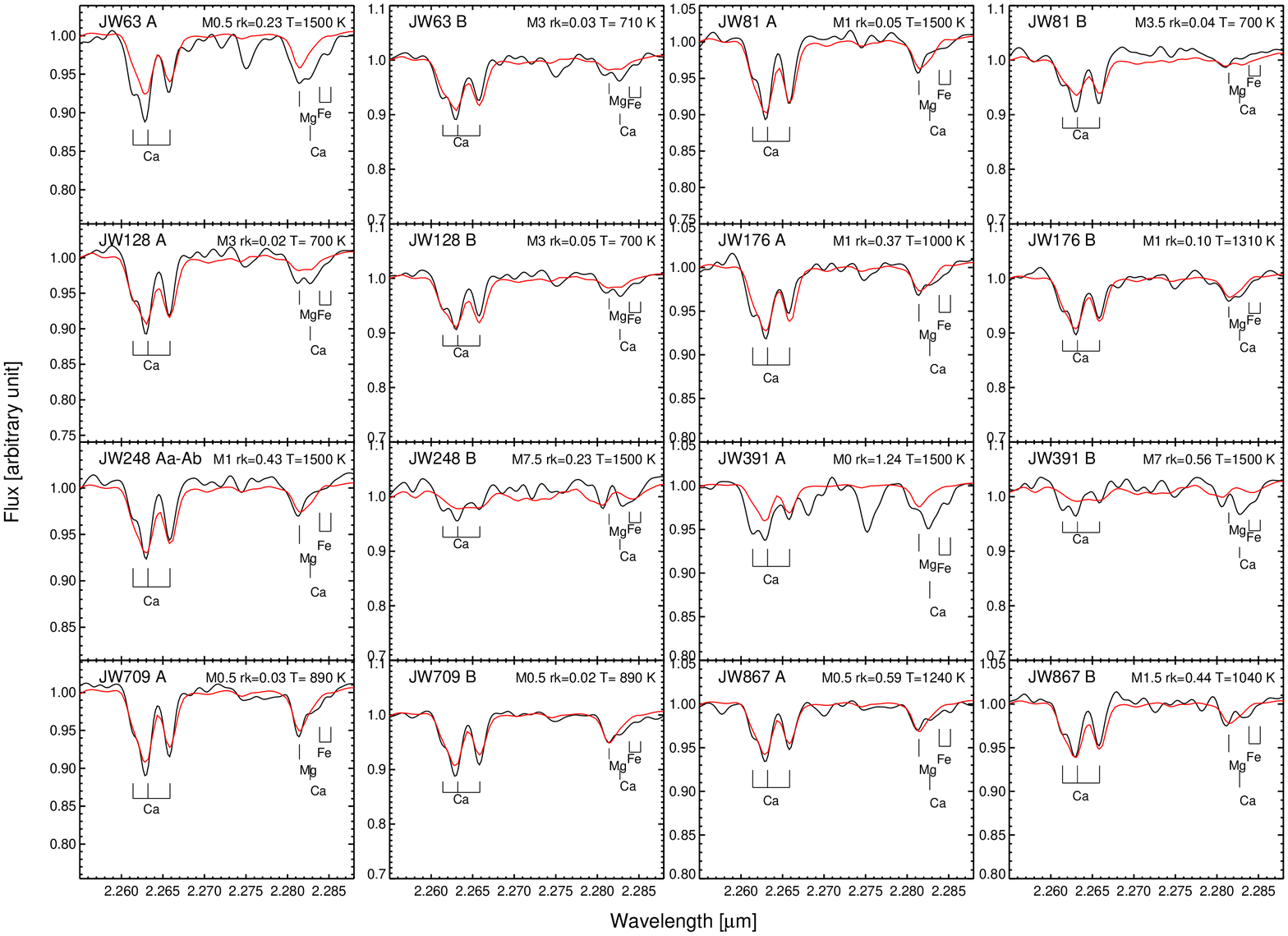}
\end{tabular}
\end{center}
\caption{Continuum-normalized spectra with best-fit models around the Mg 2.1066 and Al 2.1169\,$\mu$m features (upper panel) 
and around the Mg 2.2814 and Ca 2.2634\,$\mu$m features (lower panel).}
\label{fig: spectra with best-fit models - spectral regions}
\end{figure}

The spectral typing made use of both a line ratio analysis of temperature-sensitive photospheric features and a spectral template 
fitting scheme which additionally provided K-band excess or veiling r$_{\mathrm{K}}$ and visual extinction A$_{\mathrm{V}}$. r$_{\mathrm{K}}$ 
is defined as the ratio of the infrared excess at 2.22\,$\mu$m to the intrinsic stellar continuum flux. In addition, the spectral shape 
of the K-band excess is assumed to follow a blackbody function of temperature T. Our approach was the following. 
We obtained a first estimate of the spectral type, A$_{\mathrm{V}}$, r$_{\mathrm{K}}$ and T by using a global template fitting scheme deriving 
these four parameters simultaneously. This gave us an idea about what to expect for these parameters. We then refined these first estimates by 
using line ratios, by deriving r$_{\mathrm{K}}$ and T from continuum normalized template fitting with the spectral type fixed to its adopted value, 
and finally by obtaining A$_{\mathrm{V}}$ from template fitting with the spectral type, r$_{\mathrm{K}}$ and T fixed to their adopted values. 
We used this scheme in order to break the degeneracy that can occur when fitting all parameters simultaneously. 
Figure\,\ref{fig: spectra with best-fit models - cont-norm and non-cont-norm} shows the best-fit models together with the target spectra, 
with the spectra both continuum-normalized and not. Figure\,\ref{fig: spectra with best-fit models - spectral regions} focuses on two 
particular wavelength ranges with constraining photospheric features.

In the template fitting scheme the target spectrum was fit with a series of models of spectral type template stars which were veiled with 
infrared emission excess and reddened by extinction using a typical interstellar medium extinction law as defined in Prato et al. 
(\cite{Prato_etal2003}). We used R$\sim$\,2000 template spectra from the NASA IRTF SpeX stellar spectral 
library\footnote{http://irtfweb.ifa.hawaii.edu/$\sim$spex/IRTF\_Spectral\_Library/} (Rayner, Cushing \& Vacca \cite{Rayner_Cushing_Vacca_2009}). 
As the gravity of young low-mass stars is intermediate between that of giants and dwarfs due to their large radii, for each spectral type 
the spectral template used was a weighted average of dwarf and giant spectra with weights of 3/4 and 1/4, respectively. 
We used the locus of the binary components compared to that of the dwarf and giant templates on a Na 2.2076\,$\mu$m + Ca 2.2634\,$\mu$m 
versus CO(4-2) + CO(2-0) equivalent widths plot (Greene \& Meyer \cite{Greene_Meyer_1995}) as a guidance to define these weights 
(see Fig.\,\ref{fig: diagnostic plots}, upper left plot). Since K-band veiling could bias the use of such a plot as a gravity diagnostic, 
we restricted this procedure to the components with no obvious sign of IR-excess through K-L color, Br$\gamma$ emission, and 
a posteriori r$_{\mathrm{K}}$ determination. While extinction is in principle another source of bias for such a gravity tool, the effect on 
our low-extincted targets can be neglected. One should also note the remarkable locus of these intermediate-gravity templates with 
spectral type later than $\sim$\,M5 on this plot. The library of spectral type template spectra was constructed with a resolution of 0.5 subclass 
by averaging adjacent spectra. The determination of the best-fit model was done using a $\chi^2$ minimization search. Wavelength ranges 
encompassing strong telluric correction residuals, as well as Br$\gamma$ and the H$_2$ 2.12 micron features, were excluded from the 
$\chi^2$ computation. In the case of the continuum-normalized spectra, we found that excluding the CO lines from the $\chi^2$ computation 
gives more consistent results. This is presumably due to the strong sensitivity of the CO lines to gravity leading to small mismatches between 
the gravity of each target and the gravity assumed for the spectral templates. In both the template fitting scheme and the line ratio analysis, the binary 
component spectra were smoothed to the resolution of the spectral template library. In the template fitting scheme, the best-fit model in the minimum 
least-squares sense while fitting best the continuum shape was sometimes not giving a satisfactory fit to the relative strength of the photospheric 
features. In these cases the best-fit model was chosen by visual examination among the possible models closest to the $\chi^2$ minimum.

%
\begin{table}
\scriptsize
\caption{Determination of spectral type, extinction and K-band excess using different methods. The adopted values are reported in Table\,\ref{Table : stellar parameter}.}
\begin{center}
\renewcommand{\arraystretch}{0.5}
\setlength\tabcolsep{5pt}
\begin{tabular}{l@{\hspace{2mm}}
			r@{\,$\pm$\,}l@{\hspace{2mm}}
			l@{\hspace{1mm}}
			l@{\hspace{2mm}}
			l@{\hspace{1mm}}
			l@{\hspace{1mm}}
			l@{\hspace{1mm}}
			c@{\hspace{1mm}}
			c@{\hspace{1mm}}
			r@{\,$\pm$\,}l@{\hspace{1mm}}
			r@{\,$\pm$\,}l@{\hspace{1mm}}
			}

\hline\noalign{\smallskip}

										&  
\multicolumn{2}{c}{Line ratio}  					& 
\multicolumn{2}{c}{cont.-norm.}  				& 
										&  
\multicolumn{3}{c}{non-cont.-norm.} 				&   
										&  
\multicolumn{4}{c}{}							\\

JW	 									&  
\multicolumn{2}{c}{analysis}  					& 
\multicolumn{2}{c}{template fit}  				&
										&     
\multicolumn{3}{c}{template fit}  				&   
										&  
\multicolumn{4}{c}{Photometry}					\\

\cline{4-5}  \cline{7-9}  \cline{11-14}				\\ 

	 	   								&   
\multicolumn{2}{c}{SpT}      					& 
\multicolumn{1}{l}{SpT}						&
r$_{\mathrm{K}},\mathrm{T}$					&
										&  
\multicolumn{1}{l}{SpT}						&
A$_{\mathrm{V}}$							&
r$_{\mathrm{K}},\mathrm{T}$					&
										&  
\multicolumn{2}{l}{A$_{\mathrm{V}}$}			&   
\multicolumn{2}{c}{r$_{\mathrm{K}}$}\\

\noalign{\smallskip}
\hline

\noalign{\smallskip}
\object{63 A}   &   M1.0   &   0.6		&    K9   &   0.1,1500   	&&    M1.5   &    2.7   &    0   		&&   1.1   &   0.4   &   $-0.03$   &   0.06  \\
\noalign{\smallskip}
\object{63 B}   &   M2.6   &   1.0		&    M3   &   0.1,700   	&&    M3.5   &    0.6   &    0   		&&   1.0   &   0.5   &   $-0.14$   &   0.05  \\
\noalign{\smallskip}
\noalign{\smallskip}

\noalign{\smallskip}
\object{81 A}   &   M0.8   &   0.1		&    M1.5  &   0		   	&&   M1.5    &   0.5    &     0 		&&  0.7   &   0.4   &  $0.04$   &   0.07   \\
\noalign{\smallskip}
\object{81 B}   &    M3.7  &   0.3		&    M3     &    0.1,700     	&&   M3.5    &   0       &     0 		&&  1.1   &   0.5   &  $0.03$   &   0.05    \\
\noalign{\smallskip}
\noalign{\smallskip}

\object{128 A}   &   M1.6   &   2.0 		&    M1.5   &   0   	  	&&    M4.5   &    0   &    0   			&&   1.9   &   0.5   &   $0.04$   &   0.04  \\
\noalign{\smallskip}
\object{128 B}   &   M2.7   &   1.6 		&    M1.5   &   0.1,700   	&&    M3.5   &    1.3   &    0   		&&   1.4   &   0.6   &   $0.12$   &   0.05  \\
\noalign{\smallskip}
\noalign{\smallskip}

\object{176 A}   &   M0.8   &   0.1 		&    M0   	&   0.3,1500   	&&    M2   &    1.4   &    0		  	&&   3.9   &   0.6   &   $0.10$   &   0.04  \\
\noalign{\smallskip}
\object{176 B}   &   M0.8   &   0.2 		&    M0.5   &   0.1,1500   	&&    M3.5   &    0   &    0   			&&   2.3   &   0.4   &   $-0.13$   &   0.07  \\
\noalign{\smallskip}
\noalign{\smallskip}

\noalign{\smallskip}
\object{248 Aa}   &   M0.7   &   0.3	&   M1.5   &   0.4,1500   	&&   M2.5    &   0.8     &   0.2,850   	&&   1.8   &   1.0  &  $0.02$   &   0.06   \\
\noalign{\smallskip}
\object{248 B}   &   M6.0   &   1.5		&   M9      &   0.1,1500      	&&   M7    	   &   5.8    &    0  		&&   1.5   &   0.8  &  $-0.10$   &   0.13   \\
\noalign{\smallskip}
\noalign{\smallskip}

\object{391 A}   &   M0.8   &   0.6 		&    K7   &   0.8, 1500   	&&    M2   &    0   &    0.5, 1500   	&&   1.8   &   0.6   &   $0.46$   &   0.04  \\
\noalign{\smallskip}
\object{391 B}   &   M7.0   &   1.2 		&    M7   &   0.4,1500    	&&    M7   &    0   &    0		  	&&   2.1   &   0.6   &   $0.00$   &   0.06  \\
\noalign{\smallskip}
\noalign{\smallskip}

\object{709 A }   &   M0.7   &   0.4 	&    K9   &   0   			&&    M0.5   &    1.3   &    0   		&&   1.3   &   0.4   &   $0.05$   &   0.05  \\
\noalign{\smallskip}
\object{709 B }   &   M0.8   &   0.1 	&    M0.5   &   0   		&&    M1.5   &    0   &    0   			&&   1.1   &   0.4   &   $0.00$   &   0.07  \\
\noalign{\smallskip}
\noalign{\smallskip}

\object{867 A}   &   M0.8   &   0.2 		&    K9   &   0.5,700   		&&    M1.5   &    1.5   &    0.5, 1500   	&&   0.5   &   0.8   &   $0.26$   &   0.05  \\
\noalign{\smallskip}
\object{867 B}   &   M1.0   &   0.4 		&    M2   &   0.1, 700   	&&    M3.5   &    0.1   &    0.4, 1500   	&&   0.0   &   0.8   &   $0.22$   &   0.05  \\
\noalign{\smallskip}
\noalign{\smallskip}

\hline
\end{tabular}
\end{center}
\label{Table : stellar parameter - intermediate results}
\end{table}

%
\begin{table*}
\scriptsize
\caption{Derived stellar properties. The PMS tracks are those from Baraffe et al. (\cite{Baraffe_etal1998}) - B98, Palla \& Stahler (\cite{Palla_Stahler_1999}) - PS99, and Siess et al. (\cite{Siess_etal_2000}) - S00.}
\begin{center}
\renewcommand{\arraystretch}{0.7}
\setlength\tabcolsep{3pt}
\begin{tabular}{l@{\hspace{4mm}}
			r@{\,$\pm$\,}l@{\hspace{4mm}}
			r@{\,$\pm$\,}l@{\hspace{4mm}}
			r@{\,$\pm$\,}l@{\hspace{4mm}}
			r@{\,$\pm$\,}l@{\hspace{4mm}}
			r@{\,$\pm$\,}l@{\hspace{4mm}}
			c@{\hspace{1mm}}
			c@{\hspace{1mm}}
			c@{\hspace{1mm}}
			c@{\hspace{1mm}}
			c@{\hspace{1mm}}
			c@{\hspace{1mm}}
			c@{\hspace{4mm}}
			c@{\hspace{1mm}}
			c@{\hspace{1mm}}
			c@{\hspace{1mm}}
			c@{\hspace{1mm}}
			c@{\hspace{1mm}}
			c@{\hspace{1mm}}
			c@{\hspace{4mm}}
			r@{\,$\pm$\,}l@{\hspace{1mm}}
			}

\hline\noalign{\smallskip}

	 						&  
\multicolumn{2}{c}{}		 		& 
\multicolumn{2}{c}{}				&   
\multicolumn{2}{c}{}				&   
\multicolumn{2}{c}{}				&   
\multicolumn{2}{c}{}				& 
\multicolumn{6}{c}{Mass (M$_{\sun}$)} &  
							&
\multicolumn{6}{c}{Age (Myr)} 		&  
\multicolumn{2}{c}{}				&
							\\

\cline{12-17} \cline{19-24} 	\\ 

	 						&  
\multicolumn{2}{l}{Spectral}  		& 
\multicolumn{2}{l}{T$_{\mathrm{eff}}$}  &   
\multicolumn{2}{c}{}				&   
\multicolumn{2}{c}{}				&   
\multicolumn{2}{c}{} 				& 
\multicolumn{2}{c}{B98} 		 	&  
\multicolumn{2}{c}{PS99} 		 	&  
\multicolumn{2}{c}{S00}   			& 
							&
\multicolumn{2}{c}{B98} 		 	&  
\multicolumn{2}{c}{PS99} 		 	&  
\multicolumn{2}{c}{S00}   			& 
							&
\multicolumn{2}{c}{R$_{\star}$}		\\

Name 						&  
\multicolumn{2}{l}{Type}         		& 
\multicolumn{2}{l}{(K)}			& 
\multicolumn{2}{l}{A$_{\mathrm{V}}$} &  
\multicolumn{2}{l}{r$_{\mathrm{K}}$} &  
\multicolumn{2}{l}{log(L$_{\star}$/L$_{\sun}$)} 	& 
\multicolumn{2}{c}{} 		 		&  
\multicolumn{2}{c}{} 		 		&  
\multicolumn{2}{c}{}   			& 
							&
\multicolumn{2}{c}{}  				&  
\multicolumn{2}{c}{}  				&  
\multicolumn{2}{c}{}  				&  
							&
\multicolumn{2}{c}{(R$_{\sun}$)}  	\\

\noalign{\smallskip}
\hline

\noalign{\smallskip}

\object{JW\,63 A }   &   M0.5   &   0.5   &   3777   &    72   &    0.87   &    0.80   &   0.23   &   0.10   &   $-0.053$   &   $0.128$   &   \multicolumn{2}{c}{0.63\,$^{+0.04}_{-0.04}$}   &   \multicolumn{2}{c}{0.52\,$^{+0.06}_{-0.06}$}   &\multicolumn{2}{c}{0.51\,$^{+0.05}_{-0.05}$}   &&   \multicolumn{2}{c}{1.00\,$^{+0.30}_{-0.00}$}   &\multicolumn{2}{c}{0.86\,$^{+0.59}_{-0.59}$}   &   \multicolumn{2}{c}{1.13\,$^{+0.53}_{-0.53}$}   &&   2.2   &   0.3   \\
\noalign{\smallskip}
\object{JW\,63 B }   &   M3     &   1.0   &   3415   &   145   &    0.17   &    0.33   &   0.03   &   0.06   &   $-0.437$   &   $0.103$   &   \multicolumn{2}{c}{0.44\,$^{+0.13}_{-0.13}$}   &   \multicolumn{2}{c}{0.28\,$^{+0.09}_{-0.09}$}   &\multicolumn{2}{c}{0.32\,$^{+0.07}_{-0.07}$}   &&   \multicolumn{2}{c}{1.42\,$^{+0.99}_{-0.42}$}   &\multicolumn{2}{c}{0.85\,$^{+0.67}_{-0.67}$}   &   \multicolumn{2}{c}{1.66\,$^{+0.80}_{-0.80}$}   &&   1.7   &   0.2   \\
\noalign{\smallskip}
\noalign{\smallskip}
                    
\object{JW\,81 A }   &   M1     &   0.5   &   3705   &    72   &    0.61   &    0.92   &   0.05   &   0.08   &   $-0.162$   &   $0.137$   &   \multicolumn{2}{c}{0.60\,$^{+0.03}_{-0.03}$}   &   \multicolumn{2}{c}{0.47\,$^{+0.06}_{-0.06}$}   &\multicolumn{2}{c}{0.47\,$^{+0.04}_{-0.04}$}   &&   \multicolumn{2}{c}{1.00\,$^{+0.68}_{-0.00}$}   &\multicolumn{2}{c}{0.95\,$^{+0.46}_{-0.46}$}   &   \multicolumn{2}{c}{1.30\,$^{+0.54}_{-0.54}$}   &&   2.0   &   0.3   \\
\noalign{\smallskip}
\object{JW\,81 B }   &   M3.5   &   0.5   &   3342   &    72   &    0.00   &    0.00   &   0.04   &   0.07   &   $-0.890$   &   $0.096$   &   \multicolumn{2}{c}{0.30\,$^{+0.06}_{-0.06}$}   &   \multicolumn{2}{c}{0.23\,$^{+0.04}_{-0.04}$}   &\multicolumn{2}{c}{0.27\,$^{+0.03}_{-0.03}$}   &&   \multicolumn{2}{c}{3.38\,$^{+1.83}_{-2.38}$}   &\multicolumn{2}{c}{2.28\,$^{+0.76}_{-0.76}$}   &   \multicolumn{2}{c}{4.06\,$^{+1.03}_{-1.03}$}   &&   1.1   &   0.1   \\
\noalign{\smallskip}
\noalign{\smallskip}
                    
\object{JW\,128 A}   &   M3     &   1.0   &   3415   &   145   &    0.04   &    0.09   &   0.02   &   0.04   &   $-0.144$   &   $0.097$   &   \multicolumn{2}{c}{0.48\,$^{+0.14}_{-0.14}$}   &   \multicolumn{2}{c}{0.27\,$^{+0.09}_{-0.09}$}   &\multicolumn{2}{c}{0.32\,$^{+0.07}_{-0.07}$}   &&   \multicolumn{2}{c}{1.00\,$^{+0.33}_{-0.00}$}   &\multicolumn{2}{c}{0.41\,$^{+0.27}_{-0.31}$}   &   \multicolumn{2}{c}{0.98\,$^{+0.43}_{-0.43}$}   &&   2.4   &   0.2   \\
\noalign{\smallskip}
\object{JW\,128 B}   &   M3     &   1.0   &   3415   &   145   &    0.20   &    0.37   &   0.05   &   0.10   &   $-0.226$   &   $0.104$   &   \multicolumn{2}{c}{0.48\,$^{+0.14}_{-0.14}$}   &   \multicolumn{2}{c}{0.27\,$^{+0.08}_{-0.08}$}   &\multicolumn{2}{c}{0.32\,$^{+0.07}_{-0.07}$}   &&   \multicolumn{2}{c}{1.00\,$^{+0.35}_{-0.00}$}   &\multicolumn{2}{c}{0.47\,$^{+0.30}_{-0.37}$}   &   \multicolumn{2}{c}{1.19\,$^{+0.54}_{-0.54}$}   &&   2.2   &   0.2   \\
\noalign{\smallskip}
\noalign{\smallskip}
                    
\object{JW\,176 A}   &   M1     &   0.5   &   3705   &    72   &    0.00   &    0.00   &   0.37   &   0.11   &   $ 0.150$   &   $0.096$   &   \multicolumn{2}{c}{0.60\,$^{+0.03}_{-0.03}$}   &   \multicolumn{2}{c}{0.47\,$^{+0.05}_{-0.05}$}   &\multicolumn{2}{c}{0.46\,$^{+0.04}_{-0.04}$}   &&   \multicolumn{2}{c}{1.00}   &\multicolumn{2}{c}{0.37\,$^{+0.17}_{-0.17}$}   &   \multicolumn{2}{c}{0.70\,$^{+0.16}_{-0.16}$}   &&   2.9   &   0.2   \\
\noalign{\smallskip}
\object{JW\,176 B}   &   M1     &   0.5   &   3705   &    72   &    2.34   &    1.10   &   0.10   &   0.08   &   $ 0.230$   &   $0.151$   &   \multicolumn{2}{c}{0.60\,$^{+0.03}_{-0.03}$}   &   \multicolumn{2}{c}{0.47\,$^{+0.05}_{-0.05}$}   &\multicolumn{2}{c}{0.46\,$^{+0.04}_{-0.04}$}   &&   \multicolumn{2}{c}{1.00}   &\multicolumn{2}{c}{0.27\,$^{+0.24}_{-0.17}$}   &   \multicolumn{2}{c}{0.58\,$^{+0.27}_{-0.27}$}   &&   3.2   &   0.5   \\
\noalign{\smallskip}
\noalign{\smallskip}
                    
\object{JW\,248 Aa}   &   M1     &   0.5   &   3705   &    72   &    4.54   &    0.76   &   0.43   &   0.11   &   $ 0.296$   &   $0.126$   &   \multicolumn{2}{c}{0.60\,$^{+0.03}_{-0.03}$}   &   \multicolumn{2}{c}{0.47\,$^{+0.06}_{-0.06}$}   &\multicolumn{2}{c}{0.46\,$^{+0.04}_{-0.04}$}   &&   \multicolumn{2}{c}{1.00}   &\multicolumn{2}{c}{0.17\,$^{+0.14}_{-0.07}$}   &   \multicolumn{2}{c}{0.44\,$^{+0.26}_{-0.26}$}   &&   3.5   &   0.4   \\
\noalign{\smallskip}
\object{JW\,248 B}   &   M7.5   &   1.5   &   2795   &   295   &    1.61   &    1.70   &   0.23   &   0.16   &   $-0.494$   &   $0.204$   &   \multicolumn{2}{c}{0.04\,$^{+0.09}_{-0.09}$}   &   \multicolumn{2}{c}{$<$ 0.1}   &\multicolumn{2}{c}{0.13\,$^{+0.06}_{-0.06}$}   &&   \multicolumn{2}{c}{1.00\,$^{+1.24}_{-0.00}$}   &\multicolumn{2}{c}{}   &   \multicolumn{2}{c}{0.10\,$^{+0.64}_{-0.64}$}   &&   2.4   &   0.8   \\
\noalign{\smallskip}
\noalign{\smallskip}
                    
\object{JW\,391 A}   &   M0     &   0.5   &   3850   &    53   &    0.00   &    0.00   &   1.24   &   0.03   &   $-0.118$   &   $0.096$   &   \multicolumn{2}{c}{0.69\,$^{+0.04}_{-0.04}$}   &   \multicolumn{2}{c}{0.58\,$^{+0.05}_{-0.05}$}   &\multicolumn{2}{c}{0.56\,$^{+0.05}_{-0.05}$}   &&   \multicolumn{2}{c}{1.22\,$^{+0.50}_{-0.22}$}   &\multicolumn{2}{c}{1.39\,$^{+0.57}_{-0.57}$}   &   \multicolumn{2}{c}{1.45\,$^{+0.46}_{-0.46}$}   &&   2.0   &   0.1   \\
\noalign{\smallskip}
\object{JW\,391 B}   &   M7     &   1.0   &   2880   &   140   &    0.00   &    0.00   &   0.56   &   0.24   &   $-0.936$   &   $0.096$   &   \multicolumn{2}{c}{0.06\,$^{+0.04}_{-0.04}$}   &   \multicolumn{2}{c}{$<$ 0.1}   &\multicolumn{2}{c}{0.10\,$^{+0.02}_{-0.02}$}   &&   \multicolumn{2}{c}{1.00\,$^{+3.39}_{-0.00}$}   &\multicolumn{2}{c}{}   &   \multicolumn{2}{c}{0.44\,$^{+1.05}_{-1.05}$}   &&   1.4   &   0.2   \\
\noalign{\smallskip}
\noalign{\smallskip}
                    
\object{JW\,709 A}   &   M0.5   &   0.5   &   3777   &    72   &    0.45   &    0.37   &   0.03   &   0.07   &   $-0.209$   &   $0.104$   &   \multicolumn{2}{c}{0.64\,$^{+0.05}_{-0.05}$}   &   \multicolumn{2}{c}{0.53\,$^{+0.07}_{-0.07}$}   &\multicolumn{2}{c}{0.51\,$^{+0.05}_{-0.05}$}   &&   \multicolumn{2}{c}{1.40\,$^{+0.95}_{-0.40}$}   &\multicolumn{2}{c}{1.52\,$^{+0.66}_{-0.66}$}   &   \multicolumn{2}{c}{1.58\,$^{+0.54}_{-0.54}$}   &&   1.8   &   0.1   \\
\noalign{\smallskip}
\object{JW\,709 B}   &   M0.5   &   0.5   &   3777   &    72   &    1.11   &    0.86   &   0.02   &   0.04   &   $-0.268$   &   $0.133$   &   \multicolumn{2}{c}{0.64\,$^{+0.05}_{-0.05}$}   &   \multicolumn{2}{c}{0.53\,$^{+0.06}_{-0.06}$}   &\multicolumn{2}{c}{0.51\,$^{+0.05}_{-0.05}$}   &&   \multicolumn{2}{c}{1.77\,$^{+1.33}_{-0.77}$}   &\multicolumn{2}{c}{1.99\,$^{+0.89}_{-0.89}$}   &   \multicolumn{2}{c}{2.20\,$^{+0.88}_{-0.88}$}   &&   1.7   &   0.2   \\
\noalign{\smallskip}
\noalign{\smallskip}
                    
\object{JW\,867 A}   &   M0.5   &   0.5   &   3777   &    72   &    0.10   &    0.17   &   0.59   &   0.12   &   $-0.070$   &   $0.098$   &   \multicolumn{2}{c}{0.63\,$^{+0.04}_{-0.04}$}   &   \multicolumn{2}{c}{0.52\,$^{+0.06}_{-0.06}$}   &\multicolumn{2}{c}{0.51\,$^{+0.05}_{-0.05}$}   &&   \multicolumn{2}{c}{1.00\,$^{+0.33}_{-0.00}$}   &\multicolumn{2}{c}{0.90\,$^{+0.44}_{-0.44}$}   &   \multicolumn{2}{c}{1.18\,$^{+0.31}_{-0.31}$}   &&   2.2   &   0.1   \\
\noalign{\smallskip}
\object{JW\,867 B}   &   M1.5   &   0.5   &   3632   &    72   &    0.00   &    0.00   &   0.44   &   0.49   &   $-0.115$   &   $0.097$   &   \multicolumn{2}{c}{0.59\,$^{+0.01}_{-0.01}$}   &   \multicolumn{2}{c}{0.41\,$^{+0.05}_{-0.05}$}   &\multicolumn{2}{c}{0.42\,$^{+0.04}_{-0.04}$}   &&   \multicolumn{2}{c}{1.00\,$^{+0.26}_{-0.00}$}   &\multicolumn{2}{c}{0.71\,$^{+0.32}_{-0.32}$}   &   \multicolumn{2}{c}{1.08\,$^{+0.27}_{-0.27}$}   &&   2.2   &   0.1   \\
\noalign{\smallskip}
\noalign{\smallskip}

\hline
\end{tabular}
\end{center}
\label{Table : stellar parameter}
\end{table*}

In the line ratio analysis we measured the equivalent width of the strongest photospheric features present in the spectra of the above specified 
template library and defined several diagnostic ratios of closeby features suitable for spectral typing, i.e. ratios exhibiting a monotonic behavior in a 
certain spectral type range. For the spectrum of each binary component we measured the same diagnostic ratios and derived a 
spectral type based on the relations previously determined from the spectral templates. The spectral type derived from the line ratio analysis is a 
weighted mean of the spectral types obtained from several line ratios (notably Mg 2.1066\,/\,Al 2.1169\,$\mu$m, Mg 2.2814\,/\,Ca 2.2634\,$\mu$m, 
and Si 2.1885\,/\, Ti 2.1903\,$\mu$m), and its associated error the standard deviation of the weighted mean (see details in 
Appendix\,\ref{appendix : line ratio analysis}). Fig.\,\ref{fig: diagnostic plots} shows the behavior of some of these diagnostic line ratios. 
Even though the strong sensitivity to gravity of the ratio Mg 2.1066\,/\,Al 2.1169\,$\mu$m, the locus of the targets on these plots are well matching 
those of the spectral templates, giving us confidence on the validity of those templates.

The adopted spectral type was chosen considering the results of the line ratio analysis and of the template fitting, both continuum normalized and not. 
While the continuum-normalized template fitting is relying mainly on the strength of all absorption features, the non-normalized template fitting is 
more sensitive to the shape of the continuum which is especially constraining for late (later than $\sim$\,M2) spectral types based on the change of the spectra slope 
around 2.3\,$\mu$m. Although the determination from both analyses are found to be in fair agreement, the spectral types from the non-normalized 
template fitting are found to be systematically later than those from the continuum-normalized template fitting and the line ratio analysis 
(see Table\,\ref{Table : stellar parameter - intermediate results}). The discrepancy between continuum slope and line strengths in our spectral templates 
is probably due to the intrinsic continuum slope of the spectrum of pre-main sequence (PMS) stars which are not simply the result of an average between 
giants and dwarfs but may be bluer, i.e. closer to the continuum of a giant for a given spectral type. In consequence, we gave in our analysis more 
weight to the spectral types derived from line ratio analysis and continuum-normalized template fitting. A spectral library of PMS stars would 
be very valuable in order to address this point. Fig.\,\ref{fig: spectra with best-fit models - cont-norm and non-cont-norm} and 
\ref{fig: spectra with best-fit models - spectral regions} show that the quality of the fits degrades significantly for the latest type stars which is probably 
to be attributed to the templates. Besides the errors introduced by the templates, the values reported in 
Table\,\ref{Table : stellar parameter - intermediate results} illustrate the uncertainties involved in the determination of the parameters and the risk 
of biased results when using only a simultaneous fitting method. 
We estimated a spectral typing error based on the values derived from the line ratio analysis with 0.5 subclass as a threshold.  
The uncertainties in r$_{\mathrm{K}}$ and A$_{\mathrm{V}}$ were estimated based on the spectral type errors. The r$_{\mathrm{K}}$ values 
and uncertainties are the $\chi^2$ weighted means and standard deviations of the weighted mean of all r$_{\mathrm{K}}$ values obtained from 
continuum-normalized template fitting with the spectral type varying within its uncertainties. Similarly, A$_{\mathrm{V}}$ values and uncertainties 
were computed from the the $\chi^2$ weighted means and standard deviations of the weighted mean of all A$_{\mathrm{V}}$ values derived from 
non-continuum-normalized template fitting within the spectral type ranges.

The determination of A$_{\mathrm{V}}$ and r$_{\mathrm{K}}$ for each component based on the spectra was checked by estimating 
these quantities from photometry (Table\,\ref{Table : stellar parameter - intermediate results}). 
The extinction was derived for the components with disk signatures by dereddening to the CTTS locus  (Meyer et al. \cite{Meyer_etal_1997}) 
and for the other components from the excess in J-H color given the adopted spectral type and using tabulated colors of young stellar photospheres 
from Luhman et al. (\cite{Luhman_etal_2010}). The K-band excess r$_{\mathrm{K}}$ was estimated from the H-K color excess using the same 
tabulated photospheric colors and the adopted value of extinction (e.g. Levine et al. \cite{Levine_etal_2006}). This estimate is a lower-limit since 
the derivation assumes that r$_{\mathrm{H}}$=0. All colors were converted to the CIT system and the extinction law used is that from Cohen et al. 
(\cite{Cohen_etal_1981}). Uncertainties are the propagation of errors in colors, spectral type, and in the adopted A$_{\mathrm{V}}$ value in 
the case of the derivation of r$_{\mathrm{K}}$. The values of A$_{\mathrm{V}}$ derived from spectroscopy and from photometry agree at the 
1.5\,$\sigma$ level on average for all sources, with a better agreement for the WTTS components (1.2\,$\sigma$) compared to CTTS 
components (2.1\,$\sigma$). The K-band excesses estimated from spectroscopy and from photometry are consistent at the 1.7\,$\sigma$ 
level on average.

\subsection{Bolometric Luminosity}
\label{sect:bolometric luminosity}

The bolometric luminosity for each component was computed from the J-band magnitude and bolometric corrections from 
Kenyon \& Hartmann (\cite{Kenyon_hartmann_1995}). We assumed a distance to the ONC of 415\,pc as suggested from 
recent accurate trigonometric parallax measurements (Hirota et al. \cite{Hirota_etal_2007}, Sandstrom et al. \cite{Sandstrom_etal_2007}, 
Menten et al. \cite{Menten_etal_2007}) and 10\,\% relative error. The corresponding error in luminosity $\Delta$\,log\,L is 0.09\,dex 
which is the dominant source of error in derived luminosities. Minor source of errors are the uncertainties of 0.1\,mag in the bolometric 
correction and those from the magnitudes (either J- or K-band) which translate in $\Delta$\,log\,L of 0.04\,dex and on average 0.01\,dex. 
All errors were added quadratically. 

It has been pointed out by Levine et al. (\cite{Levine_etal_2006}) that bolometric luminosities can be more accurately determined in the 
K-band than in the J-band when the extinction uncertainties are relatively large. However, in our case the luminosity uncertainties are 
on average only marginally smaller when using K-band magnitudes ($<$$\Delta$\,log\,L$>$ = 0.110\,dex) than with J-band 
($<$$\Delta$\,log\,L$>$ = 0.117\,dex). The A$_{\mathrm{V}}$ contribution to the error in log~L is {\it on average} smaller in J-band 
(mean of 0.05\,dex) than the sum of the Av and rk contributions in K-band (0.017 and 0.04\,dex). In addition, the bolometric luminosities 
derived from K-band magnitudes are different to those derived from J-band by up to $\pm$\,0.20\,dex, but with a mean of only 0.02\,dex. 
The reason is that, given the mean A$_{\mathrm{V}}$ and r$_{\mathrm{K}}$ values of our sample (0.75\,$\pm$\,0.47 and 0.28\,$\pm$\,0.12), 
the mean contribution to the luminosity from A$_{\mathrm{V}}$ in J-band (+0.08 dex) is close to and opposite to the sum of the 
mean contributions from A$_{\mathrm{V}}$ and r$_{\mathrm{K}}$ in K-band (+0.03 and $-$0.11\,dex). We further examined the difference in 
luminosity derived from J- and K-band as a function of A$_{\mathrm{V}}$ and r$_{\mathrm{K}}$. When one splits the sample in two sub-samples 
of equal size, one with small r$_{\mathrm{K}}$ values and the other one with large r$_{\mathrm{K}}$ values, one obtains mean values 
of (log\,L)$_{\mathrm{J}}$$-$(log\,L)$_{\mathrm{K}}$ of $-$0.03 and 0.07\,dex, respectively. Therefore the correction for veiling could be 
introducing a bias on the luminosities. The same analysis performed on two sub-samples, one with high A$_{\mathrm{V}}$ values and the 
other with low A$_{\mathrm{V}}$ values, leads to a mean change in log\,L of 0.0 and 0.04\,dex, respectively, which a priori excludes the 
possibility for a bias introduced by A$_{\mathrm{V}}$. Since the uncertainties of K-band derived luminosities are not significantly smaller 
than those of J-band derived luminosities, and given the possibility to be introducing a bias in K-band with the correction for veiling, we 
used here the J-band to compute bolometric luminosities.

%
\begin{figure}
\begin{center}
\begin{tabular}{c}
\includegraphics[width=8.8cm, angle=0]{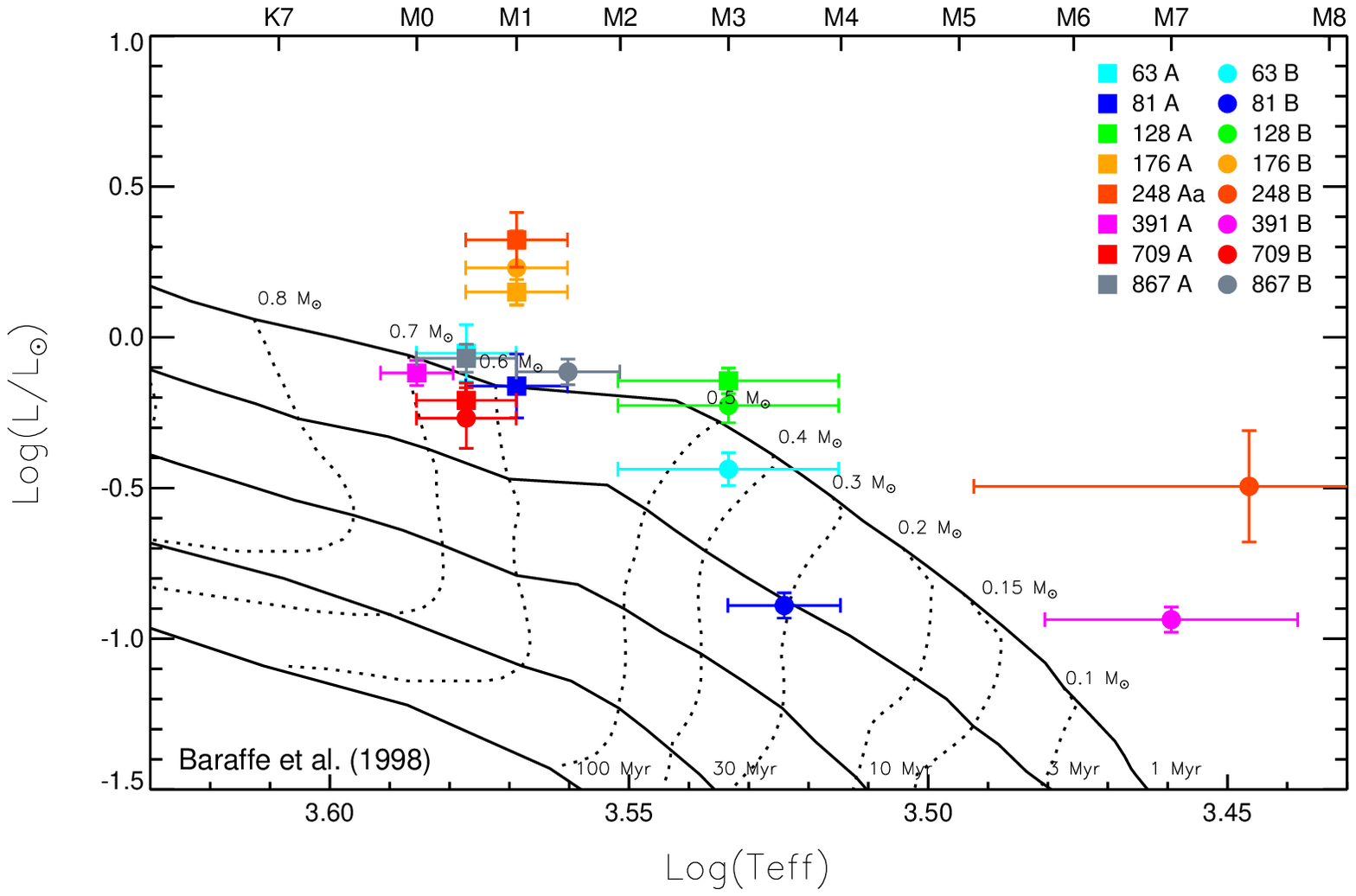} \\
\includegraphics[width=8.8cm, angle=0]{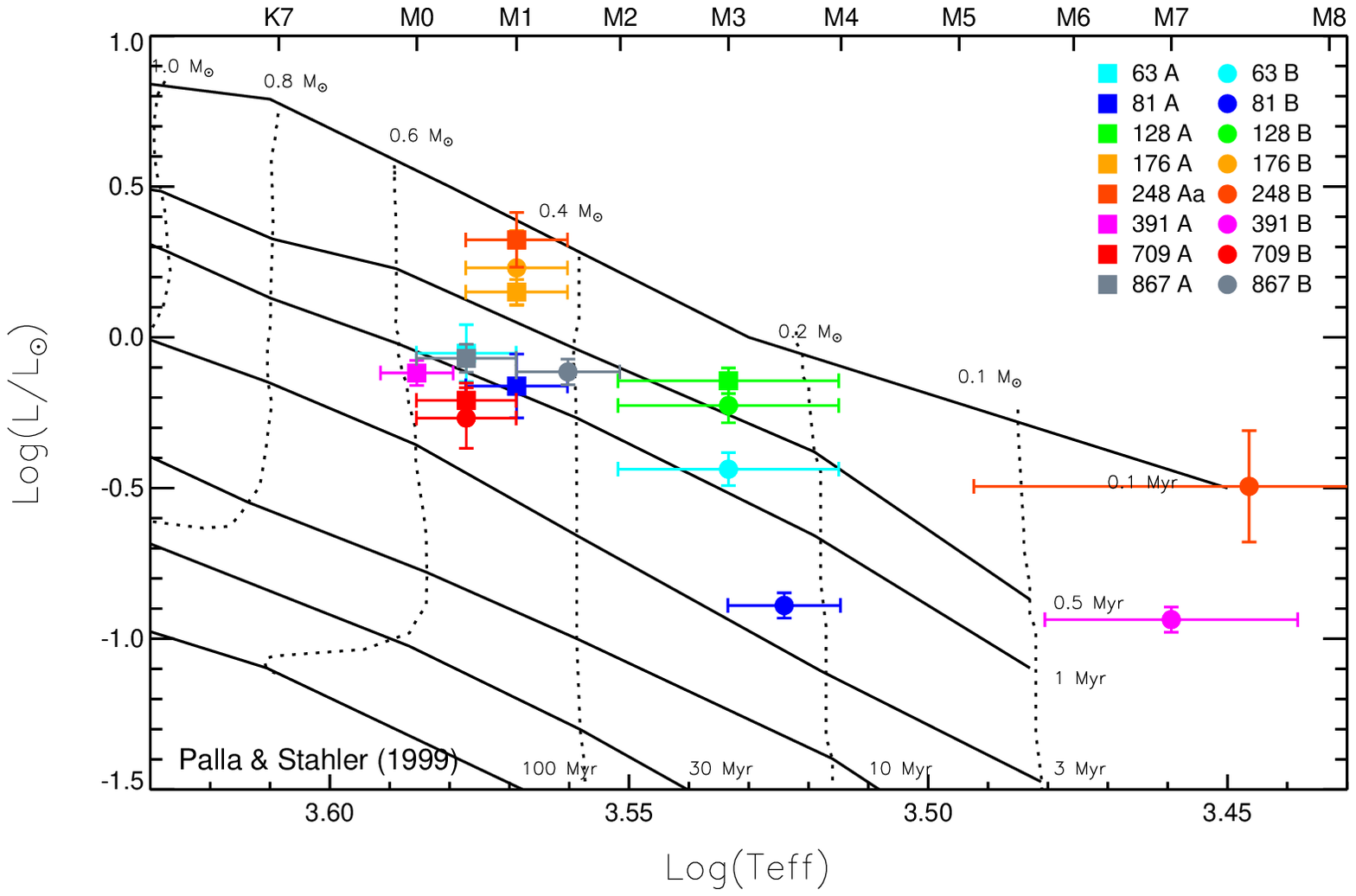}\\
\includegraphics[width=8.8cm, angle=0]{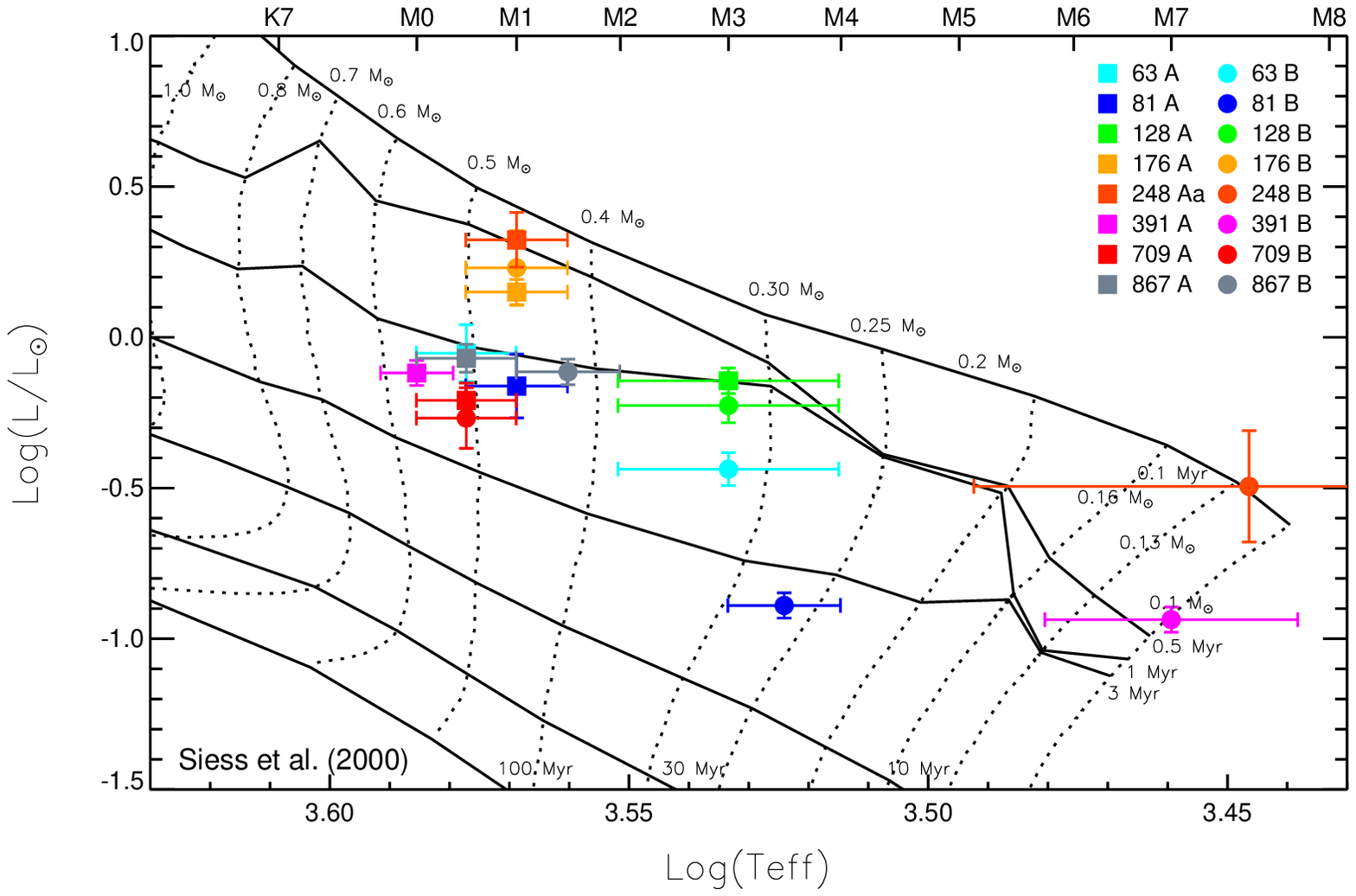}
\end{tabular}
\end{center}
\caption{H-R diagrams of the components of the T Tauri ONC binaries with the models of\,:  
(top) Baraffe et al. (\cite{Baraffe_etal1998}), (middle) Palla \& Stahler (\cite{Palla_Stahler_1999}), (bottom) Siess et al. (\cite{Siess_etal_2000}).}
\label{fig: HR diagrams}
\end{figure}

\subsection{Mass and Age}
\label{sect:mass and age}

The binaries of our sample are plotted in an H-R diagram together with the PMS evolutionary tracks of 
Baraffe et al. (\cite{Baraffe_etal1998} - B98), Palla \& Stahler (\cite{Palla_Stahler_1999} - PS99), and Siess et al. (\cite{Siess_etal_2000} - S00) 
in Figure \ref{fig: HR diagrams}. The spectral types were converted into effective temperatures using the intermediate-temperature scale 
from Luhman et al. (\cite{Luhman_etal_2003}). We derived a mass and an age for each binary component using each set of PMS tracks 
(Table\,\ref{Table : stellar parameter}). To this purpose, we used interpolated values of the irregular grid of mass and age of the PMS tracks 
around the source loci. The estimated errors are the variations in mass and age computed from varying the effective temperature 
and luminosity within their uncertainties. For the B98 evolutionary models we used $l_{mix}/H_p = 1.0$ and 1.9 at M/M$_{\sun}$\,$\leq$\,0.6 
and M/M$_{\sun}$\,$\geq$\,0.6, respectively. The mass assigned to the sources falling above the 1\,Myr isochrone of the B98 tracks 
was done by projecting down at constant temperature to the 1\,Myr isochrone.

The mass of our sources estimated from the PS99 and S00 tracks are generally in good agreement, especially above $\sim$\,0.4\,M$_{\sun}$. 
Below this value, S00 masses are slightly larger than PS99 masses by $\sim$\,15\,\%. Masses predicted by the B98 tracks are on average 20-30\,\% 
larger than those derived from both S00 and PS99 tracks, in agreement with the results of Hillenbrand \& White (\cite{Hillenbrand_White_2004}). 
The mean age of all the components of our sample is of 0.87\,$\pm$\,0.64 Myr and 1.25\,$\pm$\,0.92 Myr for the PS99 and S00 tracks, respectively. 
Compared to the averaged values found for the whole ONC cluster from the same models ($\sim$\,2\,Myr and $\sim$\,3\,Myr, respectively, 
Da Rio et al. \cite{DaRio_etal_2010}), our result shows that  the optically-bright systems of our sample are among the youngest of the 
ONC binary population. The secondaries of JW\,248 and 391 are close to being substellar with S00 masses of 0.13\,$\pm$\,0.06 and 
0.10\,$\pm$\,0.02\,M$_{\sun}$, respectively.

\subsection{Mass ratio distribution}
\label{sect:mass ratio distribution}

The differences in model-predicted masses are translating into mass ratios q=M$_{sec}$/M$_{prim}$ relatively mildly. Indeed, the mass ratio from different models are found to be quite 
unexpectedly in agreement (see Table\,\ref{Table : mass ratios and relative ages}). The uncertainties in mass ratio were estimated through a Monte-Carlo technique. 
Note that we took into account the fact that each component of a pair is at the same distance, therefore each realization in luminosity from each component of a pair was 
decomposed into a correlated term from the distance error and an uncorrelated term from the rest of the error contributors. The effect of such treatment is, however, relatively 
minor for the mass ratios as the tracks are mostly vertical.

%
\begin{table}
\scriptsize
\caption{Derived mass ratios and relative ages. The PMS tracks are those from Baraffe et al. (\cite{Baraffe_etal1998}) - B98, Palla \& Stahler (\cite{Palla_Stahler_1999}) - PS99, and Siess et al. (\cite{Siess_etal_2000}) - S00.}
\begin{center}
\renewcommand{\arraystretch}{0.7}
\setlength\tabcolsep{5pt}
\begin{tabular}{l@{\hspace{4mm}}
			r@{\,$\pm$\,}l@{\hspace{2mm}}
			r@{\,$\pm$\,}l@{\hspace{2mm}}
			r@{\,$\pm$\,}l@{\hspace{2mm}}
			c
			r@{\,$\pm$\,}l@{\hspace{2mm}}
			r@{\,$\pm$\,}l@{\hspace{2mm}}
			}

\hline\noalign{\smallskip}

	 															&  
\multicolumn{6}{c}{q=M$_{sec}$/M$_{prim}$}								&
																&
\multicolumn{4}{c}{$\Delta$\,log\,$\tau$\,=log (age$_{prim}$) - log(age$_{sec}$)} 	\\

\cline{2-7} 	\cline{9-12}	\\ 

Name 						&  
\multicolumn{2}{c}{B98} 		 	&  
\multicolumn{2}{c}{PS99} 		 	&  
\multicolumn{2}{c}{S00}   			& 
							&
\multicolumn{2}{c}{PS99} 		 	&  
\multicolumn{2}{c}{S00}   			\\

\noalign{\smallskip}
\hline
\noalign{\smallskip}

\object{JW\,63          }   &   0.69   &   0.21                    &   0.53   &   0.19                    &   0.64   &   0.13                    &   &   $ 0.01$   &   $ 0.33$              &   $-0.17$   &   $ 0.23$              \\
\noalign{\smallskip}
\noalign{\smallskip}
                    
\object{JW\,81          }   &   0.51   &   0.10                    &   0.49   &   0.11                    &   0.57   &   0.08                    &   &   $-0.38$   &   $ 0.21$              &   $-0.49$   &   $ 0.13$              \\
\noalign{\smallskip}
\noalign{\smallskip}
                    
\object{JW\,128         }   &   1.00   &   0.23                    &   1.01   &   0.20                    &   1.00   &   0.14                    &   &   $-0.06$   &   $ 0.32$              &   $-0.08$   &   $ 0.46$              \\
\noalign{\smallskip}
\noalign{\smallskip}
                    
\object{JW\,176         }   &   1.00   &   0.04                    &   1.00   &   0.08                    &   1.00   &   0.07                    &   &   $ 0.13$   &   $ 0.28$              &   $ 0.09$   &   $ 0.34$              \\
\noalign{\smallskip}
\noalign{\smallskip}
                    
\object{JW\,248         }   &   0.07   &   0.15                    &   \multicolumn{2}{c}{$<$  0.21}      &   0.27   &   0.14                    &   &   \multicolumn{2}{c}{         }      &   $ 0.66$   &   $ 0.70$              \\
\noalign{\smallskip}
\noalign{\smallskip}
                    
\object{JW\,391         }   &   0.08   &   0.06                    &   \multicolumn{2}{c}{$<$  0.17}      &   0.18   &   0.05                    &   &   \multicolumn{2}{c}{         }      &   $ 0.52$   &   $ 0.33$              \\
\noalign{\smallskip}
\noalign{\smallskip}
                    
\object{JW\,709         }   &   1.01   &   0.06                    &   1.01   &   0.08                    &   1.00   &   0.07                    &   &   $-0.12$   &   $ 0.22$              &   $-0.15$   &   $ 0.18$              \\
\noalign{\smallskip}
\noalign{\smallskip}
                    
\object{JW\,867         }   &   0.93   &   0.06                    &   0.78   &   0.12                    &   0.84   &   0.11                    &   &   $ 0.10$   &   $ 0.17$              &   $ 0.04$   &   $ 0.12$              \\
\noalign{\smallskip}
\noalign{\smallskip}

\hline
\end{tabular}
\end{center}
\label{Table : mass ratios and relative ages}
\end{table}

%
\begin{figure}
\begin{center}
\begin{tabular}{c}
\includegraphics[width=8.5cm, angle=0]{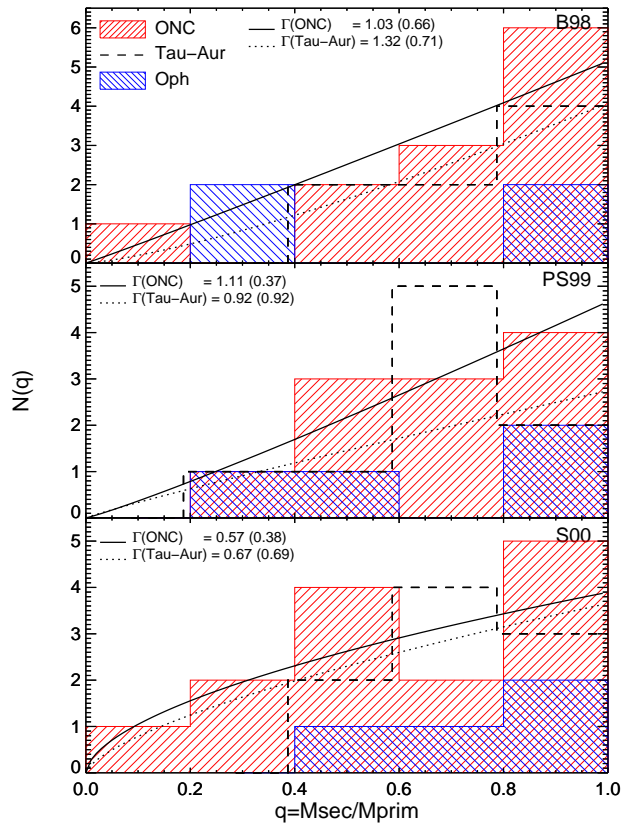} 
\end{tabular}
\end{center}
\caption{Mass ratio distribution for the ONC binaries of our extended sample in comparison with the distribution derived for wide binaries in Tau-Aur and Oph 
with primary mass range 0.15 to 0.8\,M$_{\sun}$ using the B98, PS99, and S00 PMS tracks for systems with separations from 85 to 560 AU. 
Also shown are the best-fit power-laws dN/dq\,$\varpropto$\,q$^\Gamma$ to the mass ratio distributions in ONC and Tau-Aur. 
The Tau-Aur distribution is offset along the x-axis for the purpose of clarity.}
\label{fig:mass ratio distribution}
\end{figure}

In the following we will consider an extended sample of ONC binaries consisting of our sample of 8 binaries and 7 additional 
binaries for which spectral type and stellar luminosity for both components were derived in the spectroscopic and photometric 
study of Daemgen, Correia \& Petr-Gotzens (\cite{Daemgen_etal_2012}, hereafter D12). 
Fig.\,\ref{fig:mass ratio distribution} shows the distribution of mass ratio for this extended sample of ONC binaries. We compared 
this distribution with those for Tau-Aur wide binaries from White \& Ghez (\cite{White_Ghez_2001}, hereafter WG01) and 
Hartigan \& Kenyon (\cite{Hartigan_Kenyon2003}, hereafter HK03) and for Oph wide binaries from Prato et al. (\cite{Prato_etal2003}, 
hereafter P03) spanning the separation range 85-560\,AU and with primary masses in the range 0.15 to 0.8\,M$_{\sun}$, 
i.e. the separation range of the ONC sample and the range of primary mass encompassing most of the systems in these samples. 
For all systems of this plot the component masses, hence mass ratios, were uniformly derived from the three sets of PMS tracks.  
We used the spectral types and luminosities reported in WG01, HK03, P03, and D12 to derive the component mass of Tau-Aur, Oph, 
and ONC binaries, respectively (values reported in Table\,\ref{Table : comparative samples} and \ref{Table : comparative samples : 
mass ratios and relative ages}). In our ONC sample, we excluded from the analysis the mass ratio derived from the B98 tracks for JW\,248 
and JW\,391 because of the very uncertain estimate of their secondary masses. A Kolmogorov-Smirnov (K-S) test finds a high probability 
($\gtrsim$\,40\%) that all the distributions are drawn from the same parent population. Specifically, the probability of the K-S test between 
the ONC and Tau-Aur mass ratio distributions are 57, 43, and 55\% for the B98, PS99, and S00 tracks, respectively. 
The corresponding values between ONC and Oph distributions are 55, 58, and 92\%. 
Fig.\,\ref{fig:mass ratio distribution} also shows the best-fit power-law dN/dq\,$\varpropto$\,q$^\Gamma$ to the q-distributions in ONC and Tau-Aur. 
In ONC and in the primary mass range 0.15-0.8\,M$_{\sun}$ we found $\Gamma$\,=\,1.03\,$\pm$\,0.66, 1.11\,$\pm$\,0.37 and 0.57\,$\pm$\,0.38 
for the B98, PS99 and S00 tracks, respectively. Despite the large range of $\Gamma$ encompassed by the different set of PMS tracks, the derived 
values are in agreement within the uncertainties. For each set of tracks the values of $\Gamma$ in ONC are consistent with the corresponding 
values in Tau-Aur. Although statistics are small and selection biases may be present in our sample and in the comparison samples from Tau-Aur and Oph, 
this suggests that the mass ratio distribution of wide binaries with primary mass in the range 0.15 to 0.8\,M$_{\sun}$ are in agreement 
in these three regions.

%
\begin{figure}
\begin{center}
\includegraphics[width=9cm, angle=0]{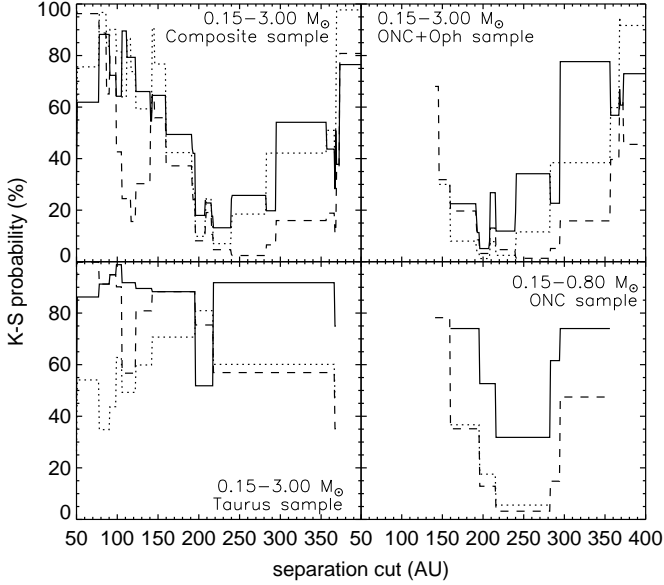}
\end{center}
\caption{ 
K-S probability that the mass ration distribution of wide and close binaries derive from the same parent distribution as a function of the separation threshold. 
The sample is composed of systems from ONC, Tau-Aur, and Oph with primary mass range 0.15 to 3.0\,M$_{\sun}$ (upper-left), 
systems from Oph and ONC with primary mass range 0.15 to 3.0\,M$_{\sun}$ (upper-right), systems from Tau-Aur with primary mass range 0.15 to 3.0\,M$_{\sun}$ (lower-left), 
and systems from ONC with primary mass range 0.15 to 0.8\,M$_{\sun}$ (lower-right). 
Continuous line is for B98 tracks, dotted lines for PS99 tracks, and dashed line for S00 tracks.} 
\label{fig:mass ratio distribution vs separation}
\end{figure}

%
\begin{figure}
\begin{center}
\begin{tabular}{c}
\includegraphics[width=8.5cm, angle=0]{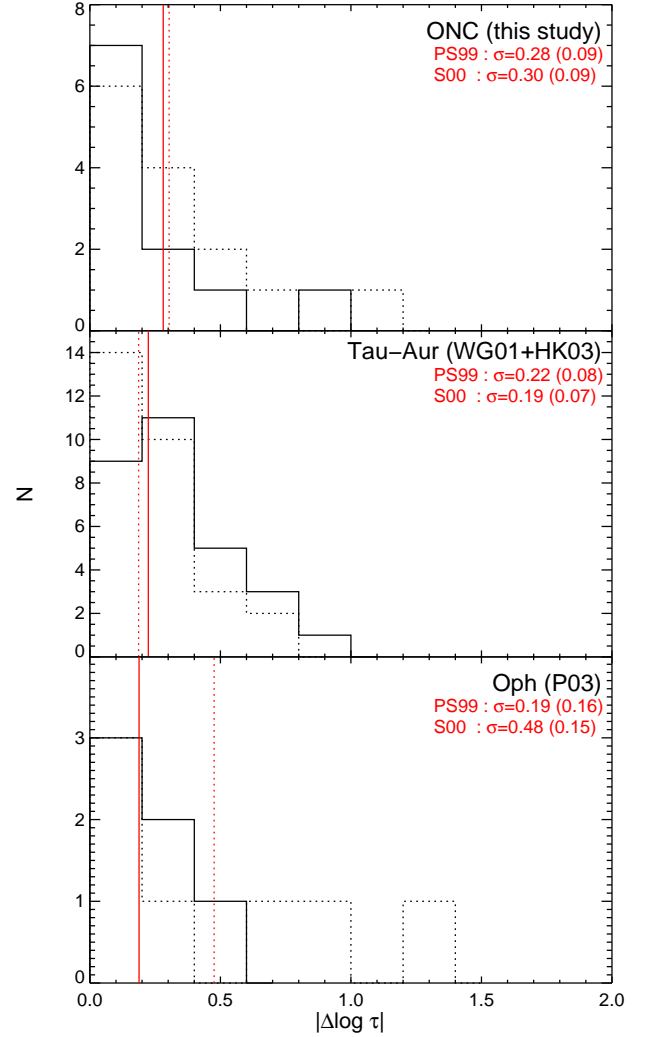}
\end{tabular}
\end{center}
\caption{Distribution of the absolute difference in logarithmic age $|\Delta$\,log\,$\tau|$ for the ONC binaries of our extended sample in comparison 
with the distribution derived for binaries in Tau-Aur and Oph using the PS99 (continous line) and S00 (dotted line) PMS tracks. The rms scatter in 
$|\Delta$\,log\,$\tau|$ for both set of tracks are indicated by red lines. 
The rms of the $|\Delta$\,log\,$\tau|$ distribution and its uncertainty are reported for each set of tracks and each sample.}
\label{fig:relative age distribution in ONC, Tau, and Oph.}
\end{figure}

\subsection{ Mass ratio vs. separation and distance from the Trapezium}
\label{sect:Mass ratio vs. separation and distance from the Trapezium}

In order to investigate the dependence of the mass ratio distribution on separation we compared the two distributions formed 
from systems with small and large separations. We used first a composite sample including all systems from ONC, Tau-Aur, and Oph 
with primary mass in the full mass range 0.15 to 3.0\,M$_{\sun}$ and computed the probability of the K-S test that both distributions are 
consistent with each other for a range of separation cut values. The upper-left panel of Fig.\,\ref{fig:mass ratio distribution vs separation} 
shows that there is a minimum of the K-S probability at a separation cut between 200 and 280\,AU for all set of tracks. A similar although 
less pronounced minimum at similar separations is seen when considering only systems with primary mass in the range 0.15 to 0.8\,M$_{\sun}$ 
(not shown). The other plots of Fig\,\ref{fig:mass ratio distribution vs separation} suggest that this feature is predominantly originating 
from the ONC and Oph samples and that systems from the comparative sample in Tau-Aur exhibit a mass ratio distribution essentially 
independent of separation. We conclude that there is a hint of a variation of the mass ratio distribution with separation in both Oph and 
ONC binaries which is not seen in Tau-Aur binaries.

While in the outer cluster regions starting at $\sim$\,3-4\,arcmin from the Trapezium the systems exhibit a mass ratio in the range 0.6-1.0, 
in the innermost cluster regions the smallest mass ratio systems of our sample are found (JW\,248, JW\,391, and [HC2000]73, 
see Fig.\,\ref{fig:mass ratio vs cluster radial distance with TT types}, upper plot). The presence of these low-mass ratio systems in the inner 
high stellar density regions of ONC is remarkable since dynamical interactions would tend to disrupt any weakly bound binaries. 
On the other hand, the formation of low-mass ratio systems is generally expected in a cluster environment for which gas and star motions 
are not correlated especially in the innermost regions (e.g. Bate\,\cite{Bate_2001}). 
In addition, in the lower plot of Fig.\,\ref{fig:mass ratio vs cluster radial distance with TT types} it can be noted that among the systems 
at distances $>$\,4\,arcmin there seems to be no trend toward a decreasing mass ratio with separation. This may suggest that the latter may be 
restricted to systems located in the inner parts of the cluster where dynamical interactions are more frequent.

\subsection{Relative age}
\label{sect:relative age}
   
The distribution of the absolute difference in logarithmic age $|\Delta$\,log\,$\tau|$\,=\,$|$log (age$_{prim}$) - log(age$_{sec}$)$|$ 
is shown in Fig.\,\ref{fig:relative age distribution in ONC, Tau, and Oph.} for the extended sample of ONC binaries and is compared with 
that for the sample of Tau-Aur and Oph binaries. The values of relative age $\Delta$\,log\,$\tau$ are reported in Table\,\ref{Table : mass 
ratios and relative ages} for our ONC sample and in Table\,\ref{Table : comparative samples : mass ratios and relative ages} for the binaries 
from WG01, HK03, P03, and D12. The uncertainties in $\Delta$\,log\,$\tau$ were estimated as in Sect.\,\ref{sect:mass ratio distribution}. In this 
case, the correlation of the luminosity errors due to the identical distance of each component of a pair resulted in a significant decrease of the 
relative age error. We excluded from our analysis the anomalous system JW\,553. The secondary of this system is underluminous with 
$\Delta$K\,$\sim$\,3\,mag with respect to the primary while both components have similar spectral types (see D12) suggesting a significant amount of flux 
from scattering due perhaps to a close to edge-on disk. 

We found a consistent rms scatter in $|\Delta$\,log\,$\tau|$ of the ONC distribution with values of 0.28\,$\pm$\,0.09 and 
0.30\,$\pm$\,0.09 dex for the PS99 and S00 tracks, respectively. This means that the ages of the binary components are essentially within a factor two of each other. 
Compared to the sample of Tau-Aur and Oph binaries, the binaries in ONC are as coeval within the uncertainties. 
The distribution of absolute relative age $|\Delta$\,log\,$\tau|$ in ONC and Tau-Aur are found to be consistent with K-S probabilities of 7.5\% 
and 71\% for the PS99 and S00 tracks, respectively. The corresponding probabilities between ONC and Oph are $\sim$\,75\%. We therefore 
conclude that the components of T Tauri binary stars in ONC are mostly as coeval as the components of T Tauri binaries in Tau-Aur and Oph. 
The rms scatter in $|\Delta$\,log\,$\tau|$ of randomly paired single components of ONC binaries is 0.70\,$\pm$\,0.08 and 0.73\,$\pm$\,0.08 dex 
for the PS99 and S00 tracks, respectively. This is significantly larger than the standard deviation of the relative age of ONC binaries and this 
demonstrates that ONC binaries are more coeval than the overall ONC population. This matches with the conclusions of WG01 and 
Kraus \& Hillenbrand (\cite{Kraus_Hillenbrand_2009b}) in Tau-Aur.

%
\begin{figure*}
\begin{center}
\begin{tabular}{cc}
\includegraphics[width=8.5cm, angle=0]{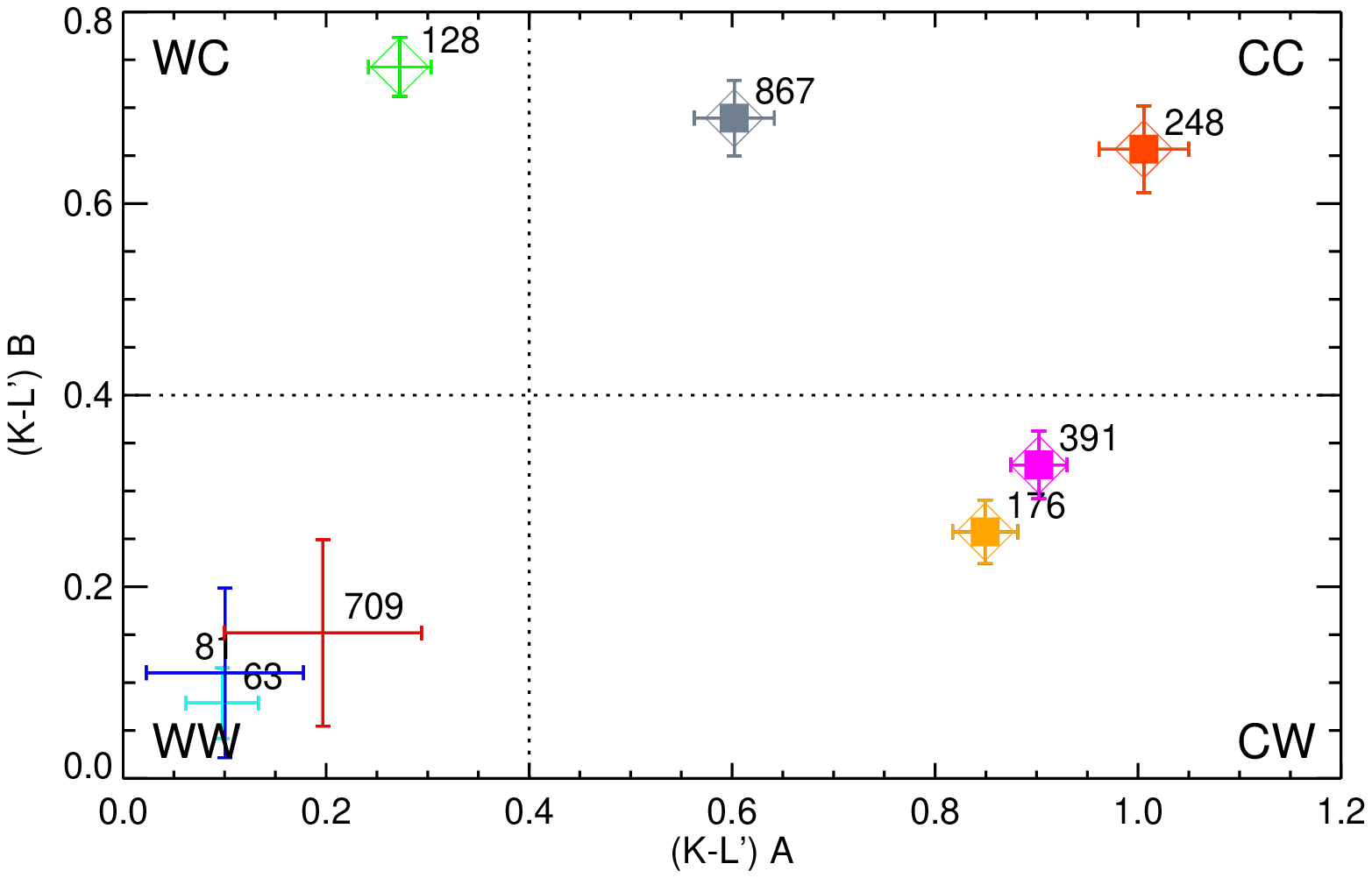} &
\includegraphics[width=8.5cm, angle=0]{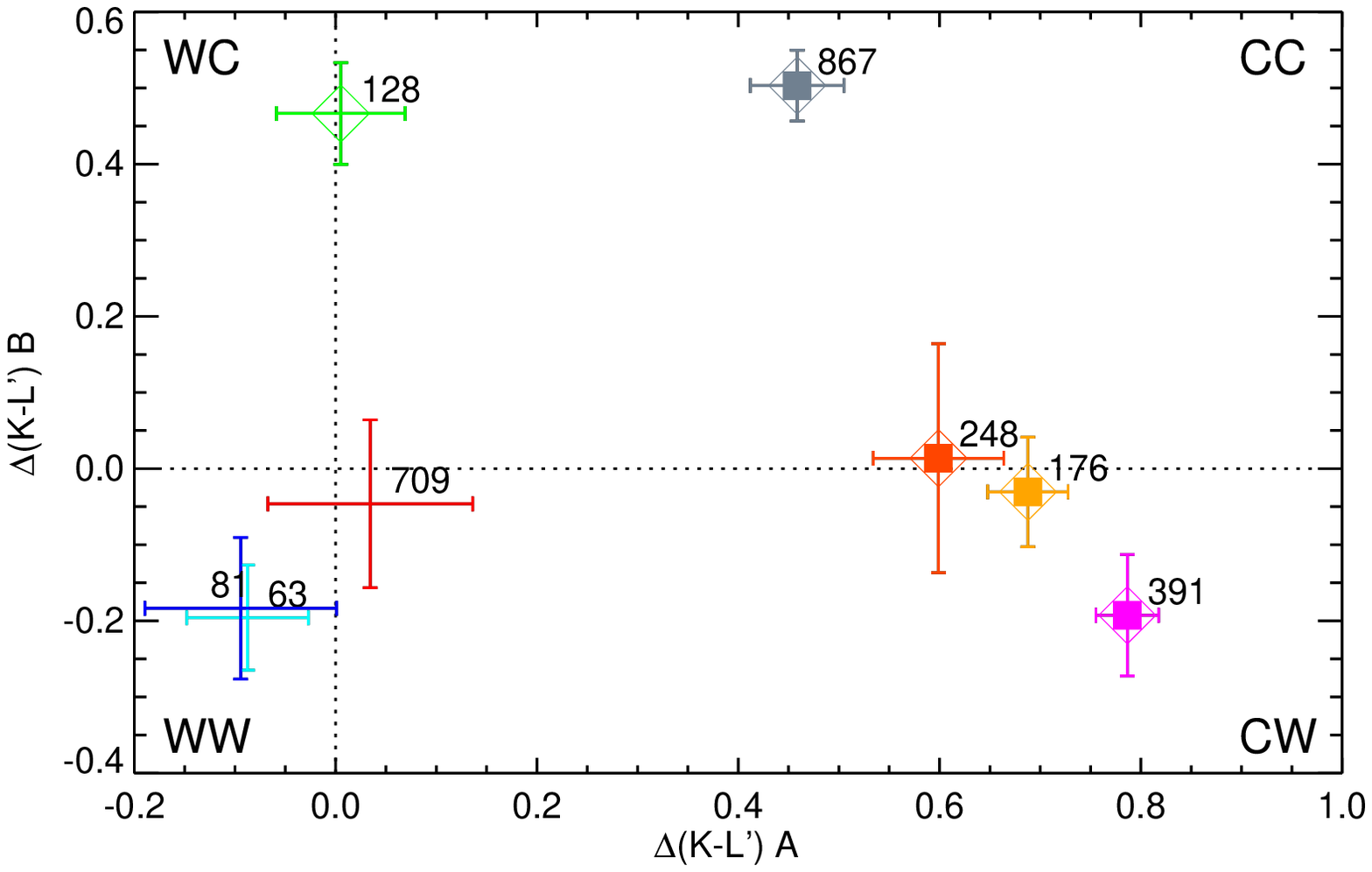}
\end{tabular}
\end{center}
\caption{K-L' color (left) and K-L' color excess (right) for the ONC binaries. The T Tauri type of each binary is named from the type of each component with C or W (CTTS or WTTS) 
indicating the presence or not of an inner disk. The dotted lines delimit the regions indicating the presence or not of an inner disk around each component, i.e. WW, CW, CC, and WC, 
with the first letter referring to the primary and the second to the secondary. 
Filled square symbols show systems with significant ($>$\,3\,$\sigma$) K-band excess r$_{\mathrm{K}}$  which are all found in the primary only. Systems classified as Class\,II from 
Spitzer IRAC colors are plotted with large open diamonds.}
\label{fig:K-L' color and color excess}
\end{figure*}

%
\begin{table*}
\scriptsize
\caption{Derived accretion disk properties.}
\begin{center}
\renewcommand{\arraystretch}{0.7}
\setlength\tabcolsep{9pt}
\begin{tabular}{l
			r@{\,$\pm$\,}l
			r@{\,$\pm$\,}l
			r@{\,$\pm$\,}l
			r@{\,$\pm$\,}l
			r@{\,$\pm$\,}l
			r@{\,$\pm$\,}l
			c}

\hline\noalign{\smallskip}

	 								&  
\multicolumn{2}{c}{EW(Br$\gamma$)} 		&  
\multicolumn{2}{c}{F(Br$\gamma$)} 			&  
\multicolumn{2}{c}{}						&   
\multicolumn{2}{c}{$\dot \mathrm{M}$}  		&   
\multicolumn{2}{c}{K$-$L' 	}				& 
\multicolumn{2}{c}{$\Delta$(K$-$L')} 		& 
T Tauri								\\

Name 								&  
\multicolumn{2}{c}{(\AA)}   				&   
\multicolumn{2}{c}{(10$^{-18}$ W\,m$^{-2}$)}  	& 
\multicolumn{2}{c}{log (L$_{acc}$/L$_{\sun}$)}&   
\multicolumn{2}{c}{(10$^{-9}$ M$_{\sun}$\,yr$^{-1}$)}	&   
\multicolumn{2}{c}{(mag)}  				&  
\multicolumn{2}{c}{(mag)}  				&  
type$^a$								\\

\noalign{\smallskip}
\hline

\noalign{\smallskip}
\object{JW\,63 A}	&    \multicolumn{2}{c}{$<$ 0.91}  	&	 \multicolumn{2}{c}{$<$ 0.7}  	&  \multicolumn{2}{c}{$<$ --2.38} & \multicolumn{2}{c}{$<$ 0.5}    &    0.10   &    0.04   &  $-0.09$	&	0.06		&  W \\
\noalign{\smallskip}
\object{JW\,63 B}	&    \multicolumn{2}{c}{$<$ 1.60}  	&	 \multicolumn{2}{c}{$<$ 2.8}  	&  \multicolumn{2}{c}{$<$ --1.65} & \multicolumn{2}{c}{$<$ 3.6}    &    0.08   &    0.04   &  $-0.20$	&	0.07		& W \\
\noalign{\smallskip}
\noalign{\smallskip}

\noalign{\smallskip}
\object{JW\,81 A}	&    \multicolumn{2}{c}{$<$ 0.72}  	&	 \multicolumn{2}{c}{$<$ 2.5}  	&  \multicolumn{2}{c}{$<$ --1.71} & \multicolumn{2}{c}{$<$ 2.7}    &   0.10   &    0.08   &  $-0.09$	&	0.10  	&  W \\
\noalign{\smallskip}
\object{JW\,81 B}	&    \multicolumn{2}{c}{$<$ 1.14}  	&	 \multicolumn{2}{c}{$<$ 1.2}  	&  \multicolumn{2}{c}{$<$ --2.10} & \multicolumn{2}{c}{$<$ 1.1}    &  0.11   &    0.09   &  $-0.18$	&	0.09 		&  W  \\
\noalign{\smallskip}
\noalign{\smallskip}

\object{JW\,128 A}	&    \multicolumn{2}{c}{$<$ 1.20}  	&	 \multicolumn{2}{c}{$<$ 7.3}  	&  \multicolumn{2}{c}{$<$ --1.12} & \multicolumn{2}{c}{$<$ 15.0}  &    0.27   &    0.03   &  0.01		&	0.06  	& W \\
\noalign{\smallskip}
\object{JW\,128 B}	&     \multicolumn{2}{c}{$<$ 1.21}  	&	  \multicolumn{2}{c}{$<$ 7.3}  	&  \multicolumn{2}{c}{$<$ --1.13}  & \multicolumn{2}{c}{$<$ 13.6} &    0.74   &    0.03   & 0.47		&	0.07  	& C \\
\noalign{\smallskip}
\noalign{\smallskip}

\object{JW\,176 A}	&     \multicolumn{2}{c}{$<$ 1.10}  	&	  \multicolumn{2}{c}{$<$ 9.5}	&   \multicolumn{2}{c}{$<$ --0.98} & \multicolumn{2}{c}{$<$ 20.2} &    0.85   &    0.03   &  0.69		&	0.04 		&  C \\
\noalign{\smallskip}
\object{JW\,176 B}	&    \multicolumn{2}{c}{$<$ 1.53}  	&	 \multicolumn{2}{c}{$<$ 10.0}  	&  \multicolumn{2}{c}{$<$ --0.95}  & \multicolumn{2}{c}{$<$ 23.8} &    0.26   &    0.03   &  $-0.03$	&	0.07  	&  W \\
\noalign{\smallskip}
\noalign{\smallskip}

\noalign{\smallskip}
\object{JW\,248 Aa}	&    \multicolumn{2}{c}{$<$ 0.71}  	&	 \multicolumn{2}{c}{$<$ 5.6}  	&  \multicolumn{2}{c}{$<$ --1.27} & \multicolumn{2}{c}{$<$ 12.7}  &   1.00   &    0.06   &  $ 0.60$	&	0.07		&  C  \\
\noalign{\smallskip}
\object{JW\,248 B}	&    \multicolumn{2}{c}{$<$ 2.14}  	&	 \multicolumn{2}{c}{$<$ 4.8}  	&  \multicolumn{2}{c}{$<$ --1.36} & \multicolumn{2}{c}{$<$ 105}   &   0.66   &    0.05   &  $ 0.01$	&	0.15 		&  W   \\
\noalign{\smallskip}
\noalign{\smallskip}

\object{JW\,391 A}	&     	8.2		& 	 0.5  			&	  	29.8		&     	2.6		& $-0.35$ &   0.12   & 50  & 14    &   0.90   &    0.03   &  0.79		&	0.03		& C \\
\noalign{\smallskip}
\object{JW\,391 B}	&    \multicolumn{2}{c}{$<$ 1.62}  	&	 \multicolumn{2}{c}{$<$ 1.8}  	&  \multicolumn{2}{c}{$<$ --1.89} & \multicolumn{2}{c}{$<$ 12.1}    	&    0.33   &    0.04   &  $-0.19$	&	0.08  	&  W  \\
\noalign{\smallskip}
\noalign{\smallskip}

\object{JW\,709 A}	&    \multicolumn{2}{c}{$<$ 0.71}  	&	 \multicolumn{2}{c}{$<$ 1.4}  	&  \multicolumn{2}{c}{$<$ --2.04}   & \multicolumn{2}{c}{$<$ 1.1} 	&    0.20   &    0.10   &  0.03		&	0.10		& W \\
\noalign{\smallskip}
\object{JW\,709 B}	&    \multicolumn{2}{c}{$<$ 0.67}  	&	 \multicolumn{2}{c}{$<$ 0.5}  	&  \multicolumn{2}{c}{$<$ --2.61}   & \multicolumn{2}{c}{$<$ 0.2}   	&    0.15   &    0.10   &  $-0.05$	&	0.11	 	& W \\
\noalign{\smallskip}
\noalign{\smallskip}

\object{JW\,867 A}	&   	0.6		&    	0.2			&	  1.1			&	1.0		& $-2.16$   &  0.51 &    0.9  &   1.1 		&    0.60   &    0.04   &  0.46	&	0.05  	& C \\
\noalign{\smallskip}
\object{JW\,867 B}	&     	1.8		&    	0.2			&	  6.1			&     	3.4		& $-1.22$   &  0.41  &   9.1  &   6.8   	  	&    0.69   &    0.04   &  0.50	&	0.05		& C \\
\noalign{\smallskip}

\hline
\end{tabular}
\end{center}
\label{Table : disks measurements}
\begin{minipage}[position]{18cm}
Note\,: Equivalents width given as negative values. \\
$^a$\,: Based on the value of $\Delta$(K$-$L') such as the presence of an inner disk (T Tauri type C) corresponds to $\Delta$(K$-$L')\,$>$\,0 at the 3\,$\sigma$ level. \\  
\end{minipage}
\end{table*}


\section{Circumstellar Disks}
\label{sect:circumstellar disks}

\subsection{Disk Census, Fraction of mixed pairs}
\label{sect:disk census, fraction of mixed pairs}

The left panel of Fig.\,\ref{fig:K-L' color and color excess} shows the observed K-L' color of the secondary versus that of the primary 
for each ONC binary system. The T Tauri type of each binary is named from the type of each component with C or W (CTTS or WTTS) indicating the presence or not of an inner disk 
\footnote{Throughout the paper, we extended the CTTS and WTTS nomenclature to the presence or not of a circumstellar disk based on several measurement and not only the 
original meaning of this nomenclature based on H$\alpha$.}. 
While K-L'\,$\gtrsim$\,0.4, which is roughly the photospheric K-L' color of a M5 dwarf, suggests an IR-excess hence the presence of an inner disk, 
it is necessary to take into account reddening and spectral type to properly assess the presence or not of an inner circumstellar disk around each component.
We therefore computed the K-L' color excess of the individual components of ONC binary systems.
The K-L' color excess $\Delta$(K$-$L') is defined as the difference between the observed dereddened color and the photospheric 
color expected based on the spectral type. The photospheric K-L' colors as a function of spectral type for M-dwarfs were derived 
from data of Leggett (\cite{Leggett_1992}) for M0-M4, and of Leggett et al. (\cite{Leggett_etal_2002}) for M4-M9.5, converting the K-band data of Leggett (\cite{Leggett_1992}) from 
the CIT to MKO system. The resulting K-L' scale is a 3rd order polynomial fit of the data and agrees well with the M-dwarf colors reported in Bessell \& Brett (\cite{Bessel_Brett_1988}) 
for M0-M6 when converted to the MKO system. The uncertainties of $\Delta$(K$-$L') are those propagating photometric, spectral type, and reddening errors 
using a Monte-Carlo technique for the colors.
The resulting excesses and their errors are listed in Table\,\ref{Table : disks measurements} and plotted in the right panel of Fig.\,\ref{fig:K-L' color and color excess}.  
According to that analysis, JW\,128\,B, JW\,176\,A, JW\,248\,A, JW\,391\,A, and JW\,867\,A and B are consistent with having inner disks. 
Of these, JW\,391\,A, JW\,867\,A and B show Br$\gamma$ emission. 
All the binary components with significant K-band excess r$_{\mathrm{K}}$ exhibit K-L' color excess (Fig.\,\ref{fig:K-L' color and color excess}). 
The binary component JW\,128\,B shows K-L' excess but no significant K-band veiling which indicates the presence of a disk with an inner hole. 
The systems with at least one component harboring an inner disk as revealed by the K-L' color excess are the same systems classified as Class II from 
Spitzer IRAC colors (see Table\,\ref{Table : sample} and Fig.\,\ref{fig:K-L' color and color excess}).

From these measurements, JW\,128, JW\,176, JW\,248, and JW\,391 are mixed pairs, i.e. systems with a disk around only one component. 
In order to compute the fraction of mixed pairs and its uncertainties, we used a Monte-Carlo technique similar to that described in Liu et al. 
(\cite{Liu_etal_2003}) that take into account the color errors. This yields a fraction of mixed pairs of 42\,$\pm$\,18\,\%. When systems from 
D12 are also included, the fraction rises to 57\,$\pm$\,12\,\%. This can be compared to the fraction of 19\,$\pm$\,7\,\% found for binaries in 
Tau-Aur (Monin et al. \cite{Monin_etal_2007}) or that of 21\,$\pm$\,8\,\% (5/24 systems) in the combined WG01 + HK03 sample. However, 
it can be noticed from the comparative samples in Tau-Aur and Oph that the fraction of mixed pairs is markedly larger in systems with a high-mass 
primary (see Table\,\ref{Table : comparative samples}). Restricting to wide binaries with separation $>$\,85\,AU and to a common primary mass 
range of 0.15-0.8\,M$_{\sun}$ the fraction of mixed pairs is 20\,$\pm$\,18\,\% (1/5 systems) in both Tau-Aur and Oph comparative samples. 
Altogether there might be a hint that the fraction of mixed pairs is larger in ONC compared to Tau-Aur and Oph.

\subsection{Accretion activity}
\label{sect:accretion activity}

Active accretion, revealed by the presence of Br$\gamma$ emission, is another disk tracer although less reliable than K-L' color excess. 
Indeed Br$\gamma$ emission is seen in only a small fraction ($\sim$\,20\,\%) of the components with K-L' color excess. In 
Table\,\ref{Table : disks measurements} the accretion luminosity and mass accretion rate derived for these sources, as well as upper 
limits for the other components, are reported. The accretion luminosity is derived from the flux of the Br$\gamma$ emission-line 
which is computed from the flux-scaled spectra, the distance, and the relation from Muzerolle et al. (\cite{Muzerolle_etal_1998}, Eq.\,2). 
The spectra are flux-scaled using dereddened and de-veiled magnitudes previously converted to the 2MASS photometric system. 
The absolute calibration from 2MASS is used and the spectra are also multiplied by the transmission curve of the 2MASS Ks filter. 
Errors include uncertainties from the line flux measurement, errors from dereddening and de-veiling, and distance errors. 
The mass accretion rate is estimated from the accretion luminosity, stellar mass derived from B98 PMS tracks and stellar radius both 
reported in Table\,\ref{Table : stellar parameter}, using standard magnetospheric model (Gullbring et al. \cite{Gullbring_etal_1998}, Eq.\,8). 
Mass accretion uncertainties include those from the accretion luminosity and the mass and radius uncertainties. Both derived accretion 
luminosities and mass accretion rates are found to be consistent with typical values for T Tauri stars (e.g. Muzerolle et al. \cite{Muzerolle_etal_1998}), 
as well as with sources of similar age and mass in ONC (Manara et al. \cite{Manara_etal_2012}).

%
\begin{figure}
\begin{center}
\begin{tabular}{c}
\includegraphics[width=8.8cm, angle=0]{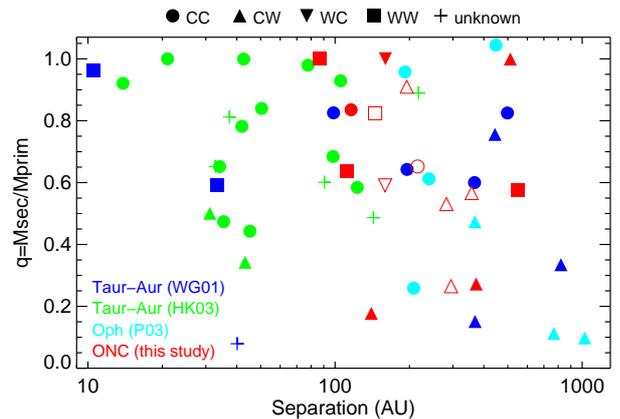}
\end{tabular}
\end{center}
\caption{Mass ratio vs separation for binaries in ONC (this study), in Tau-Aur (WG01, HK03), and in Oph (P03) with T Tauri types (CTTS and WTTS) based on K-L color excess. 
For ONC systems, open symbols are systems taken from D12 with T Tauri types defined from a conjunction of r$_{\mathrm{K}}$, H-K color excess, and Br$\gamma$ emission.
Mass ratios derived from S00 tracks are plotted.}
\label{fig:mass ratio vs separation with TT types}
\end{figure}

\subsection{ Relation with mass ratio, separation, and component mass}
\label{sect:disk relation with mass ratio, separation, and component mass}

The locus of the ONC mixed pairs in a mass ratio vs separation plot is shown in Fig.\,\ref{fig:mass ratio vs separation with TT types}. 
For comparison the same samples of binaries with T Tauri types from Tau-Aur and Oph from WG01, HK03, and P03 as before are plotted.  
The T Tauri types are assigned from K-L color excess measured in this study, WG01, HK03, and in McCabe et al. (\cite{McCabe_etal2006}). 
For the ONC systems from D12 we assigned the T Tauri type based on a conjunction of r$_{\mathrm{K}}$, H-K color excess, and 
Br$\gamma$ emission\,\footnote{In most cases we required a minimum of two positive indicators to assign a T Tauri type C.}. 
While such assignment is less secure than those based on K-L color excess, this gives us a rough estimate of the presence/absence of disks in this sample. 
Also, while including the D12 sample in the analysis of the presence of disks lowers the statistical uncertainties for ONC systems, it could introduce biases. 
For this reason, we will in the following mention also numbers without the D12 sample. We nevertheless note that such biases are likely to be small. 
If we compare the classification from D12 for the 8 sources of this study (therefore with K-L), only 2 out of the 16 components have their classification changed\,: 
JW\,128\,B (goes from W to C with the L band photometry) and JW\,176\,A (clear excess at L, not as marked in H-K). 
As found in the wide ($\gtrsim$\,100\,AU) binaries from the Tau-Aur sample, 
and to a smaller extent from the Oph sample, mixed pairs in wide ONC binaries are present in both high- and low-mass ratio systems. 
This suggests that the disk dissipation timescale is independent of the system's mass ratio for such wide binaries. Conversely, the mixed pairs 
in close ($\lesssim$\,100\,AU) binaries from Tau-Aur are mainly found in low mass ratio systems which is consistent with an early 
exhaustion of the secondary disk due to a smaller tidal truncation radius compared to the primary disk (Armitage et al. \cite{Armitage_etal_1999}). 
Overall the mass ratio distributions of mixed and unmixed pairs are found to be significantly different. A K-S test gives probabilities of 7\%, 2.7\%, 
and 0.05\% that the two distributions are drawn from the same parent distribution for the B98, PS99, and S00 tracks. Similar values are found when the ONC systems are not 
included therefore confirming the result of Monin et al. (\cite{Monin_etal_2007}) and showing that ONC systems follow the same pattern. 
At separations $\gtrsim$\,100\,AU, the difference between these distributions are greatly reduced, especially for the B98 and PS99 tracks with 
K-S probabilities of 48\%, 26\%, and 0.6\%. The mass ratio distribution of mixed pairs is therefore significantly different from that of unmixed pairs, especially at large separations.

Although the ONC and Oph samples are not covering the small separations, it seems that mixed pairs are mainly found at large separations 
which hints that early exhaustion due to disk truncation is not the main process for these systems to form. This is most apparent in the 
composite sample of binaries including all 3 regions but can also be seen within the limited separation range of ONC and Oph binaries. 
A K-S test on the separation distribution of mixed and unmixed pairs in the composite sample gives probabilities of 8\%, 9\%, and 3\% that the 
two distributions are drawn from the same parent distribution for the B98, PS99, and S00 tracks. These probabilities are respectively 12\%, 8\%, and 14\% when only considering 
ONC systems (including D12 systems). This is consistent with a more synchronized evolution of disks in close pairs and confirms the result based 
on accretion indicators from the ONC sample of D12.

%
\begin{figure}
\begin{center}
\begin{tabular}{c}
\includegraphics[width=8.8cm, angle=0]{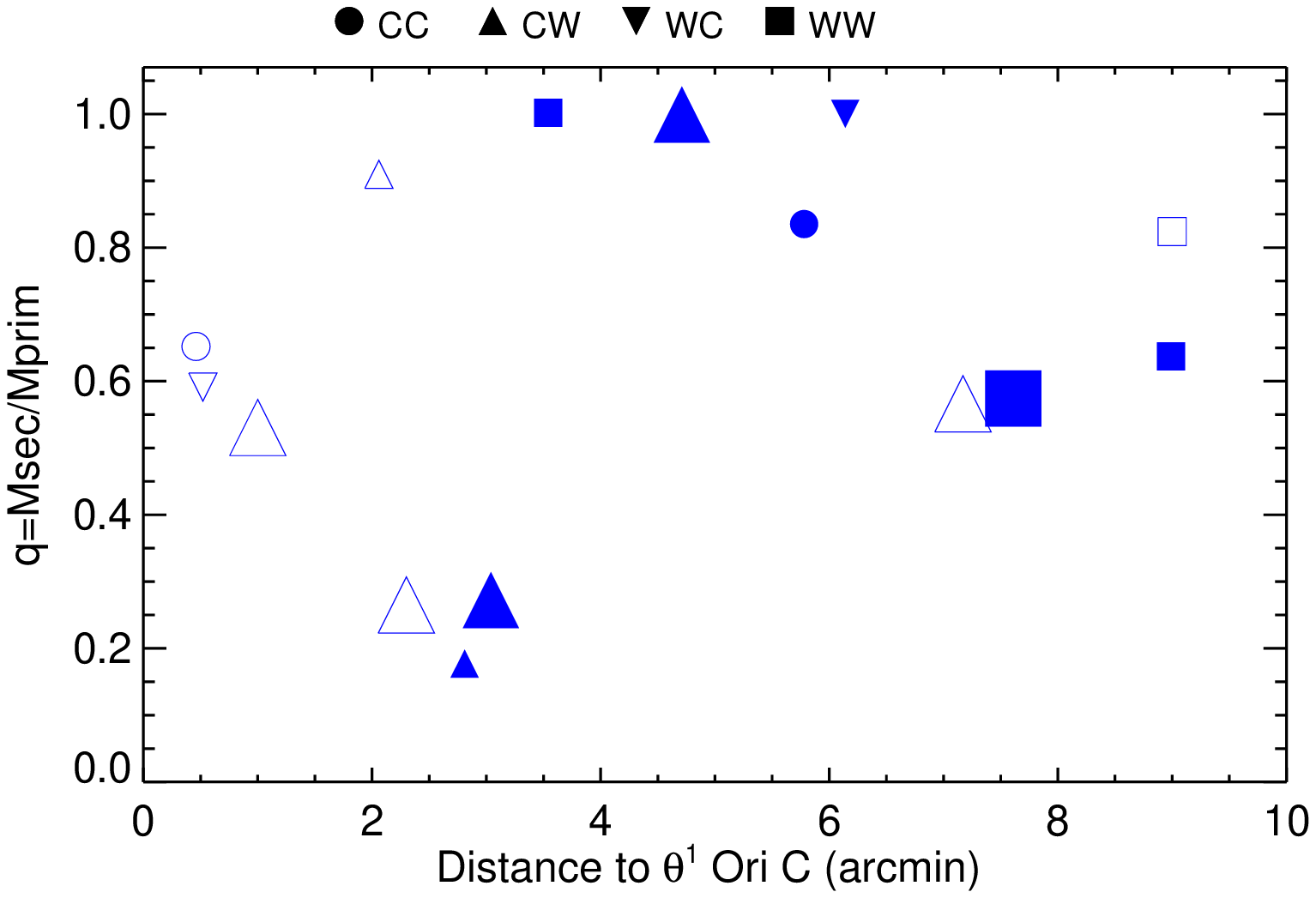} \\
\includegraphics[width=8.8cm, angle=0]{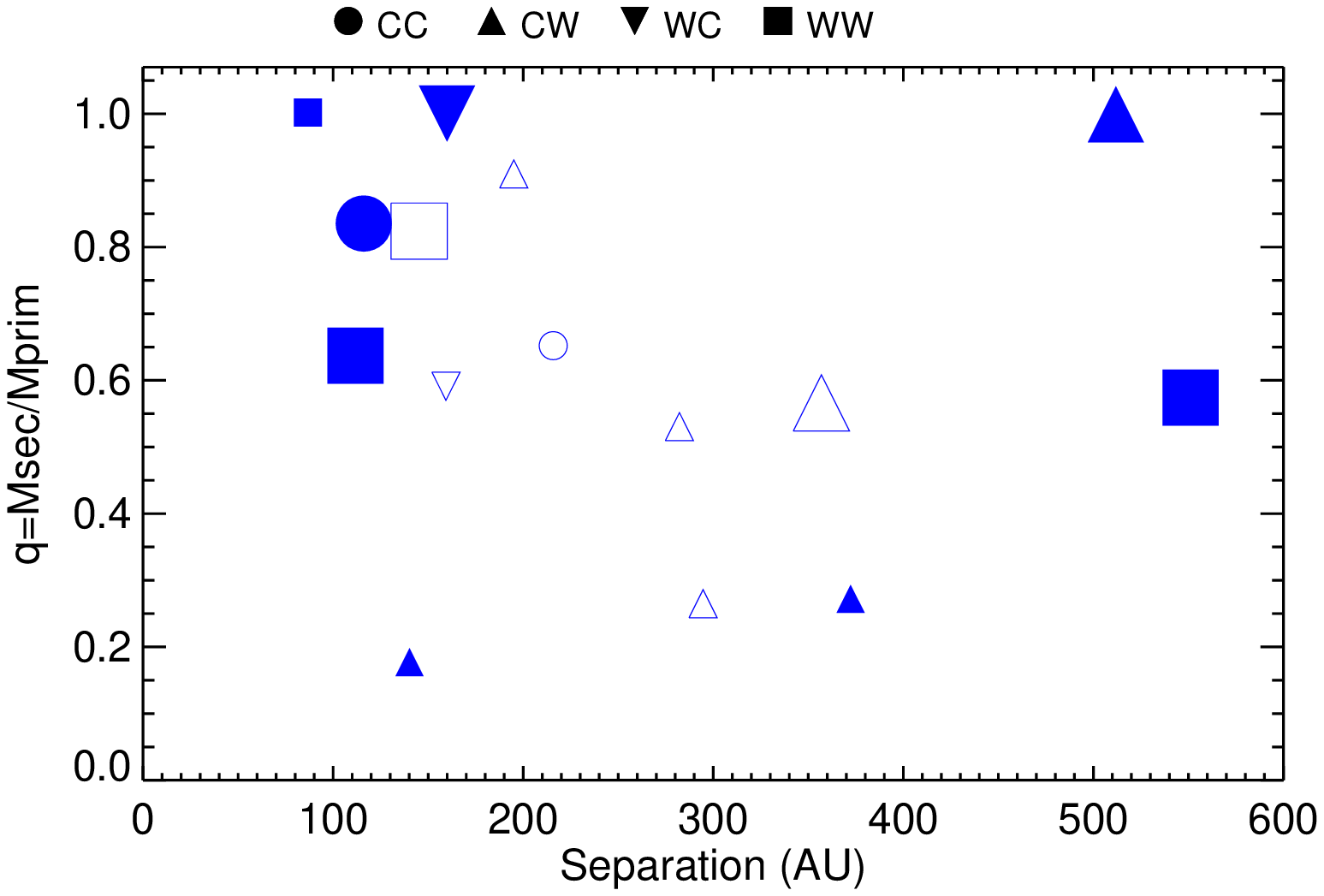}
\end{tabular}
\end{center}
\caption{Upper plot\,: Mass ratio vs the distance from $\theta^1$\,Ori\,C with T Tauri types (CTTS and WTTS). 
Small symbols denote systems with separation $<$\,220\,AU and large symbols separation $>$\,220\,AU. 
Lower plot\,: Mass ratio vs binary separation with T Tauri types. 
Small symbols correspond to systems at a distance to $\theta^1$\,Ori\,C $<$\,4\,arcmin and large symbols a distance $>$\,4\,arcmin. 
As in Fig.\,\ref{fig:mass ratio vs separation with TT types} open symbols denote ONC systems from D12. 
Mass ratios derived from S00 tracks are plotted.}
\label{fig:mass ratio vs cluster radial distance with TT types}
\end{figure}

The disk in mixed systems is mostly found around the primary for all binary samples. The only WC system identified from our sample (JW\,128) 
is an equal-mass binary component, therefore the definition of the primary is arbitrary. The same applies for the equal-mass binary system JW\,176.
Finally, there seems to be a dependence of the presence of a disk on stellar mass. Including K-L measurement errors, the disk fraction of binary 
components with S00 masses\,$>$\,0.4\,M$_{\sun}$ is 65\,$\pm$\,15\% while it is 35\,$\pm$\,20\% for components with S00 masses\,$<$\,0.4\,M$_{\sun}$. 
It is unlikely that this can be explained exclusively by a lower sensitivity to disks around cooler components. 
While the amplitude of the K-L excess likely depends on the component mass, 4 out of the 6 components with S00 masses\,$<$\,0.4\,M$_{\sun}$ have 
a spectral type earlier than M4 and therefore assuming that the amplitude of IR-excess in binaries is similar to that in T~Tauri singles, any K-L excess in 
these components should be detectable given our uncertainties (e.g. Kenyon \& Hartmann \cite{Kenyon_hartmann_1995}). For the coolest components, 
Liu et al. (\cite{Liu_etal_2003}) studied young brown dwarfs of M5.5-M9.5 type and found a K-L excess in the range $\sim$0.15-0.4 which could have 
been missed in some of our cases. There is therefore a hint of mass dependence on the presence of disks such that higher mass components are more likely to harbor a disk.

\subsection{Relation with distance from the Trapezium}
\label{sect:disk relation with distance from the Trapezium}

We also investigated the relation between the presence of an inner disk and the distance of the system from the Trapezium. 
Although the plot of Fig.\,\ref{fig:mass ratio vs cluster radial distance with TT types} shows no strong correlation, there is a trend toward a difference in the distance distribution of 
systems with and without disks such that systems with disks are found mainly in the inner regions at distance $\lesssim$\,7\,arcmin from $\theta^1$\,Ori\,C. 
Including the ONC systems T Tauri types from D12 leads to a K-S probability of 9\% that the distance distribution of systems with disks (CC, CW, and WC) and without (WW) 
are drawn from the same distribution. While the explanation for such a difference is unclear, increasing further the sample size will tell us whether this trend is real or not.


\section{Discussion}
\label{sect:Discussion}

Given the small sample size and the presence of potential selection biases, one should be cautious in interpreting the results. 
Nevertheless, a few interesting trends, which need to be confirmed with a larger sample size, are identified. 
They can be compared qualitatively with theoretical expectations and with results from other environments.

\subsection{ A steep mass ratio distribution in ONC and Taurus ?}
\label{sect:A steep mass ratio distribution in ONC and Taurus ?}

In the most complete sample of multiple systems in Tau-Aur to date, Kraus et al. (\cite{Kraus_etal_2011}) derived a mass ratio 
distribution in the primary mass range 0.25-0.7\,M$_{\sun}$ which is much flatter than that we obtained in both our Tau-Aur comparative 
sample and in our ONC sample. The exponent of the power-law they derived from their mass ratio distribution and using B98 tracks ($\Gamma$\,=\,0.2\,$\pm$\,0.4) 
is, however, still consistent with what we found here in our Tau-Aur comparison sample (see Sect.\,\ref{sect:mass ratio distribution}). 
An important difference between our study and that of Kraus et al. (\cite{Kraus_etal_2011}) is that we used samples of binaries 
with stellar properties derived for each component from spectra (this study, HK03, P03, D12) and/or visual photometry (WG01). 
This has the advantage of taking into account differential extinction and veiling. When using flux ratio and system spectral type like it is done 
in many instances in Kraus et al. (\cite{Kraus_etal_2011}) the mass ratio can be biased especially with flux ratios in K-band. Systems from our 
study, as well as from P03 and D12 studies, show that primaries are often significantly more veiled than their secondaries and that the visual 
extinction of primaries and secondaries are within a few magnitudes of each other. If the same holds for the systems in Tau-Aur then a method 
based on K-band flux ratios would lead to an overabundance of small mass ratio systems and therefore a flatter distribution. This could explain 
the fact that our Tau-Aur distribution is less flat than that derived in Kraus et al. (\cite{Kraus_etal_2011}) in the same primary mass range, 
although a combination with selection effects cannot be excluded. 

In order to check this assertion, we compared the mass ratio distribution for systems with mass ratios derived from Kraus et al. (\cite{Kraus_etal_2011}) 
and in the primary mass range 0.15-0.8\,M$_{\sun}$ using the two methods employed in this study. 
First, for systems with only K-band flux ratio and system spectral type (39 out of the 89 pairs of the study), and second, for systems with component 
spectral types taken from WG01 and HK03 and therefore in common with our Tau-Aur comparative sample (15 systems). The distribution derived from 
flux ratio is much flatter with $\Gamma$\,=\,0.4\,$\pm$\,0.5 than that using component spectral types for which $\Gamma$\,=\,1.1\,$\pm$\,0.7. Although 
the systems with only flux ratio have on average smaller separations than those with component spectral type (median of 20\,mas vs 80\,mas), this 
difference in mass ratio distribution is presumably not a selection effect since Kraus et al. (\cite{Kraus_etal_2011}) did not find evidence for a variation 
of mass ratio distribution with separation. It is also unlikely that the difference arises from a better sensitivity to low mass ratio systems at small separations 
($\Delta$K\,$\sim$\,5\,mag at $\lambda$/D, i.e. $\sim$\,70mas vs $\sim$\,2.5\,mag at $\sim$\,0$\farcs$1 for speckle surveys in the 90s) since most of the 
close new systems have $\Delta$K\,$<$\,2.5\,mag. Nevertheless, possible selection effects are that (1) the systems with only flux ratio are due to the 
use of a more complete census of cloud members including sources not only selected from H$\alpha$ surveys but also from criteria such as IR-excess 
and lithium test, and (2) the systems with component spectral types taken from WG01 and HK03 are not representative of the Taurus binary population. 

Our result is consistent with the study of mass ratio from Woitas et al. (\cite{Woitas_etal_2001}). Although they are estimating masses from system spectral type 
and flux ratio, the latter are in J-band which a priori minimizes biases from IR-excess. While their sample is composed of binaries from several star-forming regions 
(Tau-Aur, Cha~I, Lupus, Upper Scorpius) most systems are from Tau-Aur. Both the primary mass range and the separation range are similar to that 
of our samples. The power-law exponent we computed from the mass ratio distribution they derived with the B98 set of tracks is 0.8\,$\pm$\,0.2, in agreement 
with what we found in the ONC and in the comparative Tau-Aur sample. 

Although selection effects are a complex issue, we would like to understand to which extent those could affect our results. 
To do so, one has to consider the completeness of the multiplicity survey on which the binary sample is based (i.e. the observational 
technique and target list used) and how the observed binaries were selected from this sample. 
Since R07 searched for the multiplicity of all ONC members present in their H$\alpha$ imaging survey which covered most of the cluster, 
their sample can be considered as volume-limited. A possible source of bias in the binary sample of R07 is the strong and localized nebulosity 
emission in H$\alpha$ which could prevent from detecting faint companions although the survey is complete down to 5\,mag flux difference. 
The sensitivity to faint companions is also lower in the optical than at NIR wavelengths due to the steeper stellar mass-luminosity relation. 
Thus, in comparison with the NIR multiplicity surveys in Tau-Aur and Oph, the R07 survey is presumably more incomplete for small (q $\lesssim$ 0.2) 
mass ratio binaries. 
The sample of ONC binaries for which we investigated the mass ratio distribution includes the optically brightest systems of the R07's list\footnote{excluding 
the high-mass binary JW\,945 whose spectral type is B6.}, i.e. the 8 binaries of this study. The additional 7 systems from D12 are among the brightest systems 
found in the NIR surveys of Petr et al. (\cite{Petr_etal_1998}) and K\"ohler et al. (\cite{Koehler_etal_2006}). 
The natural question that arises is whether or not this selection of bright objects favors equal-mass systems. 
In order to check that, we compared the distribution of flux ratio in H$\alpha$ for our extended sample with that of the binary population from R07. 
For the 10 systems present in R07 (all from this study and 2 from D12), the distribution of flux ratio in H$\alpha$ presents a similar extension to the 
R07 distribution and is statistically indistinguishable with a K-S probability of 98\%. Assuming that the flux ratio in H$\alpha$ is indicative of the 
ratio of photospheric fluxes, this indicates that our extended ONC binary sample should not be significantly biased towards equal-mass systems. 
Since the selection of binaries in WG01 and HK03 is not based on flux ratios but on the at that time current binary census for WG01 and on T Tauri type and separation 
(CTTS and subarcsecond binaries) for HK03, a similar conclusion should be reached concerning the comparative sample we used in Tau-Aur. 
This suggests that selection biases are probably not a significant effect when analyzing the mass ratio distribution from the ONC and Tau-Aur samples. 
Whether selection criteria have led to a somewhat biased mass ratio distribution, studying a larger sample of ONC binaries is needed to reduce 
the statistical uncertainties, which are still too large to allow for decisive conclusions to be reached, relative both to other star-forming regions or to field stars.

\subsection{The mass ratio distribution in ONC and in other environments}
\label{sect:The mass ratio distribution in ONC and in other environments}

We found that the mass ratio distribution of wide ($\gtrsim$\,100\,AU) low-mass binary stars appears to be independent of the environment 
when comparing ONC, Tau-Aur, and Oph distributions. Bate (\cite{Bate_2000}, \cite{Bate_2001}) predicted that the decoupling of gas and stars 
would imply a flatter q-distribution in dense clusters than in isolated protostellar accretion for which the gas has a higher specific angular momentum. 
Accordingly, this would mean that the mass ratio distribution in ONC should be flatter than in Tau-Aur in contrast to what is observed. 
However, systems located in the outer cluster regions have been subject to little or no dynamical interactions compared to the inner cluster 
regions. If we only consider the systems closer than 3-4\,arcmin from the Trapezium, then the q-distribution is much flatter. With a dividing distance 
set to 3.5\,arcmin, the two q-distributions formed from S00 tracks and for systems in the primary mass range 0.15-0.8\,M$_{\sun}$ have 
$\Gamma$\,=\,0.0\,$\pm$\,0.2 and 0.3\,$\pm$\,0.6 for the inner and outer regions, respectively. While these numbers are not yet statistically significant, 
it could be an indication for a variation of the mass ratio distribution with the distance to the cluster center. 

There is an indication for a decreasing mass ratio with increasing separation in ONC binaries which would be 
consistent with theoretical predictions (Bate \cite{Bate_2001}, Bate \cite{Bate_2009}). Further there are hints that this trend is only present for systems 
located in the inner regions of the cluster with a dividing distance at $\approx$\,4\,arcmin. The mass ratio distribution of our comparison sample in Tau-Aur 
shows no significant variation with separation which is in agreement with the results from other studies (Woitas et al. \cite{Woitas_etal_2001}, 
Kraus et al. \cite{Kraus_etal_2011}). More generally, the trend with separation in the ONC is in contrast with the situation in the field and in other young 
associations and clusters (Reggiani \& Meyer \cite{Reggiani_Meyer_2011}) where no significant variation of the mass ratio distribution with binary 
separation was found.

In a more speculative way, the mass ratio distribution in the ONC could be composed of two regimes. A first regime for the systems in the inner cluster regions where 
dynamical interactions tend to favor the formation of low-mass ratio systems leading to a mass ratio distribution which is flatter than in Tau-Aur and 
is a decreasing function of the binary separation. A second regime in the outer regions of the cluster where the dynamical interactions are much reduced 
and is reminiscent of isolated star formation like in Tau-Aur, i.e. the mass ratio distribution tends to peak towards unity and is independent of separation.

\subsection{Circumstellar disks around wide binary components}
\label{sect:Circumstellar disks around wide binary components}

Two aspects can be considered regarding the evolution of disks in young binary components. First, one can compare the occurrence of circumstellar 
disks around wide binary components with that around apparently single stars for a given environment. Second, the relative evolution of disks around 
these binary components can be investigated as a function of environment as well. \\

A comparison of the fraction of disks in binary components (54\,$\pm$\,13\% and 52\,$\pm$\,9\% with D12 systems, including K-L measurement errors) 
with that in the population of single stars in ONC (80\,$\pm$\,7\%, Lada et al. \cite{Lada_etal_2000}) would indicate that there is a significantly lower fraction 
of disks in binaries compared to singles. In order to remove any possibility for accelerated disk evolution due to disk truncation, we could further restrict to 
systems with separation $>$\,200\,AU and would obtain 50\,$\pm$\,20\% and 56\,$\pm$\,13\% with D12 systems, including K-L measurement errors, i.e. 
the same result with a slightly lower statistical significance. However, since the survey of Lada et al. (\cite{Lada_etal_2000}) only covered the inner regions 
of the ONC (6$\times$6\,arcmin), we should take into account only systems located at a distance up to $\sim$\,3\,arcmin from $\theta^1$\,Ori\,C. There are only 2 
systems at such a distance in our sample (7 with D12 systems) and the corresponding fraction of disks in binary components becomes 72\,$\pm$\,22\% 
and 66\,$\pm$\,14\% with D12 systems, including K-L measurement errors, therefore not statistically different from that of singles. 
There is therefore no evidence for a difference in the frequency of disks, hence disk lifetime, around wide binary components and around singles in ONC.

The situation seems to be similar in Tau-Aur and Oph. In Tau-Aur there is a fraction of disks in binary components of  75\,$\pm$\,13\% (9 disks/12 components 
in systems with separation $>$200\,AU) in the WG01\,+\,HK03 sample. If one considers the Tau-Aur sample studied by McCabe et al. (\cite{McCabe_etal2006}) the 
fraction is 83\,$\pm$\,9\%, (15/18). These numbers agree with the disk fraction in Tau-Aur singles of 63\,$\pm$\,9\% derived from Spitzer measurements 
(Hern\'andez et al. \cite{Hernandez_etal_2007}, Hartmann et al. \cite{Hartmann_etal_2005}). 
Kraus et al. (\cite{Kraus_etal_2012}) reached a similar conclusion using a larger sample size and with binaries and singles identified at smaller spatial scales, 
although their result could have been biased by the use of unresolved disk indicators. This bias actually depends on the fraction of mixed pairs and their separation 
distribution. As already noted in D12, the effect could be substantial if mixed pairs are predominantly distributed at larger separations.
In Oph the fraction of disks in binary components is 79\,$\pm$\,11\% 
(11/14 components in systems with separation $\gtrsim$200\,AU) from the sample of P03. Wilking et al. (\cite{Wilking_etal_2008}) counted 93 Class\,II and 38 Class\,III 
objects among the L1688 cloud members based on Spitzer IRAC colors which implies a disk fraction of 71\,$\pm$\,4\%.
Therefore, in both these star-forming regions the disk fraction of singles and of wide binary components are in agreement as it is found in the inner regions of the 
ONC cluster. \\

The frequency of disks around primaries and secondaries agree within the uncertainties with values of 67\,$\pm$\,15\% and 47\,$\pm$\,16\%, 
respectively, and 69\,$\pm$\,11\% and 38\,$\pm$\,12\% when D12 systems are included. In this frequency the systems with mass ratio unity 
JW\,128 and JW\,176 are counted both in primaries and secondaries. 
On the other hand there is a significant fraction of mixed pairs (see Sect.\,\ref{sect:disk census, fraction of mixed pairs}) and a hint that this fraction 
could be higher than the frequency of mixed pairs in wide binaries in Taur-Aur and Oph within the same primary mass range. 

Mixed systems are more frequently found in wider pairs, their occurrence is mostly dependent of mass ratio (most strictly in Oph and more marginally in ONC), 
is increasing with primary mass, and the disk is generally found around the primary component. The fact that the presence of mixed pairs is relatively dependent 
of mass ratio is rather expected. It is reasonable to think that systems with exhausted secondary disks are systems with small mass ratio for which less material 
was accreted by the proto-secondary compared to the proto-primary hence leading to both low mass ratio and smaller secondary disk mass. 
In this framework, systems with similar stellar masses should have similar disk masses as well. A possible explanation for the presence of mixed pairs with 
high mass ratio could be that the disk around the secondary has an inner hole. McCabe et al. (\cite{McCabe_etal2006}) found that 14\,$\pm$\,5\,\% 
(7/51 binary components) of binary components have disks without K-L excess. We can therefore expect 2.6 binary components with disks without K-L excess in 
our sample which is similar to the 2 binary components (JW\,128\,A and JW\,176\,B) without K-L excess found in high mass ratio (q\,$>$\,0.9) mixed-systems. 
Alternatively, these systems could be high-order hierarchical multiples (triples, quadruples) for which the close (presumably a few AU to a few tens of AU depending 
on eccentricity) yet undetected companion would tidally truncate one of the component disk leading to its earlier dissipation. Incidentally, $\sim$\,25\,\% of wide T Tauri 
binaries are found to be triples or quadruples (Correia et al. \cite{Correia_etal_2006}), which is similar to the fraction of mixed pairs found in Taurus and Oph. 
The discovery of a hierarchical triple candidate in our small sample size (JW\,248), for which the separation of the additional companion is however likely too large 
to lead to an early disk dissipation at a $\sim$\,1\,Myr age, strengthens the possibility that the origin of mixed pairs in ONC and perhaps also in Taurus and Oph 
could be in part due to hierarchical multiples.

In addition to the above considerations, it could be that, as for the mass ratio distribution, the systems in the inner regions are dominated by mixed pairs 
due to the fact that accretion is occurring mainly on the primary star and disk. As for the mass ratio, studies of protostellar accretion suggest that most 
of the accreted material should be accreted by the primaries in a cluster environment because of the small specific angular momentum of this material. 
Therefore, disks should be significantly less massive and long-lived around secondaries than around primaries. As for mass ratios, it would be expected that the effect is 
stronger in the inner cluster regions implying a larger number of mixed pairs in these regions and that mixed pairs occur primarily in low-mass ratio systems. 
This is what is found for the 2 systems at distance $\lesssim$\,3.5\,arcmin from $\theta^1$\,Ori\,C with known K-L colors, and for 4 out 7 systems in these regions 
when including systems from D12.

Finally, a high fraction of mixed pairs in dense clusters, and further the fact that the lone disk is usually found around the primary, would be consistent 
with a small fraction of extrasolar planets around the secondary component of binaries. Although the statistics of close hot Jupiter-like planets and Super-Earths 
around the components of mature binary stars is clearly biased against their detection around secondaries, the scarcity of planets around secondaries\,\footnote{to 
the best of our knowledge there are only five planet host secondaries known so far\,: GJ\,667\,C (e.g. Delfosse et al. \cite{Delfosse_etal_2012}), 
83\,Leo\,B (Mugrauer et al. \cite{Mugrauer_etal_2007}), ADS\,16402 B (Bakos et al. \cite{Bakos_etal_2007}), 16\,Cyg\,B (Cochran et al. \cite{Cochran_etal_1997}), 
and $\alpha$\,Cen\,B (Dumusque et al. \cite{Dumusque_etal_2012})}. indicates that their frequency is probably rather small compared to that around primaries. 
This could be explained by the absence of the inner circumsecondary disk in mixed pairs at an evolutionary stage as early as a few Myr because 
the suggested short lifetime of these disks would a priori preclude the formation of planets.


\section{Summary}
\label{sect:summary}

We have derived the stellar and circumstellar properties for the components of 8 young low-mass wide binaries in the ONC through 
spatially-resolved K-band R\,$\sim$\,5000 spectroscopy and near-IR JHKL' photometry. The sample is complemented with 7 additional systems 
for which the stellar properties were recently derived using similar methods (Daemgen, Correia \& Petr-Gotzens \cite{Daemgen_etal_2012}) and 
comparison is made with samples from Tau-Aur and Oph in order to investigate the dependence on the environment.  
The main results are as follows\,: \\

(1) The mass ratio distribution of ONC binaries is indistinguishable from that of Tau-Aur and to some extent from that of Oph. 
There is an indication for a variation of the mass ratio distribution with separation for ONC binaries which is not seen in Tau-Aur binaries. 
Such variation may be limited to systems located in the inner cluster regions. \\ 

(2) The components of ONC binaries are significantly more coeval than the overall ONC population and 
are found to be as coeval as components of binaries in Tau-Aur and Oph within the uncertainties. \\

(3) The presence of inner disks around wide ONC binary components is found to be less frequent than the presence of inner disks around ONC single stars. 
However, when considering only systems located in the inner cluster regions, the fractions of disks around wide binary components and singles are 
not statistically different in ONC, as it is found in Tau-Aur and Oph. \\

(4) There is a hint of a larger fraction of mixed pairs in wide ONC binaries in comparison to wide binaries in Tau-Aur and Oph within the same primary mass range. 
The mass ratio distribution of mixed and unmixed pairs are found to be significantly different for all systems of the three star-forming regions. \\

(5) Among the 8 ONC binaries of our sample we uncovered a new hierarchical triple candidate. This supports the idea that hierarchical triples could be a 
non-negligible cause for the large fraction of mixed pairs in the ONC.


\begin{acknowledgements}
We thank the referee for detailed and helpful comments on this paper. 
The authors wish to recognize and acknowledge the very significant cultural role and reverence that the summit of Mauna Kea has always had within 
the indigenous Hawaiian community. We are most fortunate to have the opportunity to conduct observations from this mountain. This material is based 
upon work supported by the National Aeronautics and Space Administration through the NASA Astrobiology Institute under Cooperative Agreement 
No. NNA09DA77A issued through the Office of Space Science. This research made use of data products from the Two Micron All Sky Survey and of the 
SIMBAD database, operated at CDS, Strasbourg, France.

\end{acknowledgements}


\begin{appendix} 

%
\section{Line ratio analysis}
\label{appendix : line ratio analysis}

%
\begin{figure*}
\begin{center}
\begin{tabular}{cc}
\includegraphics[width=9cm, angle=0]{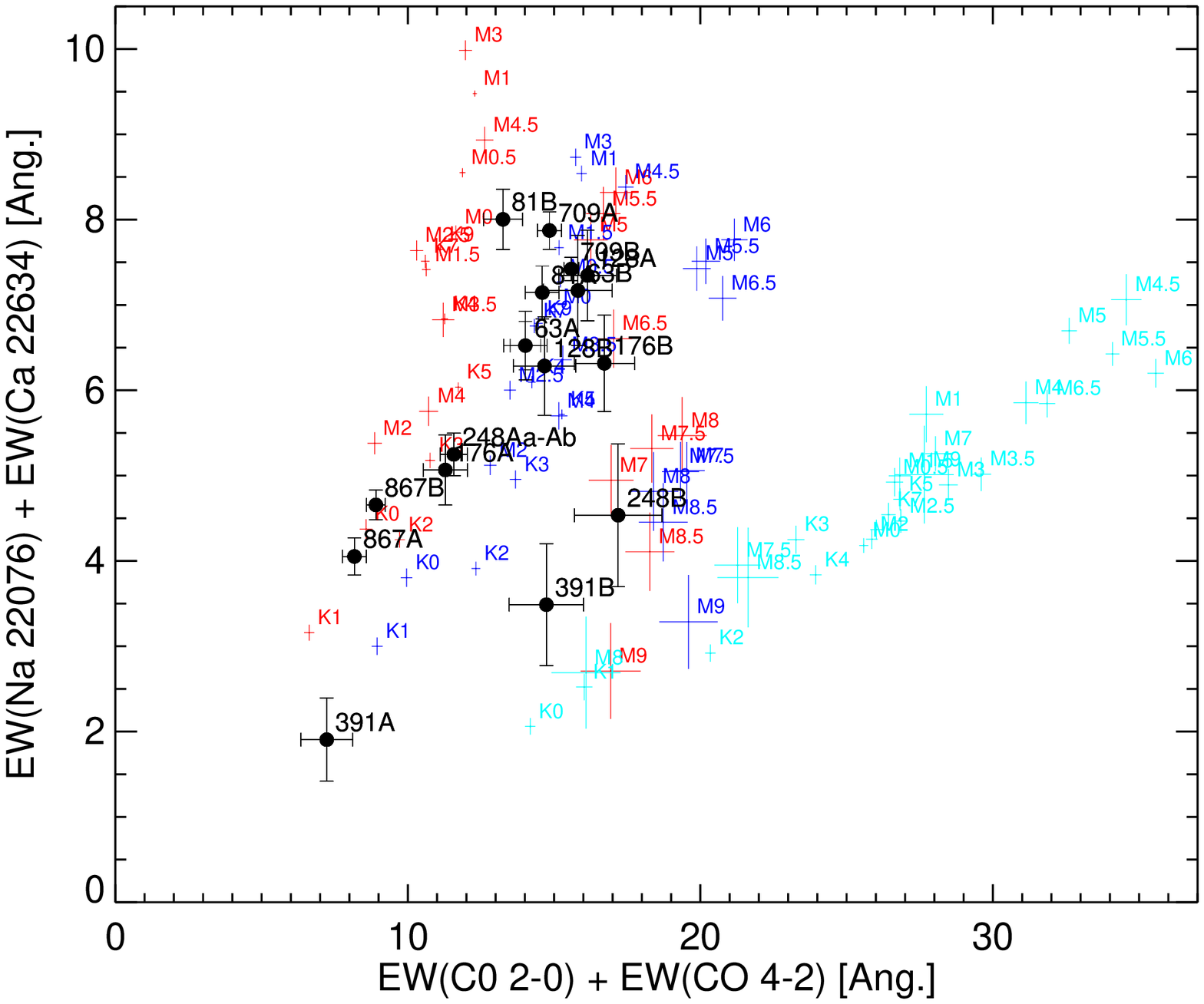} 
\includegraphics[width=9cm, angle=0]{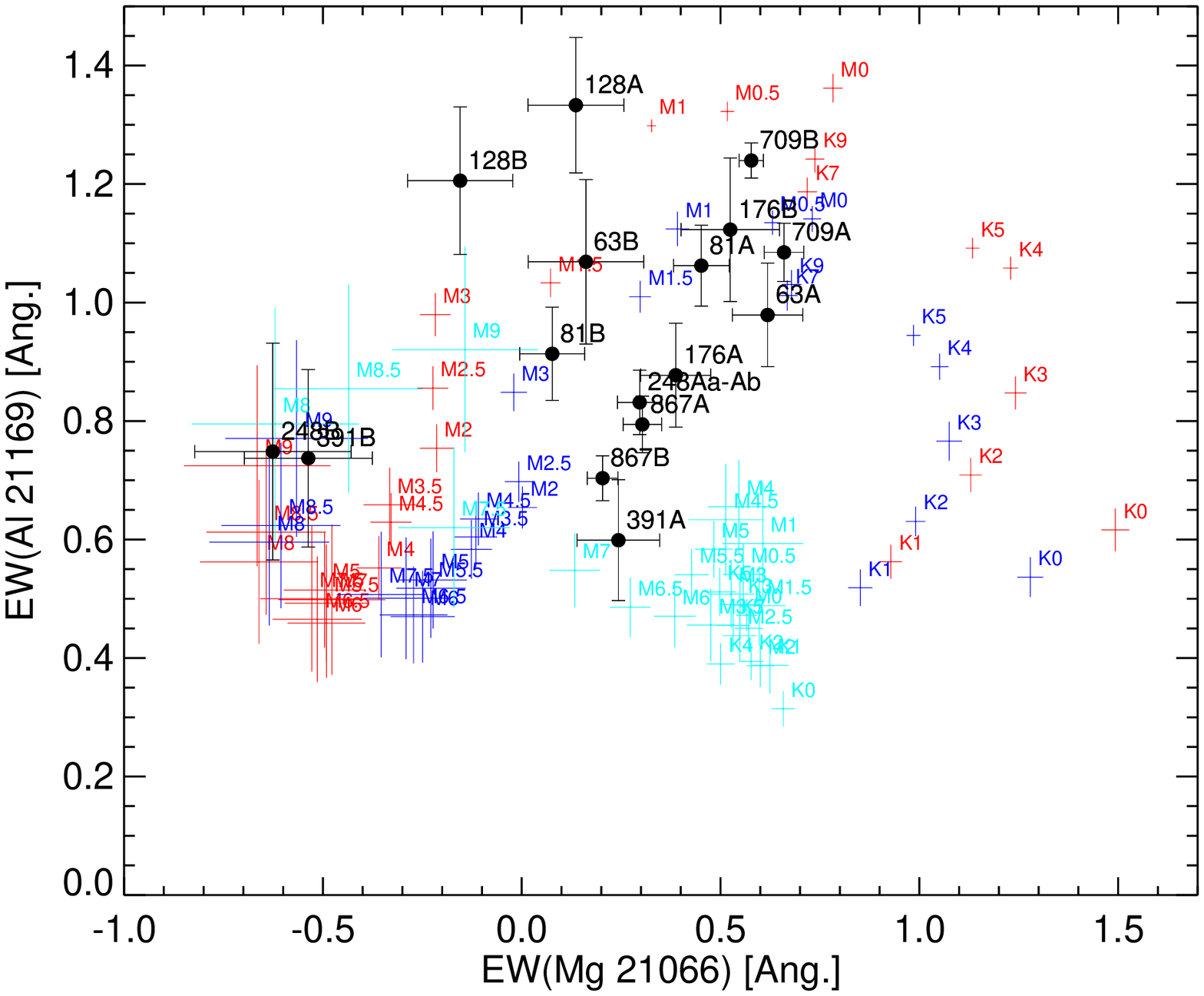} \\
\includegraphics[width=9cm, angle=0]{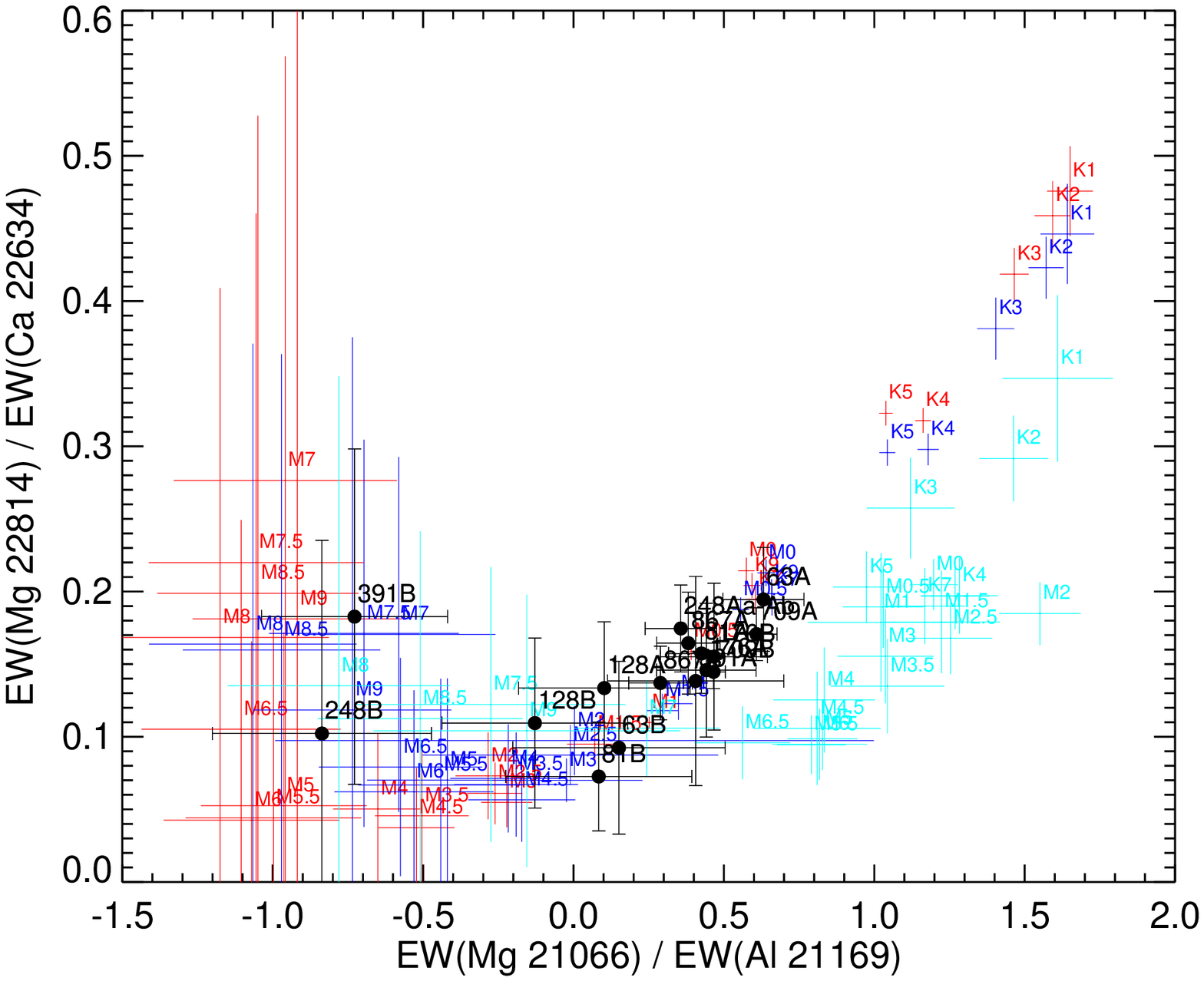}
\includegraphics[width=9cm, angle=0]{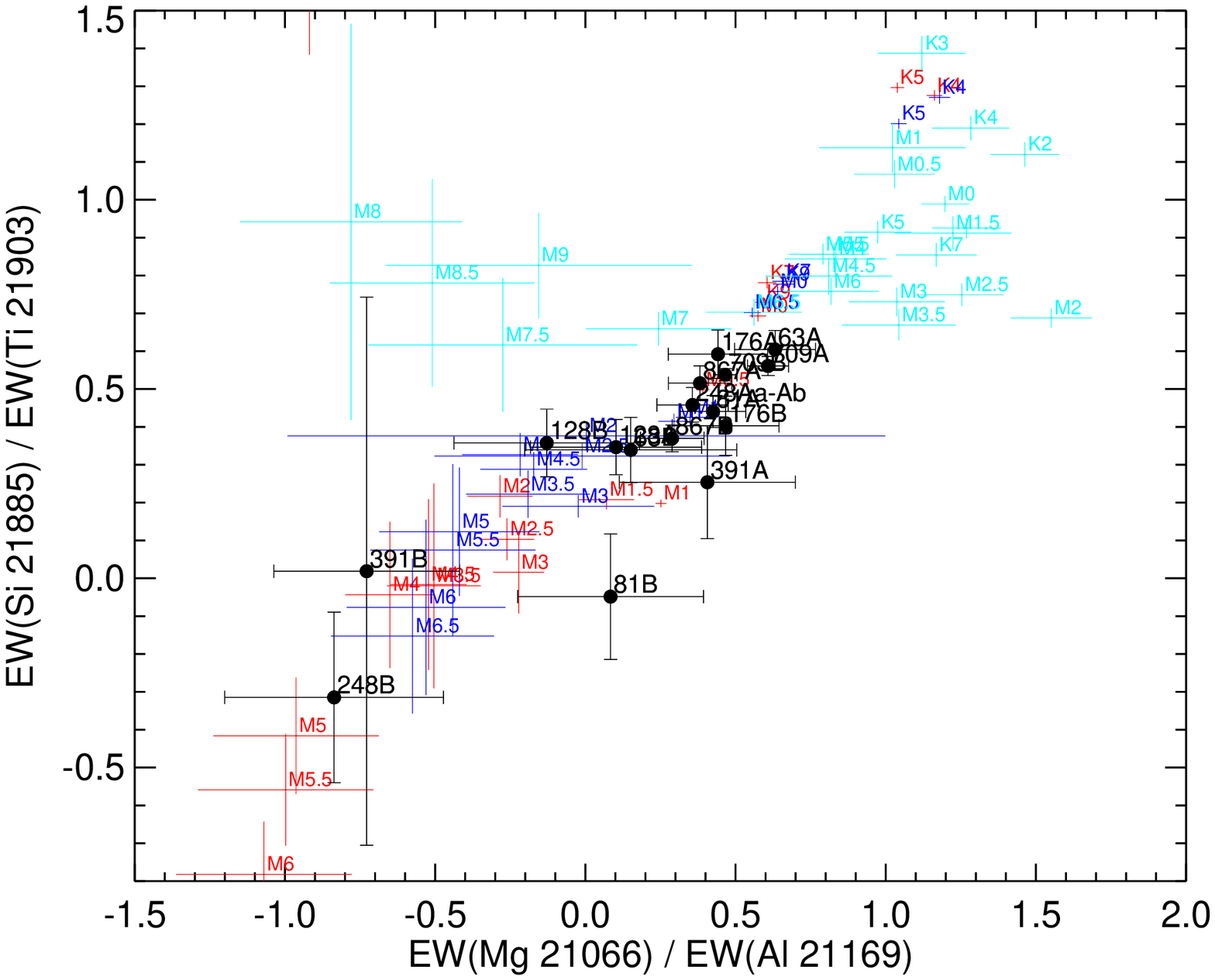} \\
\end{tabular}
\end{center}
\caption{Locus of the binary components in several equivalent width diagnostic plots. These plots are the Na 22076\,$\mu$m + Ca 2.2634\,$\mu$m versus CO(4-2) + CO(2-0) (upper right), 
Al 2.1169\,$\mu$m versus Mg 2.1066\,$\mu$m (upper left), Mg 2.2814\,/\,Ca 2.2634\,$\mu$m versus Mg 2.1066\,/\,Al 2.1169\,$\mu$m (lower left), 
and Si 2.1885\,/\, Ti 21903\,$\mu$m versus Mg 2.1066\,/\,Al 2.1169\,$\mu$m (lower right). The ONC binary components (in black) are labeled, as well as the spectral type of the dwarf (red), 
giant (turquoise), and intermediate-gravity (blue) spectral type templates.}
\label{fig: diagnostic plots}
\end{figure*}

%
\begin{landscape}
\begin{table}[htbp]
\scriptsize
\caption{Determination of spectral type using several line ratios. The final result considered in this paper and reported in Table\,\ref{Table : stellar parameter - intermediate results} 
is the value in the column "excluding outliers". }
\begin{center}
\renewcommand{\arraystretch}{0.7}
\setlength\tabcolsep{7pt}
\begin{tabular}{l@{\hspace{1mm}}
			c@{\hspace{1mm}}
			c@{\hspace{1mm}}
			c@{\hspace{1mm}}
			c@{\hspace{1mm}}
			c@{\hspace{1mm}}
			c@{\hspace{1mm}}
			c@{\hspace{1mm}}
			c@{\hspace{1mm}}
			c@{\hspace{1mm}}
			c@{\hspace{1mm}}
			c@{\hspace{1mm}}
			c@{\hspace{3mm}}
			c@{\hspace{3mm}}
			c@{\hspace{3mm}}
			}

\hline\noalign{\smallskip}

JW										&
Si 20922									&
Mg 21066									&
Mg 21066									&
Si 21825									&
Si 21885									&
Si 21825									&
Fe 22392									&
Fe 22392									&
(Ti 22217+Ti 22238)							&
Mg 22814									&
Ca 22634									&
										&
										&
										\\

										&
/										&
/										&
/										&
/										&
/										&
/										&
/										&
/										&
/										&
/										&
/										&
result									&
excluding									&
excluding									\\

										&
Mg 21066									&
Al 21098									&
Al 21169									&
Ti 22010									&
Ti 21903									&
Si,Ti 21787								&
Ca 22634									&
(Ti 22217+Ti 22238)							&
Ca 22634									&
Ca 22634									&
CO 2-0 22935								&
										&
outliers									&
 Mg 22814 / Ca 22634						\\

\noalign{\smallskip}
\hline

\noalign{\smallskip}
\object{63 A}   &	not constraining & K5.95\,$\pm$\,0.95$^*$  & K8.65\,$\pm$\,1.85 & $^a$ & M0.67\,$\pm$\,0.14 & $^b$ & M1.05\,$\pm$\,0.45 & M1.57\,$\pm$\,0.18 & M3.98\,$\pm$\,0.91$^*$ & K8.75\,$\pm$\,1.75 & ...  & M0.9\,$\pm$\,1.0 & M1.0\,$\pm$\,0.6 &	  ... \\
\noalign{\smallskip}
\object{63 B}   &	... & ... & ... & $^a$ & M2.81\,$\pm$\,1.81 & $^b$ & M3.28\,$\pm$\,1.48 & M3.00\,$\pm$\,1.00 & M4.76\,$\pm$\,1.30 & M2.25\,$\pm$\,0.45 & ... & M2.6\,$\pm$\,1.0 & no deviants & M3.5\,$\pm$\,0.9 \\
\noalign{\smallskip}
\noalign{\smallskip}

\object{81 A}   & not constraining &  M0.75\,$\pm$\,0.25 &  M0.82\,$\pm$\,0.11 &  $^a$ &  M1.36\,$\pm$\,0.48 & M0.5\,$\pm$\,1.5 & M0.55\,$\pm$\,0.25 & M1.75\,$\pm$\,0.15 & M6.0\,$\pm$\,0.5$^*$ & M0.73\,$\pm$\,0.10 & ... & M0.9\,$\pm$\,0.9 &  M0.8\,$\pm$\,0.1 &  ... \\
\noalign{\smallskip}
\object{81 B}   & ...  & ...  & ...  & M3.25\,$\pm$\,1.25 & ...  & not constraining & M1.0\,$\pm$\,0.5$^*$ & M5.27\,$\pm$\,3.65 & M3.75\,$\pm$\,0.25 & M3.38\,$\pm$\,0.63 & ... & M2.0\,$\pm$\,2.0 & M3.7\,$\pm$\,0.3 & M3.7\,$\pm$\,0.5 \\
\noalign{\smallskip}
\noalign{\smallskip}

\object{128 A}   & ... & ... & ... & $^a$ &	 M2.86\,$\pm$\,1.69 & $^b$ & M3.30\,$\pm$\,1.70 & M5.23\,$\pm$\,3.34 & M4.55\,$\pm$\,1.15 &	M1.15\,$\pm$\,0.35 & ... &	 M1.6\,$\pm$\,2.0 & no deviants & M3.9\,$\pm$\,1.0 \\
\noalign{\smallskip}
\object{128 B}   & ... & ... & ... & $^a$ &	 M2.76\,$\pm$\,1.82 & $^b$ & M3.25\,$\pm$\,1.55 &	M5.52\,$\pm$\,2.99 & M4.70\,$\pm$\,1.30 & M1.75\,$\pm$\,0.75 & ... & M2.7\,$\pm$\,1.6 & no deviants & M3.9\,$\pm$\,1.1  \\
\noalign{\smallskip}
\noalign{\smallskip}

\object{176 A}   & ... &	M1.0\,$\pm$\,0.5 & M0.78\,$\pm$\,0.18 & $^a$ & M0.70\,$\pm$\,0.18 & $^b$ & M3.30\,$\pm$\,1.70$^*$ & M5.30\,$\pm$\,3.51$^*$ & M3.98\,$\pm$\,0.94$^*$ & M0.82\,$\pm$\,0.10 &	 ... & M0.8\,$\pm$\,0.4 & M0.8\,$\pm$\,0.1 & ...  \\
\noalign{\smallskip}
\object{176 B}   & ... & K5.2\,$\pm$\,1.7$^*$ & M0.72\,$\pm$\,0.20 & $^a$ & M1.7\,$\pm$\,0.8 & $^b$ & K9.25\,$\pm$\,2.25 & M5.27\,$\pm$\,3.65$^*$ & M5.0\,$\pm$\,1.5$^*$ & M0.85\,$\pm$\,0.10 & ... & M0.8\,$\pm$\,0.7 &	M0.8\,$\pm$\,0.2 &  ... \\
\noalign{\smallskip}
\noalign{\smallskip}

\object{248 Aa-Ab}   & not constraining & M0.75\,$\pm$\,0.25 & M1.11\,$\pm$\,0.23 & K6.5\,$\pm$\,3.5$^*$ & M1.15\,$\pm$\,0.35  & M0.75\,$\pm$\,1.75 &  K7.75\,$\pm$\,2.75$^*$ & M1.65\,$\pm$\,0.15$^*$ & not constraining & M0.6\,$\pm$\,0.1 & ... & M0.9\,$\pm$\,0.6 & M0.7\,$\pm$\,0.3 & ... \\
\noalign{\smallskip}
\object{248 B}   & ... & ... & ... & ... & ... & ... & ... & ... & M7.62\,$\pm$\,1.38 & M7.65\,$\pm$\,1.35 & M5.37\,$\pm$\,0.63 & M6.0\,$\pm$\,1.5 & no deviants & M5.8\,$\pm$\,1.6 \\
\noalign{\smallskip}
\noalign{\smallskip}

\object{391 A}   & ... &	K8.75\,$\pm$\,2.75 & M1.04\,$\pm$\,0.48 & ... & ... &	... & ... & ... & M5.0\,$\pm$\,1.5$^*$ &	 M0.88\,$\pm$\,0.15 & ... & M0.8\,$\pm$\,1.1 & M0.8\,$\pm$\,0.6 &  ... \\
\noalign{\smallskip}
\object{391 B}   & ... & ... & ... & ... & ... & ... &	 ... & ... & M7.72\,$\pm$\,0.89 & M7.65\,$\pm$\,0.85 & M5.65\,$\pm$\,0.85 & M7.0\,$\pm$\,1.2 & no deviants &	 M6.6\,$\pm$\,1.5 \\
\noalign{\smallskip}
\noalign{\smallskip}

\object{709 A}   & not constraining & K8.75\,$\pm$\,1.75 & K8.75\,$\pm$\,1.75 &	 M1.10\,$\pm$\,0.60 & M0.76\,$\pm$\,0.07 & M3.52\,$\pm$\,3.09 & K7.75\,$\pm$\,2.75 & M1.25\,$\pm$\,0.25 & M4.5\,$\pm$\,1.0$^*$ & M0.59\,$\pm$\,0.10 & ... & M0.7\,$\pm$\,0.3 & M0.7\,$\pm$\,0.4 &	...   \\
\noalign{\smallskip}
\object{709 B}   & M0.60\,$\pm$\,0.05 & M0.59\,$\pm$\,0.08 & M0.72\,$\pm$\,0.04 & K8.75\,$\pm$\,1.25$^*$ & M0.80\,$\pm$\,0.04 &	 K3.75\,$\pm$\,1.25$^*$ & K8.75\,$\pm$\,1.75$^*$ & M1.21\,$\pm$\,0.17$^*$ & M4.13\,$\pm$\,0.76$^*$ & M0.81\,$\pm$\,0.03 & ... & M0.8\,$\pm$\,0.2 &	M0.8\,$\pm$\,0.1 &  ...  \\
\noalign{\smallskip}
\noalign{\smallskip}

\object{867 A}   & not constraining & K8.9\,$\pm$\,1.9 & M0.95\,$\pm$\,0.13 & K2.8\,$\pm$\,0.3$^*$ & M0.84\,$\pm$\,0.10 & not reliable &	 M1.13\,$\pm$\,0.38 & M4.71\,$\pm$\,2.86$^*$ & M6.5\,$\pm$\,0.2$^*$ &	M0.68\,$\pm$\,0.08 & ... & M0.9\,$\pm$\,2.5 &	M0.8\,$\pm$\,0.2 &  ...  \\
\noalign{\smallskip}
\object{867 B}   & ... &	K6.0\,$\pm$\,0.5$^*$ & M1.39\,$\pm$\,0.17 & not constraining & M2.01\,$\pm$\,0.38 & not constraining & M1.3\,$\pm$\,0.3 & M5.32\,$\pm$\,3.28$^*$ & M7.69\,$\pm$\,1.05$^*$ &	 M0.89\,$\pm$\,0.05 & ... &	M0.9\,$\pm$\,1.2 & M1.0\,$\pm$\,0.4 & M1.5\,$\pm$\,0.3 \\
\noalign{\smallskip}
\noalign{\smallskip}

\hline
\end{tabular}
\end{center}
\label{Table : line ratio analysis}
\begin{minipage}[position]{18cm}
Note\,: \\
$^*$\,: outlier. \\   
$^a$\,: Ti 22010 not reliable. \\   
$^b$\,: Si 21825 contaminated. \\   
\end{minipage}
\end{table}
\end{landscape}

%
\section{Comparative samples of T Tauri binaries}
\label{appendix : Comparative samples of T Tauri binaries}

We used several samples of T Tauri binaries to study the dependency of the properties of T Tauri binaries on the environment. In order to minimize 
potential biases introduced by relative extinction and veiling, we prefer to use systems with a significant amount of observational data especially 
including spectroscopy and/or visual photometry. The considered samples are those from WG01 (20 systems) and HK03 (19 systems) in Tau-Aur, 
from P03 in Oph (8 systems), and from D12 in ONC (7 systems). 
We used the spectral types and luminosities reported in these studies to derive the component mass and age from the B98, PS99, and S00 PMS tracks. 
For those 9 systems from WG01 found in HK03 we used the stellar parameters derived in HK03. Errors in spectral type were assumed to be $\pm$\,1\,subclass 
for WG01, HK03, and P03 and those reported in the case of D12. For HK03 systems uncertainties in log L were assumed to be 0.2\,dex. The T~Tauri types 
(C for CTTS and W for WTTS) were assigned from resolved IR-excess measurements, notably the K-L measurements from McCabe et al. (\cite{McCabe_etal2006}). 
For the D12 sample, we made use of criteria such as r$_{\mathrm{K}}$, H-K color excess, and Br$\gamma$ emission. The component masses and 
ages are reported in Table\,\ref{Table : comparative samples} and the system mass ratios and relative ages are shown in 
Table\,\ref{Table : comparative samples : mass ratios and relative ages}.

\begin{table*}
\scriptsize
\caption{Derived stellar properties of comparative samples of T Tauri binaries. The PMS tracks are those from Baraffe et al. (\cite{Baraffe_etal1998}) - B98, Palla \& Stahler (\cite{Palla_Stahler_1999}) - PS99, and Siess et al. (\cite{Siess_etal_2000}) - S00.}
\begin{center}
\renewcommand{\arraystretch}{1.0}
\setlength\tabcolsep{2pt}
\begin{tabular}{l@{\hspace{2mm}}
			r@{\hspace{2mm}}
			l
			l
			c
			c@{\hspace{1mm}}
			c@{\hspace{1mm}}
			c@{\hspace{1mm}}
			c@{\hspace{1mm}}
			c@{\hspace{1mm}}
			c@{\hspace{1mm}}
			c@{\hspace{2mm}}
			c@{\hspace{1mm}}
			c@{\hspace{1mm}}
			c@{\hspace{1mm}}
			c@{\hspace{1mm}}
			c@{\hspace{1mm}}
			c@{\hspace{1mm}}
			c@{\hspace{2mm}}
			c@{\hspace{1mm}}
			c@{\hspace{1mm}}
			c@{\hspace{1mm}}
			c@{\hspace{1mm}}
			c@{\hspace{1mm}}
			c@{\hspace{1mm}}
			c@{\hspace{2mm}}
			c@{\hspace{1mm}}
			c@{\hspace{1mm}}
			c@{\hspace{1mm}}
			c@{\hspace{1mm}}
			c@{\hspace{1mm}}
			c@{\hspace{1mm}}
			}

\hline\noalign{\smallskip}

	 							&  
	 							&  
					 			& 
								&   
								&   
\multicolumn{6}{c}{Mass primary (M$_{\sun}$)} 	&  
								&
\multicolumn{6}{c}{Age primary (Myr)} 		&
								&  
\multicolumn{6}{c}{Mass secondary (M$_{\sun}$)} 	&  
								&
\multicolumn{6}{c}{Age secondary (Myr)} 		\\

\multicolumn{32}{c}{\vspace{-3mm}} \\

\cline{6-11} \cline{13-18} 	\cline{20-25}  \cline{27-32}  \\ 

\multicolumn{32}{c}{\vspace{-3mm}} \\

	 						&  
Sep.							&							
SpT							& 
SpT							& 
T Tauri						&							
\multicolumn{2}{c}{B98} 		 	&  
\multicolumn{2}{c}{PS99} 		 	&  
\multicolumn{2}{c}{S00}   			& 
							&
\multicolumn{2}{c}{B98} 		 	&  
\multicolumn{2}{c}{PS99} 		 	&  
\multicolumn{2}{c}{S00}			&
							&
\multicolumn{2}{c}{B98} 		 	&  
\multicolumn{2}{c}{PS99} 		 	&  
\multicolumn{2}{c}{S00}   			& 
							&
\multicolumn{2}{c}{B98} 		 	&  
\multicolumn{2}{c}{PS99} 		 	&  
\multicolumn{2}{c}{S00}			\\

Name 						&  
(AU)       						& 
(p)							& 
(s)							& 
Type							&
\multicolumn{2}{c}{} 		 		&  
\multicolumn{2}{c}{} 		 		&  
\multicolumn{2}{c}{}   			& 
							&
\multicolumn{2}{c}{}  				&  
\multicolumn{2}{c}{}  				&  
\multicolumn{2}{c}{}  				&  
							&
\multicolumn{2}{c}{} 		 		&  
\multicolumn{2}{c}{} 		 		&  
\multicolumn{2}{c}{}   			& 
							&
\multicolumn{2}{c}{}  				&  
\multicolumn{2}{c}{}  				&  
\multicolumn{2}{c}{}  				\\

\noalign{\smallskip}
\hline
\noalign{\smallskip}
\multicolumn{32}{c}{Tau-Aur from WG01} \\
\noalign{\smallskip}
\hline
\noalign{\smallskip}
\object{FW Tau A-B     }   &     10   &   M5.5   &   M5.5   &   WW   &   \multicolumn{2}{c}{ 0.14\,$^{+ 0.07}_{- 0.07}$}   &   \multicolumn{2}{c}{ 0.11\,$^{+ 0.05}_{- 0.05}$}   &   \multicolumn{2}{c}{ 0.14\,$^{+ 0.05}_{- 0.05}$}   &   &\multicolumn{2}{c}{ 1.62\,$^{+  5.41}_{-  0.62}$}   &   \multicolumn{2}{c}{ 1.01\,$^{+  3.47}_{-  0.15}$}   &   \multicolumn{2}{c}{ 4.04\,$^{+  1.67}_{-  0.69}$}   &   &\multicolumn{2}{c}{ 0.14\,$^{+ 0.06}_{- 0.06}$}   &   \multicolumn{2}{c}{ 0.11\,$^{+ 0.05}_{- 0.05}$}   &   \multicolumn{2}{c}{ 0.13\,$^{+ 0.04}_{- 0.04}$}   &   &\multicolumn{2}{c}{ 1.81\,$^{+  4.34}_{-  0.81}$}   &   \multicolumn{2}{c}{ 1.26\,$^{+  3.67}_{-  0.17}$}   &   \multicolumn{2}{c}{ 4.33\,$^{+  2.34}_{-  0.81}$}       \\
\noalign{\smallskip}
      
\object{332 G1         }   &     33   &   K7     &   M1     &   WW   &   \multicolumn{2}{c}{ 0.82\,$^{+ 0.07}_{- 0.07}$}   &   \multicolumn{2}{c}{ 0.79\,$^{+ 0.08}_{- 0.08}$}   &   \multicolumn{2}{c}{ 0.78\,$^{+ 0.10}_{- 0.10}$}   &   &\multicolumn{2}{c}{ 3.10\,$^{+  6.05}_{-  2.10}$}   &   \multicolumn{2}{c}{ 4.01\,$^{+  3.13}_{-  0.73}$}   &   \multicolumn{2}{c}{ 4.37\,$^{+  1.24}_{-  0.68}$}   &   &\multicolumn{2}{c}{ 0.60\,$^{+ 0.07}_{- 0.07}$}   &   \multicolumn{2}{c}{ 0.47\,$^{+ 0.12}_{- 0.12}$}   &   \multicolumn{2}{c}{ 0.46\,$^{+ 0.09}_{- 0.09}$}   &   &\multicolumn{2}{c}{ 1.00\,$^{+  1.50}_{-  0.00}$}   &   \multicolumn{2}{c}{ 0.75\,$^{+  0.35}_{-  0.14}$}   &   \multicolumn{2}{c}{ 1.05\,$^{+  0.52}_{-  0.19}$}       \\
\noalign{\smallskip}
      
\object{V410 Tau A-C   }   &     40   &   K4     &   M5.5   &        &   \multicolumn{2}{c}{ 1.31\,$^{+ 0.16}_{- 0.16}$}   &   \multicolumn{2}{c}{ 1.50\,$^{+ 0.15}_{- 0.15}$}   &   \multicolumn{2}{c}{ 1.50\,$^{+ 0.20}_{- 0.20}$}   &   &\multicolumn{2}{c}{ 1.43\,$^{+  1.25}_{-  0.43}$}   &   \multicolumn{2}{c}{ 2.53\,$^{+  0.35}_{-  0.37}$}   &   \multicolumn{2}{c}{ 2.69\,$^{+  0.44}_{-  0.36}$}   &   &\multicolumn{2}{c}{ 0.13\,$^{+ 0.05}_{- 0.05}$}   &   \multicolumn{2}{c}{ 0.11\,$^{+ 0.06}_{- 0.06}$}   &   \multicolumn{2}{c}{ 0.12\,$^{+ 0.03}_{- 0.03}$}   &   &\multicolumn{2}{c}{ 2.56\,$^{+  3.30}_{-  1.56}$}   &   \multicolumn{2}{c}{ 2.16\,$^{+  0.76}_{-  0.34}$}   &   \multicolumn{2}{c}{ 5.02\,$^{+  0.21}_{-  0.53}$}       \\
\noalign{\smallskip}
      
\object{UZ Tau A-Ba    }   &    498   &   M1     &   M2     &   CC   &   \multicolumn{2}{c}{ 0.60\,$^{+ 0.08}_{- 0.08}$}   &   \multicolumn{2}{c}{ 0.48\,$^{+ 0.11}_{- 0.11}$}   &   \multicolumn{2}{c}{ 0.47\,$^{+ 0.08}_{- 0.08}$}   &   &\multicolumn{2}{c}{ 3.83\,$^{+  5.93}_{-  2.83}$}   &   \multicolumn{2}{c}{ 2.77\,$^{+  3.18}_{-  0.23}$}   &   \multicolumn{2}{c}{ 3.27\,$^{+  2.22}_{-  0.30}$}   &   &\multicolumn{2}{c}{ 0.56\,$^{+ 0.10}_{- 0.10}$}   &   \multicolumn{2}{c}{ 0.36\,$^{+ 0.11}_{- 0.11}$}   &   \multicolumn{2}{c}{ 0.38\,$^{+ 0.07}_{- 0.07}$}   &   &\multicolumn{2}{c}{ 2.54\,$^{+  4.69}_{-  1.54}$}   &   \multicolumn{2}{c}{ 1.61\,$^{+  0.31}_{-  0.27}$}   &   \multicolumn{2}{c}{ 1.98\,$^{+  0.53}_{-  0.32}$}       \\
\noalign{\smallskip}
      
\object{FV Tau A-B     }   &     98   &   K5     &   K6     &   CC   &   \multicolumn{2}{c}{ 1.06\,$^{+ 0.15}_{- 0.15}$}   &   \multicolumn{2}{c}{ 1.05\,$^{+ 0.15}_{- 0.15}$}   &   \multicolumn{2}{c}{ 1.13\,$^{+ 0.19}_{- 0.19}$}   &   &\multicolumn{2}{c}{ 2.78\,$^{+  1.58}_{-  1.78}$}   &   \multicolumn{2}{c}{ 4.05\,$^{+  2.38}_{-  0.53}$}   &   \multicolumn{2}{c}{ 4.20\,$^{+  2.33}_{-  0.61}$}   &   &\multicolumn{2}{c}{ 0.96\,$^{+ 0.07}_{- 0.07}$}   &   \multicolumn{2}{c}{ 0.83\,$^{+ 0.09}_{- 0.09}$}   &   \multicolumn{2}{c}{ 0.93\,$^{+ 0.10}_{- 0.10}$}   &   &\multicolumn{2}{c}{ 7.62\,$^{+  4.38}_{-  4.38}$}   &   \multicolumn{2}{c}{ 8.20\,$^{+  0.00}_{-  0.63}$}   &   \multicolumn{2}{c}{ 9.72\,$^{+  0.00}_{-  0.61}$}       \\
\noalign{\smallskip}
      
\object{RW Aur A-B     }   &    195   &   K1     &   K5     &   CC   &   \multicolumn{2}{c}{ 1.40\,$^{+ 0.09}_{- 0.09}$}   &   \multicolumn{2}{c}{ 1.25\,$^{+ 0.16}_{- 0.16}$}   &   \multicolumn{2}{c}{ 1.40\,$^{+ 0.20}_{- 0.20}$}   &   &\multicolumn{2}{c}{ 7.38\,$^{+  3.82}_{-  6.38}$}   &   \multicolumn{2}{c}{ 8.71\,$^{+  0.00}_{-  0.84}$}   &   \multicolumn{2}{c}{10.10\,$^{+  0.09}_{-  0.84}$}   &   &\multicolumn{2}{c}{ 0.95\,$^{+ 0.05}_{- 0.05}$}   &   \multicolumn{2}{c}{ 0.80\,$^{+ 0.07}_{- 0.07}$}   &   \multicolumn{2}{c}{ 0.90\,$^{+ 0.09}_{- 0.09}$}   &   &\multicolumn{2}{c}{12.77\,$^{+  7.46}_{-  7.46}$}   &   \multicolumn{2}{c}{14.69\,$^{+  0.00}_{-  0.40}$}   &   \multicolumn{2}{c}{17.90\,$^{+  0.00}_{-  0.50}$}       \\
\noalign{\smallskip}
      
\object{GG Tau Ba-Bb   }   &    212   &   M5.5   &   M7.5   &   CC   &   \multicolumn{2}{c}{ 0.14\,$^{+ 0.06}_{- 0.06}$}   &   \multicolumn{2}{c}{ 0.11\,$^{+ 0.06}_{- 0.06}$}   &   \multicolumn{2}{c}{ 0.13\,$^{+ 0.04}_{- 0.04}$}   &   &\multicolumn{2}{c}{ 2.11\,$^{+  5.24}_{-  1.11}$}   &   \multicolumn{2}{c}{ 1.63\,$^{+  0.76}_{-  0.12}$}   &   \multicolumn{2}{c}{ 4.57\,$^{+  0.36}_{-  0.60}$}   &   &\multicolumn{2}{c}{ 0.04\,$^{+ 0.03}_{- 0.03}$}   &   \multicolumn{2}{c}{$<$ 0.1  }   &                     \multicolumn{2}{c}{$<$ 0.1  }   &                     &\multicolumn{2}{c}{ 1.00\,$^{+  6.63}_{- 0.00}$}   &   \multicolumn{2}{c}{         }   &                       \multicolumn{2}{c}{         }                           \\
\noalign{\smallskip}
      
\object{Haro 6-37 A-B  }   &    366   &   K7     &   M1     &   CC   &   \multicolumn{2}{c}{ 0.81\,$^{+ 0.07}_{- 0.07}$}   &   \multicolumn{2}{c}{ 0.79\,$^{+ 0.09}_{- 0.09}$}   &   \multicolumn{2}{c}{ 0.77\,$^{+ 0.11}_{- 0.11}$}   &   &\multicolumn{2}{c}{ 2.43\,$^{+  5.58}_{-  1.43}$}   &   \multicolumn{2}{c}{ 3.21\,$^{+  0.30}_{-  0.46}$}   &   \multicolumn{2}{c}{ 3.25\,$^{+  0.32}_{-  0.43}$}   &   &\multicolumn{2}{c}{ 0.60\,$^{+ 0.08}_{- 0.08}$}   &   \multicolumn{2}{c}{ 0.48\,$^{+ 0.10}_{- 0.10}$}   &   \multicolumn{2}{c}{ 0.46\,$^{+ 0.10}_{- 0.10}$}   &   &\multicolumn{2}{c}{ 7.19\,$^{+ 17.56}_{-  6.19}$}   &   \multicolumn{2}{c}{ 4.47\,$^{+  0.13}_{-  0.52}$}   &   \multicolumn{2}{c}{ 5.47\,$^{+  0.21}_{-  0.53}$}       \\
\noalign{\smallskip}
      
\object{UX Tau A-B     }   &    820   &   K5     &   M2     &   CW   &   \multicolumn{2}{c}{ 1.07\,$^{+ 0.14}_{- 0.14}$}   &   \multicolumn{2}{c}{ 1.04\,$^{+ 0.13}_{- 0.13}$}   &   \multicolumn{2}{c}{ 1.12\,$^{+ 0.17}_{- 0.17}$}   &   &\multicolumn{2}{c}{ 3.12\,$^{+  2.12}_{-  2.12}$}   &   \multicolumn{2}{c}{ 4.51\,$^{+  0.14}_{-  0.46}$}   &   \multicolumn{2}{c}{ 4.64\,$^{+  0.03}_{-  0.44}$}   &   &\multicolumn{2}{c}{ 0.52\,$^{+ 0.11}_{- 0.11}$}   &   \multicolumn{2}{c}{ 0.36\,$^{+ 0.10}_{- 0.10}$}   &   \multicolumn{2}{c}{ 0.38\,$^{+ 0.08}_{- 0.08}$}   &   &\multicolumn{2}{c}{ 5.80\,$^{+  5.27}_{-  4.80}$}   &   \multicolumn{2}{c}{ 2.92\,$^{+  0.24}_{-  0.40}$}   &   \multicolumn{2}{c}{ 4.65\,$^{+  0.24}_{-  0.53}$}       \\
\noalign{\smallskip}
      
\object{UX Tau A-C     }   &    368   &   K5     &   M5     &   CW   &   \multicolumn{2}{c}{ 1.07\,$^{+ 0.13}_{- 0.13}$}   &   \multicolumn{2}{c}{ 1.04\,$^{+ 0.14}_{- 0.14}$}   &   \multicolumn{2}{c}{ 1.12\,$^{+ 0.20}_{- 0.20}$}   &   &\multicolumn{2}{c}{ 3.12\,$^{+  2.38}_{-  2.38}$}   &   \multicolumn{2}{c}{ 4.51\,$^{+  6.30}_{-  0.08}$}   &   \multicolumn{2}{c}{ 4.64\,$^{+ 16.44}_{-  0.00}$}   &   &\multicolumn{2}{c}{ 0.18\,$^{+ 0.07}_{- 0.07}$}   &   \multicolumn{2}{c}{ 0.14\,$^{+ 0.06}_{- 0.06}$}   &   \multicolumn{2}{c}{ 0.17\,$^{+ 0.06}_{- 0.06}$}   &   &\multicolumn{2}{c}{ 1.77\,$^{+  6.59}_{-  0.77}$}   &   \multicolumn{2}{c}{ 1.21\,$^{+  3.16}_{-  0.02}$}   &   \multicolumn{2}{c}{ 3.75\,$^{+  9.77}_{-  0.13}$}       \\
\noalign{\smallskip}
      
\object{V710 Tau A-B   }   &    443   &   M0.5   &   M2     &   CW   &   \multicolumn{2}{c}{ 0.64\,$^{+ 0.07}_{- 0.07}$}   &   \multicolumn{2}{c}{ 0.54\,$^{+ 0.10}_{- 0.10}$}   &   \multicolumn{2}{c}{ 0.51\,$^{+ 0.09}_{- 0.09}$}   &   &\multicolumn{2}{c}{ 1.95\,$^{+  1.76}_{-  0.95}$}   &   \multicolumn{2}{c}{ 2.03\,$^{+  4.54}_{-  0.24}$}   &   \multicolumn{2}{c}{ 2.04\,$^{+  5.50}_{-  0.14}$}   &   &\multicolumn{2}{c}{ 0.58\,$^{+ 0.10}_{- 0.10}$}   &   \multicolumn{2}{c}{ 0.36\,$^{+ 0.10}_{- 0.10}$}   &   \multicolumn{2}{c}{ 0.39\,$^{+ 0.07}_{- 0.07}$}   &   &\multicolumn{2}{c}{ 1.03\,$^{+  1.14}_{-  0.03}$}   &   \multicolumn{2}{c}{ 0.73\,$^{+  1.00}_{-  0.05}$}   &   \multicolumn{2}{c}{ 1.17\,$^{+  0.74}_{-  0.06}$}       \\
\noalign{\smallskip}
      
\noalign{\smallskip}
\hline
\noalign{\smallskip}
\multicolumn{32}{c}{Tau-Aur from HK03} \\
\noalign{\smallskip}
\hline
\noalign{\smallskip}
\object{DD Tau         }   &     77   &   M3.5   &   M3.5   &   CC   &   \multicolumn{2}{c}{ 0.38\,$^{+ 0.14}_{- 0.14}$}   &   \multicolumn{2}{c}{ 0.23\,$^{+ 0.08}_{- 0.08}$}   &   \multicolumn{2}{c}{ 0.29\,$^{+ 0.05}_{- 0.05}$}   &   &\multicolumn{2}{c}{ 1.00\,$^{+  1.11}_{-  0.00}$}   &   \multicolumn{2}{c}{ 0.63\,$^{+  0.35}_{-  0.08}$}   &   \multicolumn{2}{c}{ 1.48\,$^{+  0.35}_{-  0.23}$}   &   &\multicolumn{2}{c}{ 0.34\,$^{+ 0.13}_{- 0.13}$}   &   \multicolumn{2}{c}{ 0.23\,$^{+ 0.08}_{- 0.08}$}   &   \multicolumn{2}{c}{ 0.29\,$^{+ 0.05}_{- 0.05}$}   &   &\multicolumn{2}{c}{ 1.67\,$^{+  3.15}_{-  0.67}$}   &   \multicolumn{2}{c}{ 1.06\,$^{+  0.56}_{-  0.14}$}   &   \multicolumn{2}{c}{ 2.22\,$^{+  0.73}_{-  0.27}$}       \\
\noalign{\smallskip}
      
\object{DF Tau         }   &     13   &   M2     &   M2.5   &   CC   &   \multicolumn{2}{c}{ 0.58\,$^{+ 0.09}_{- 0.09}$}   &   \multicolumn{2}{c}{ 0.36\,$^{+ 0.10}_{- 0.10}$}   &   \multicolumn{2}{c}{ 0.39\,$^{+ 0.07}_{- 0.07}$}   &   &\multicolumn{2}{c}{ 1.38\,$^{+  1.43}_{-  0.38}$}   &   \multicolumn{2}{c}{ 0.90\,$^{+  0.48}_{-  0.13}$}   &   \multicolumn{2}{c}{ 1.37\,$^{+  0.46}_{-  0.14}$}   &   &\multicolumn{2}{c}{ 0.57\,$^{+ 0.10}_{- 0.10}$}   &   \multicolumn{2}{c}{ 0.32\,$^{+ 0.10}_{- 0.10}$}   &   \multicolumn{2}{c}{ 0.36\,$^{+ 0.07}_{- 0.07}$}   &   &\multicolumn{2}{c}{ 1.00\,$^{+  0.77}_{-  0.00}$}   &   \multicolumn{2}{c}{ 0.49\,$^{+  0.29}_{-  0.08}$}   &   \multicolumn{2}{c}{ 1.07\,$^{+  0.28}_{-  0.15}$}       \\
\noalign{\smallskip}
      
\object{FO Tau         }   &     20   &   M3.5   &   M3.5   &   CC   &   \multicolumn{2}{c}{ 0.36\,$^{+ 0.14}_{- 0.14}$}   &   \multicolumn{2}{c}{ 0.23\,$^{+ 0.08}_{- 0.08}$}   &   \multicolumn{2}{c}{ 0.29\,$^{+ 0.05}_{- 0.05}$}   &   &\multicolumn{2}{c}{ 1.33\,$^{+  1.88}_{-  0.33}$}   &   \multicolumn{2}{c}{ 0.82\,$^{+  0.38}_{-  0.10}$}   &   \multicolumn{2}{c}{ 1.84\,$^{+  0.40}_{-  0.24}$}   &   &\multicolumn{2}{c}{ 0.36\,$^{+ 0.13}_{- 0.13}$}   &   \multicolumn{2}{c}{ 0.23\,$^{+ 0.08}_{- 0.08}$}   &   \multicolumn{2}{c}{ 0.29\,$^{+ 0.05}_{- 0.05}$}   &   &\multicolumn{2}{c}{ 1.33\,$^{+  2.29}_{-  0.33}$}   &   \multicolumn{2}{c}{ 0.82\,$^{+  0.42}_{-  0.10}$}   &   \multicolumn{2}{c}{ 1.84\,$^{+  0.95}_{-  0.26}$}       \\
\noalign{\smallskip}
      
\object{FQ Tau         }   &    105   &   M3     &   M3.5   &   CC   &   \multicolumn{2}{c}{ 0.38\,$^{+ 0.13}_{- 0.13}$}   &   \multicolumn{2}{c}{ 0.28\,$^{+ 0.08}_{- 0.08}$}   &   \multicolumn{2}{c}{ 0.31\,$^{+ 0.07}_{- 0.07}$}   &   &\multicolumn{2}{c}{ 2.73\,$^{+  5.42}_{-  1.73}$}   &   \multicolumn{2}{c}{ 1.90\,$^{+  0.93}_{-  0.28}$}   &   \multicolumn{2}{c}{ 2.82\,$^{+  1.39}_{-  0.26}$}   &   &\multicolumn{2}{c}{ 0.36\,$^{+ 0.14}_{- 0.14}$}   &   \multicolumn{2}{c}{ 0.23\,$^{+ 0.09}_{- 0.09}$}   &   \multicolumn{2}{c}{ 0.29\,$^{+ 0.05}_{- 0.05}$}   &   &\multicolumn{2}{c}{ 1.43\,$^{+  1.86}_{-  0.43}$}   &   \multicolumn{2}{c}{ 0.88\,$^{+  0.46}_{-  0.12}$}   &   \multicolumn{2}{c}{ 1.94\,$^{+  0.44}_{-  0.21}$}       \\
\noalign{\smallskip}
      
\object{FS Tau         }   &     35   &   M0     &   M3.5   &   CC   &   \multicolumn{2}{c}{ 0.71\,$^{+ 0.07}_{- 0.07}$}   &   \multicolumn{2}{c}{ 0.60\,$^{+ 0.07}_{- 0.07}$}   &   \multicolumn{2}{c}{ 0.59\,$^{+ 0.09}_{- 0.09}$}   &   &\multicolumn{2}{c}{19.19\,$^{+ 23.81}_{- 18.19}$}   &   \multicolumn{2}{c}{15.23\,$^{+  9.03}_{-  2.13}$}   &   \multicolumn{2}{c}{16.09\,$^{+ 14.59}_{-  2.06}$}   &   &\multicolumn{2}{c}{ 0.32\,$^{+ 0.11}_{- 0.11}$}   &   \multicolumn{2}{c}{ 0.23\,$^{+ 0.08}_{- 0.08}$}   &   \multicolumn{2}{c}{ 0.28\,$^{+ 0.06}_{- 0.06}$}   &   &\multicolumn{2}{c}{ 2.41\,$^{+  3.92}_{-  1.41}$}   &   \multicolumn{2}{c}{ 1.74\,$^{+  0.84}_{-  0.24}$}   &   \multicolumn{2}{c}{ 3.07\,$^{+  1.38}_{-  0.30}$}       \\
\noalign{\smallskip}
      
\object{FVTau c        }   &     98   &   M2.5   &   M3.5   &   CC   &   \multicolumn{2}{c}{ 0.44\,$^{+ 0.11}_{- 0.11}$}   &   \multicolumn{2}{c}{ 0.32\,$^{+ 0.09}_{- 0.09}$}   &   \multicolumn{2}{c}{ 0.34\,$^{+ 0.08}_{- 0.08}$}   &   &\multicolumn{2}{c}{ 4.45\,$^{+  6.01}_{-  3.45}$}   &   \multicolumn{2}{c}{ 2.57\,$^{+  1.88}_{-  0.30}$}   &   \multicolumn{2}{c}{ 3.84\,$^{+  2.13}_{-  0.41}$}   &   &\multicolumn{2}{c}{ 0.30\,$^{+ 0.10}_{- 0.10}$}   &   \multicolumn{2}{c}{ 0.23\,$^{+ 0.08}_{- 0.08}$}   &   \multicolumn{2}{c}{ 0.23\,$^{+ 0.07}_{- 0.07}$}   &   &\multicolumn{2}{c}{ 9.56\,$^{+ 22.76}_{- 22.76}$}   &   \multicolumn{2}{c}{ 6.15\,$^{+  4.28}_{-  0.86}$}   &   \multicolumn{2}{c}{ 8.73\,$^{+  8.62}_{-  1.04}$}       \\
\noalign{\smallskip}
      
\object{GG Tau         }   &     34   &   M0     &   M2     &   CC   &   \multicolumn{2}{c}{ 0.72\,$^{+ 0.08}_{- 0.08}$}   &   \multicolumn{2}{c}{ 0.60\,$^{+ 0.07}_{- 0.07}$}   &   \multicolumn{2}{c}{ 0.58\,$^{+ 0.09}_{- 0.09}$}   &   &\multicolumn{2}{c}{ 4.02\,$^{+  5.39}_{-  5.39}$}   &   \multicolumn{2}{c}{ 3.60\,$^{+  1.61}_{-  0.42}$}   &   \multicolumn{2}{c}{ 4.08\,$^{+  2.78}_{-  0.53}$}   &   &\multicolumn{2}{c}{ 0.53\,$^{+ 0.11}_{- 0.11}$}   &   \multicolumn{2}{c}{ 0.36\,$^{+ 0.11}_{- 0.11}$}   &   \multicolumn{2}{c}{ 0.38\,$^{+ 0.08}_{- 0.08}$}   &   &\multicolumn{2}{c}{ 5.44\,$^{+  5.35}_{-  4.44}$}   &   \multicolumn{2}{c}{ 2.82\,$^{+  2.10}_{-  0.35}$}   &   \multicolumn{2}{c}{ 3.86\,$^{+  1.26}_{-  0.37}$}       \\
\noalign{\smallskip}
      
\object{GH Tau         }   &     42   &   M2     &   M2     &   CC   &   \multicolumn{2}{c}{ 0.54\,$^{+ 0.11}_{- 0.11}$}   &   \multicolumn{2}{c}{ 0.36\,$^{+ 0.09}_{- 0.09}$}   &   \multicolumn{2}{c}{ 0.38\,$^{+ 0.07}_{- 0.07}$}   &   &\multicolumn{2}{c}{ 3.42\,$^{+  4.54}_{-  2.42}$}   &   \multicolumn{2}{c}{ 2.13\,$^{+  0.63}_{-  0.30}$}   &   \multicolumn{2}{c}{ 2.69\,$^{+  1.41}_{-  0.23}$}   &   &\multicolumn{2}{c}{ 0.54\,$^{+ 0.10}_{- 0.10}$}   &   \multicolumn{2}{c}{ 0.36\,$^{+ 0.11}_{- 0.11}$}   &   \multicolumn{2}{c}{ 0.38\,$^{+ 0.07}_{- 0.07}$}   &   &\multicolumn{2}{c}{ 3.69\,$^{+  6.06}_{-  2.69}$}   &   \multicolumn{2}{c}{ 2.21\,$^{+  1.26}_{-  0.30}$}   &   \multicolumn{2}{c}{ 2.70\,$^{+  0.99}_{-  0.25}$}       \\
\noalign{\smallskip}
      
\object{HBC 358        }   &    217   &   M3.5   &   M4     &        &   \multicolumn{2}{c}{ 0.30\,$^{+ 0.12}_{- 0.12}$}   &   \multicolumn{2}{c}{ 0.23\,$^{+ 0.08}_{- 0.08}$}   &   \multicolumn{2}{c}{ 0.26\,$^{+ 0.07}_{- 0.07}$}   &   &\multicolumn{2}{c}{ 3.72\,$^{+  8.07}_{-  2.72}$}   &   \multicolumn{2}{c}{ 2.43\,$^{+  2.52}_{-  0.33}$}   &   \multicolumn{2}{c}{ 4.36\,$^{+  1.58}_{-  0.49}$}   &   &\multicolumn{2}{c}{ 0.26\,$^{+ 0.10}_{- 0.10}$}   &   \multicolumn{2}{c}{ 0.19\,$^{+ 0.07}_{- 0.07}$}   &   \multicolumn{2}{c}{ 0.24\,$^{+ 0.06}_{- 0.06}$}   &   &\multicolumn{2}{c}{ 2.73\,$^{+  7.93}_{-  1.73}$}   &   \multicolumn{2}{c}{ 1.92\,$^{+  1.41}_{-  0.25}$}   &   \multicolumn{2}{c}{ 3.85\,$^{+  1.24}_{-  0.39}$}       \\
\noalign{\smallskip}
      
\object{IS Tau         }   &     31   &   M0     &   M3.5   &   CW   &   \multicolumn{2}{c}{ 0.70\,$^{+ 0.06}_{- 0.06}$}   &   \multicolumn{2}{c}{ 0.59\,$^{+ 0.10}_{- 0.10}$}   &   \multicolumn{2}{c}{ 0.57\,$^{+ 0.10}_{- 0.10}$}   &   &\multicolumn{2}{c}{ 1.64\,$^{+  2.04}_{-  0.64}$}   &   \multicolumn{2}{c}{ 1.95\,$^{+  0.65}_{-  0.27}$}   &   \multicolumn{2}{c}{ 1.81\,$^{+  0.99}_{-  0.19}$}   &   &\multicolumn{2}{c}{ 0.33\,$^{+ 0.12}_{- 0.12}$}   &   \multicolumn{2}{c}{ 0.23\,$^{+ 0.08}_{- 0.08}$}   &   \multicolumn{2}{c}{ 0.28\,$^{+ 0.07}_{- 0.07}$}   &   &\multicolumn{2}{c}{ 1.90\,$^{+  2.62}_{-  0.90}$}   &   \multicolumn{2}{c}{ 1.34\,$^{+  0.80}_{-  0.20}$}   &   \multicolumn{2}{c}{ 2.66\,$^{+  0.89}_{-  0.28}$}       \\
\noalign{\smallskip}
      
\object{LkCa 7         }   &    142   &   M0     &   M3.5   &        &   \multicolumn{2}{c}{ 0.72\,$^{+ 0.08}_{- 0.08}$}   &   \multicolumn{2}{c}{ 0.60\,$^{+ 0.10}_{- 0.10}$}   &   \multicolumn{2}{c}{ 0.58\,$^{+ 0.09}_{- 0.09}$}   &   &\multicolumn{2}{c}{ 3.26\,$^{+  4.84}_{-  2.26}$}   &   \multicolumn{2}{c}{ 3.08\,$^{+  1.15}_{-  0.44}$}   &   \multicolumn{2}{c}{ 3.09\,$^{+  2.23}_{-  0.27}$}   &   &\multicolumn{2}{c}{ 0.32\,$^{+ 0.12}_{- 0.12}$}   &   \multicolumn{2}{c}{ 0.23\,$^{+ 0.08}_{- 0.08}$}   &   \multicolumn{2}{c}{ 0.28\,$^{+ 0.05}_{- 0.05}$}   &   &\multicolumn{2}{c}{ 2.26\,$^{+  3.18}_{-  1.26}$}   &   \multicolumn{2}{c}{ 1.63\,$^{+  1.30}_{-  0.23}$}   &   \multicolumn{2}{c}{ 2.91\,$^{+  0.75}_{-  0.33}$}       \\
\noalign{\smallskip}
      
\object{UY Aur         }   &    122   &   M0     &   M2.5   &   CC   &   \multicolumn{2}{c}{ 0.72\,$^{+ 0.08}_{- 0.08}$}   &   \multicolumn{2}{c}{ 0.60\,$^{+ 0.08}_{- 0.08}$}   &   \multicolumn{2}{c}{ 0.58\,$^{+ 0.09}_{- 0.09}$}   &   &\multicolumn{2}{c}{ 5.33\,$^{+  6.36}_{-  6.36}$}   &   \multicolumn{2}{c}{ 4.33\,$^{+  3.12}_{-  0.49}$}   &   \multicolumn{2}{c}{ 5.08\,$^{+  2.69}_{-  0.64}$}   &   &\multicolumn{2}{c}{ 0.44\,$^{+ 0.11}_{- 0.11}$}   &   \multicolumn{2}{c}{ 0.32\,$^{+ 0.09}_{- 0.09}$}   &   \multicolumn{2}{c}{ 0.34\,$^{+ 0.08}_{- 0.08}$}   &   &\multicolumn{2}{c}{ 4.83\,$^{+  6.87}_{-  6.87}$}   &   \multicolumn{2}{c}{ 2.69\,$^{+  1.69}_{-  0.33}$}   &   \multicolumn{2}{c}{ 4.25\,$^{+  3.31}_{-  0.44}$}       \\
\noalign{\smallskip}
      
\object{UZ Tau W       }   &     50   &   M2     &   M3     &   CC   &   \multicolumn{2}{c}{ 0.57\,$^{+ 0.11}_{- 0.11}$}   &   \multicolumn{2}{c}{ 0.36\,$^{+ 0.10}_{- 0.10}$}   &   \multicolumn{2}{c}{ 0.39\,$^{+ 0.07}_{- 0.07}$}   &   &\multicolumn{2}{c}{ 1.79\,$^{+  1.60}_{-  0.79}$}   &   \multicolumn{2}{c}{ 1.10\,$^{+  0.50}_{-  0.14}$}   &   \multicolumn{2}{c}{ 1.60\,$^{+  0.47}_{-  0.18}$}   &   &\multicolumn{2}{c}{ 0.48\,$^{+ 0.14}_{- 0.14}$}   &   \multicolumn{2}{c}{ 0.27\,$^{+ 0.08}_{- 0.08}$}   &   \multicolumn{2}{c}{ 0.32\,$^{+ 0.06}_{- 0.06}$}   &   &\multicolumn{2}{c}{ 1.00\,$^{+  0.67}_{-  0.00}$}   &   \multicolumn{2}{c}{ 0.39\,$^{+  0.16}_{-  0.06}$}   &   \multicolumn{2}{c}{ 0.89\,$^{+  0.20}_{-  0.15}$}       \\
\noalign{\smallskip}
      
\object{V807 Tau       }   &     43   &   K5     &   M2     &   CW   &   \multicolumn{2}{c}{ 1.00\,$^{+ 0.19}_{- 0.19}$}   &   \multicolumn{2}{c}{ 1.16\,$^{+ 0.22}_{- 0.22}$}   &   \multicolumn{2}{c}{ 1.13\,$^{+ 0.25}_{- 0.25}$}   &   &\multicolumn{2}{c}{ 1.00\,$^{+  1.16}_{-  0.00}$}   &   \multicolumn{2}{c}{ 1.08\,$^{+  0.67}_{-  0.15}$}   &   \multicolumn{2}{c}{ 1.26\,$^{+  0.66}_{-  0.15}$}   &   &\multicolumn{2}{c}{ 0.58\,$^{+ 0.08}_{- 0.08}$}   &   \multicolumn{2}{c}{ 0.36\,$^{+ 0.11}_{- 0.11}$}   &   \multicolumn{2}{c}{ 0.39\,$^{+ 0.07}_{- 0.07}$}   &   &\multicolumn{2}{c}{ 1.00\,$^{+  0.50}_{-  0.00}$}   &   \multicolumn{2}{c}{ 0.26\,$^{+  0.25}_{-  0.04}$}   &   \multicolumn{2}{c}{ 0.77\,$^{+  0.13}_{-  0.12}$}       \\
\noalign{\smallskip}
      
\object{V927 Tau       }   &     37   &   M3     &   M3.5   &        &   \multicolumn{2}{c}{ 0.42\,$^{+ 0.13}_{- 0.13}$}   &   \multicolumn{2}{c}{ 0.28\,$^{+ 0.09}_{- 0.09}$}   &   \multicolumn{2}{c}{ 0.32\,$^{+ 0.06}_{- 0.06}$}   &   &\multicolumn{2}{c}{ 1.62\,$^{+  1.73}_{-  0.62}$}   &   \multicolumn{2}{c}{ 0.96\,$^{+  0.37}_{-  0.13}$}   &   \multicolumn{2}{c}{ 1.84\,$^{+  0.76}_{-  0.26}$}   &   &\multicolumn{2}{c}{ 0.30\,$^{+ 0.11}_{- 0.11}$}   &   \multicolumn{2}{c}{ 0.23\,$^{+ 0.08}_{- 0.08}$}   &   \multicolumn{2}{c}{ 0.26\,$^{+ 0.06}_{- 0.06}$}   &   &\multicolumn{2}{c}{ 4.15\,$^{+  5.42}_{-  3.15}$}   &   \multicolumn{2}{c}{ 2.49\,$^{+  2.05}_{-  0.32}$}   &   \multicolumn{2}{c}{ 4.76\,$^{+  1.67}_{-  0.49}$}       \\
\noalign{\smallskip}
      
\object{V955 Tau       }   &     45   &   K7     &   M2.5   &   CC   &   \multicolumn{2}{c}{ 0.80\,$^{+ 0.08}_{- 0.08}$}   &   \multicolumn{2}{c}{ 0.65\,$^{+ 0.07}_{- 0.07}$}   &   \multicolumn{2}{c}{ 0.72\,$^{+ 0.09}_{- 0.09}$}   &   &\multicolumn{2}{c}{20.97\,$^{+ 22.47}_{- 19.97}$}   &   \multicolumn{2}{c}{20.67\,$^{+  8.70}_{-  3.33}$}   &   \multicolumn{2}{c}{22.40\,$^{+ 16.38}_{-  2.98}$}   &   &\multicolumn{2}{c}{ 0.44\,$^{+ 0.11}_{- 0.11}$}   &   \multicolumn{2}{c}{ 0.32\,$^{+ 0.09}_{- 0.09}$}   &   \multicolumn{2}{c}{ 0.32\,$^{+ 0.09}_{- 0.09}$}   &   &\multicolumn{2}{c}{11.68\,$^{+ 25.21}_{- 25.21}$}   &   \multicolumn{2}{c}{ 6.52\,$^{+  7.11}_{-  0.90}$}   &   \multicolumn{2}{c}{ 8.52\,$^{+  6.06}_{-  0.93}$}       \\
\noalign{\smallskip}
      
\object{V999 Tau       }   &     32   &   M0.5   &   M2.5   &        &   \multicolumn{2}{c}{ 0.67\,$^{+ 0.07}_{- 0.07}$}   &   \multicolumn{2}{c}{ 0.55\,$^{+ 0.10}_{- 0.10}$}   &   \multicolumn{2}{c}{ 0.52\,$^{+ 0.10}_{- 0.10}$}   &   &\multicolumn{2}{c}{ 6.86\,$^{+ 10.68}_{-  5.86}$}   &   \multicolumn{2}{c}{ 4.77\,$^{+  2.90}_{-  0.53}$}   &   \multicolumn{2}{c}{ 5.72\,$^{+  6.27}_{-  0.72}$}   &   &\multicolumn{2}{c}{ 0.44\,$^{+ 0.12}_{- 0.12}$}   &   \multicolumn{2}{c}{ 0.32\,$^{+ 0.09}_{- 0.09}$}   &   \multicolumn{2}{c}{ 0.34\,$^{+ 0.07}_{- 0.07}$}   &   &\multicolumn{2}{c}{ 5.66\,$^{+  8.10}_{-  4.66}$}   &   \multicolumn{2}{c}{ 2.94\,$^{+  5.19}_{-  0.34}$}   &   \multicolumn{2}{c}{ 5.04\,$^{+  1.96}_{-  0.55}$}       \\
\noalign{\smallskip}
      
\object{V1026 Tau      }   &     90   &   M2     &   M3.5   &        &   \multicolumn{2}{c}{ 0.51\,$^{+ 0.11}_{- 0.11}$}   &   \multicolumn{2}{c}{ 0.37\,$^{+ 0.08}_{- 0.08}$}   &   \multicolumn{2}{c}{ 0.35\,$^{+ 0.08}_{- 0.08}$}   &   &\multicolumn{2}{c}{16.74\,$^{+ 20.27}_{- 15.74}$}   &   \multicolumn{2}{c}{ 9.28\,$^{+  7.94}_{-  1.39}$}   &   \multicolumn{2}{c}{10.19\,$^{+ 10.96}_{-  1.15}$}   &   &\multicolumn{2}{c}{ 0.26\,$^{+ 0.10}_{- 0.10}$}   &   \multicolumn{2}{c}{ 0.24\,$^{+ 0.08}_{- 0.08}$}   &   \multicolumn{2}{c}{ 0.21\,$^{+ 0.06}_{- 0.06}$}   &   &\multicolumn{2}{c}{20.36\,$^{+ 31.05}_{- 19.36}$}   &   \multicolumn{2}{c}{18.42\,$^{+ 12.40}_{-  2.73}$}   &   \multicolumn{2}{c}{20.63\,$^{+ 10.50}_{-  2.43}$}       \\
\noalign{\smallskip}
      
\object{XZ Tau         }   &     41   &   M2     &   M3.5   &   CC   &   \multicolumn{2}{c}{ 0.51\,$^{+ 0.12}_{- 0.12}$}   &   \multicolumn{2}{c}{ 0.36\,$^{+ 0.10}_{- 0.10}$}   &   \multicolumn{2}{c}{ 0.37\,$^{+ 0.08}_{- 0.08}$}   &   &\multicolumn{2}{c}{ 6.72\,$^{+  8.06}_{-  5.72}$}   &   \multicolumn{2}{c}{ 3.37\,$^{+  2.61}_{-  0.36}$}   &   \multicolumn{2}{c}{ 5.47\,$^{+  4.99}_{-  0.65}$}   &   &\multicolumn{2}{c}{ 0.37\,$^{+ 0.13}_{- 0.13}$}   &   \multicolumn{2}{c}{ 0.23\,$^{+ 0.09}_{- 0.09}$}   &   \multicolumn{2}{c}{ 0.29\,$^{+ 0.05}_{- 0.05}$}   &   &\multicolumn{2}{c}{ 1.29\,$^{+  1.94}_{-  0.29}$}   &   \multicolumn{2}{c}{ 0.81\,$^{+  0.50}_{-  0.10}$}   &   \multicolumn{2}{c}{ 1.79\,$^{+  0.34}_{-  0.25}$}       \\
\noalign{\smallskip}
      
\noalign{\smallskip}
\hline
\noalign{\smallskip}
\multicolumn{32}{c}{Oph from P03} \\
\noalign{\smallskip}
\hline
\noalign{\smallskip}
\object{AS 205         }   &    182   &   K5     &   M3     &   CC   &   \multicolumn{2}{c}{ 1.00\,$^{+ 0.19}_{- 0.19}$}   &   \multicolumn{2}{c}{ 1.14\,$^{+ 0.29}_{- 0.29}$}   &   \multicolumn{2}{c}{ 1.24\,$^{+ 0.32}_{- 0.32}$}   &   &\multicolumn{2}{c}{ 1.00}   &   \multicolumn{2}{c}{ 0.10\,$^{+  0.04}_{-  0.00}$}   &   \multicolumn{2}{c}{ 0.41\,$^{+  0.10}_{-  0.04}$}   &   &\multicolumn{2}{c}{ 0.48\,$^{+ 0.15}_{- 0.15}$}   &   \multicolumn{2}{c}{ 0.26\,$^{+ 0.09}_{- 0.09}$}   &   \multicolumn{2}{c}{ 0.32\,$^{+ 0.06}_{- 0.06}$}   &   &\multicolumn{2}{c}{ 1.00}   &   \multicolumn{2}{c}{ 0.10\,$^{+  0.01}_{-  0.00}$}   &   \multicolumn{2}{c}{ 0.10\,$^{+  0.07}_{-  0.00}$}       \\
\noalign{\smallskip}
      
\object{WSB 4          }   &    392   &   M3     &   M3     &   CC   &   \multicolumn{2}{c}{ 0.36\,$^{+ 0.11}_{- 0.11}$}   &   \multicolumn{2}{c}{ 0.28\,$^{+ 0.09}_{- 0.09}$}   &   \multicolumn{2}{c}{ 0.29\,$^{+ 0.07}_{- 0.07}$}   &   &\multicolumn{2}{c}{ 6.02\,$^{+  7.73}_{-  7.73}$}   &   \multicolumn{2}{c}{ 3.28\,$^{+  2.15}_{-  0.36}$}   &   \multicolumn{2}{c}{ 5.87\,$^{+  1.38}_{-  0.63}$}   &   &\multicolumn{2}{c}{ 0.37\,$^{+ 0.12}_{- 0.12}$}   &   \multicolumn{2}{c}{ 0.28\,$^{+ 0.09}_{- 0.09}$}   &   \multicolumn{2}{c}{ 0.30\,$^{+ 0.06}_{- 0.06}$}   &   &\multicolumn{2}{c}{ 3.88\,$^{+  5.27}_{-  2.88}$}   &   \multicolumn{2}{c}{ 2.42\,$^{+  1.05}_{-  0.29}$}   &   \multicolumn{2}{c}{ 3.89\,$^{+  0.67}_{-  0.34}$}       \\
\noalign{\smallskip}
      
\object{WSB 19         }   &    210   &   M3     &   M5     &   CC   &   \multicolumn{2}{c}{ 0.41\,$^{+ 0.12}_{- 0.12}$}   &   \multicolumn{2}{c}{ 0.28\,$^{+ 0.10}_{- 0.10}$}   &   \multicolumn{2}{c}{ 0.32\,$^{+ 0.06}_{- 0.06}$}   &   &\multicolumn{2}{c}{ 1.81\,$^{+  1.67}_{-  0.81}$}   &   \multicolumn{2}{c}{ 1.11\,$^{+  0.45}_{-  0.12}$}   &   \multicolumn{2}{c}{ 2.05\,$^{+  0.35}_{-  0.19}$}   &   &\multicolumn{2}{c}{ 0.15\,$^{+ 0.07}_{- 0.07}$}   &   \multicolumn{2}{c}{ 0.13\,$^{+ 0.06}_{- 0.06}$}   &   \multicolumn{2}{c}{ 0.20\,$^{+ 0.05}_{- 0.05}$}   &   &\multicolumn{2}{c}{ 1.00\,$^{+  1.81}_{-  0.00}$}   &   \multicolumn{2}{c}{ 0.57\,$^{+  0.23}_{-  0.06}$}   &   \multicolumn{2}{c}{ 2.23\,$^{+  0.24}_{-  0.39}$}       \\
\noalign{\smallskip}
      
\object{WSB 28         }   &    714   &   M3     &   M7     &   WW   &   \multicolumn{2}{c}{ 0.48\,$^{+ 0.13}_{- 0.13}$}   &   \multicolumn{2}{c}{ 0.27\,$^{+ 0.09}_{- 0.09}$}   &   \multicolumn{2}{c}{ 0.32\,$^{+ 0.06}_{- 0.06}$}   &   &\multicolumn{2}{c}{ 1.00\,$^{+  0.23}_{-  0.00}$}   &   \multicolumn{2}{c}{ 0.44\,$^{+  0.09}_{-  0.06}$}   &   \multicolumn{2}{c}{ 1.03\,$^{+  0.07}_{-  0.18}$}   &   &\multicolumn{2}{c}{ 0.06\,$^{+ 0.03}_{- 0.03}$}   &   \multicolumn{2}{c}{$<$ 0.1  }   &                     \multicolumn{2}{c}{$<$ 0.1  }   &                     &\multicolumn{2}{c}{ 1.00\,$^{+  4.40}_{-  0.00}$}   &   \multicolumn{2}{c}{         }   &                       \multicolumn{2}{c}{         }                           \\
\noalign{\smallskip}
      
\object{DoAr 26        }   &    322   &   M4     &   M6     &   CW   &   \multicolumn{2}{c}{ 0.29\,$^{+ 0.13}_{- 0.13}$}   &   \multicolumn{2}{c}{ 0.19\,$^{+ 0.08}_{- 0.08}$}   &   \multicolumn{2}{c}{ 0.27\,$^{+ 0.05}_{- 0.05}$}   &   &\multicolumn{2}{c}{ 1.00\,$^{+  0.57}_{-  0.00}$}   &   \multicolumn{2}{c}{ 0.51\,$^{+  0.17}_{-  0.06}$}   &   \multicolumn{2}{c}{ 1.65\,$^{+  0.26}_{-  0.29}$}   &   &\multicolumn{2}{c}{ 0.10\,$^{+ 0.06}_{- 0.06}$}   &   \multicolumn{2}{c}{ 0.07\,$^{+ 0.06}_{- 0.06}$}   &   \multicolumn{2}{c}{ 0.13\,$^{+ 0.03}_{- 0.03}$}   &   &\multicolumn{2}{c}{ 1.00\,$^{+  3.68}_{-  0.00}$}   &   \multicolumn{2}{c}{ 0.90\,$^{+  2.24}_{-  0.14}$}   &   \multicolumn{2}{c}{ 0.71\,$^{+  0.71}_{-  0.06}$}       \\
\noalign{\smallskip}
      
\object{ROX 15         }   &    168   &   M3     &   M3     &   CC   &   \multicolumn{2}{c}{ 0.48\,$^{+ 0.14}_{- 0.14}$}   &   \multicolumn{2}{c}{ 0.26\,$^{+ 0.09}_{- 0.09}$}   &   \multicolumn{2}{c}{ 0.33\,$^{+ 0.06}_{- 0.06}$}   &   &\multicolumn{2}{c}{ 1.00}   &   \multicolumn{2}{c}{ 0.10\,$^{+  0.00}_{-  0.00}$}   &   \multicolumn{2}{c}{ 0.10\,$^{+  0.03}_{-  0.00}$}   &   &\multicolumn{2}{c}{ 0.48\,$^{+ 0.13}_{- 0.13}$}   &   \multicolumn{2}{c}{ 0.26\,$^{+ 0.08}_{- 0.08}$}   &   \multicolumn{2}{c}{ 0.32\,$^{+ 0.05}_{- 0.05}$}   &   &\multicolumn{2}{c}{ 1.00}   &   \multicolumn{2}{c}{ 0.10\,$^{+  0.05}_{-  0.00}$}   &   \multicolumn{2}{c}{ 0.10\,$^{+  0.12}_{-  0.00}$}       \\
\noalign{\smallskip}
      
\object{SR 21          }   &    896   &   G2.5   &   M4     &   CW   &   \multicolumn{2}{c}{$>$ 1.4  }   &                     \multicolumn{2}{c}{ 2.71\,$^{+ 0.28}_{- 0.28}$}   &   \multicolumn{2}{c}{ 2.75\,$^{+ 0.27}_{- 0.27}$}   &   &\multicolumn{2}{c}{         }   &                       \multicolumn{2}{c}{ 0.93\,$^{+  0.17}_{-  0.14}$}   &   \multicolumn{2}{c}{ 2.45\,$^{+  0.28}_{-  0.17}$}   &   &\multicolumn{2}{c}{ 0.29\,$^{+ 0.16}_{- 0.16}$}   &   \multicolumn{2}{c}{ 0.18\,$^{+ 0.07}_{- 0.07}$}   &   \multicolumn{2}{c}{ 0.27\,$^{+ 0.05}_{- 0.05}$}   &   &\multicolumn{2}{c}{ 1.00\,$^{+  0.39}_{-  0.00}$}   &   \multicolumn{2}{c}{ 0.30\,$^{+  0.06}_{-  0.04}$}   &   \multicolumn{2}{c}{ 0.38\,$^{+  0.20}_{-  0.06}$}       \\
\noalign{\smallskip}
      
\object{Ha 71          }   &    672   &   K2     &   M6     &   CW   &   \multicolumn{2}{c}{ 1.40\,$^{+ 0.16}_{- 0.16}$}   &   \multicolumn{2}{c}{ 1.50\,$^{+ 0.13}_{- 0.13}$}   &   \multicolumn{2}{c}{ 1.60\,$^{+ 0.14}_{- 0.14}$}   &   &\multicolumn{2}{c}{ 4.38\,$^{+  1.98}_{-  3.38}$}   &   \multicolumn{2}{c}{ 4.14\,$^{+  0.82}_{-  0.35}$}   &   \multicolumn{2}{c}{ 5.28\,$^{+  8.05}_{-  0.48}$}   &   &\multicolumn{2}{c}{ 0.10\,$^{+ 0.04}_{- 0.04}$}   &   \multicolumn{2}{c}{$<$ 0.1  }   &                     \multicolumn{2}{c}{ 0.18\,$^{+ 0.04}_{- 0.04}$}   &   &\multicolumn{2}{c}{ 1.00}   &   \multicolumn{2}{c}{         }   &                       \multicolumn{2}{c}{ 0.25\,$^{+  0.26}_{-  0.03}$}       \\
\noalign{\smallskip}
      
\noalign{\smallskip}
\hline
\noalign{\smallskip}
\multicolumn{32}{c}{ONC from D12} \\
\noalign{\smallskip}
\hline
\noalign{\smallskip}
\object{JW 260         }   &    145   &   G0     &   F7     &   WW     &   \multicolumn{2}{c}{$>$ 1.4  }   &                     \multicolumn{2}{c}{ 2.29\,$^{+ 0.23}_{- 0.23}$}   &   \multicolumn{2}{c}{ 2.25\,$^{+ 0.23}_{- 0.23}$}   &   &\multicolumn{2}{c}{         }   &                       \multicolumn{2}{c}{ 1.89\,$^{+  0.35}_{-  0.18}$}   &   \multicolumn{2}{c}{ 4.35\,$^{+  0.39}_{-  0.29}$}   &   &\multicolumn{2}{c}{$>$ 1.4 }   &                     \multicolumn{2}{c}{ 2.00\,$^{+ 0.19}_{- 0.19}$}   &   \multicolumn{2}{c}{ 1.85\,$^{+ 0.20}_{- 0.20}$}   &   &\multicolumn{2}{c}{         }   &                       \multicolumn{2}{c}{ 3.29\,$^{+  0.39}_{-  0.23}$}   &   \multicolumn{2}{c}{ 7.44\,$^{+ 20.70}_{-  0.58}$}       \\
\noalign{\smallskip}
      
\object{JW 553         }   &    159   &   K7     &   M1     &   WC     &   \multicolumn{2}{c}{ 0.78\,$^{+ 0.12}_{- 0.12}$}   &   \multicolumn{2}{c}{ 0.79\,$^{+ 0.23}_{- 0.23}$}   &   \multicolumn{2}{c}{ 0.77\,$^{+ 0.21}_{- 0.21}$}   &   &\multicolumn{2}{c}{ 1.00}   &   \multicolumn{2}{c}{ 0.10\,$^{+  0.04}_{-  0.00}$}   &   \multicolumn{2}{c}{ 0.35\,$^{+  0.05}_{-  0.06}$}   &   &\multicolumn{2}{c}{ 0.60\,$^{+ 0.19}_{- 0.19}$}   &   \multicolumn{2}{c}{ 0.48\,$^{+ 0.18}_{- 0.18}$}   &   \multicolumn{2}{c}{ 0.46\,$^{+ 0.17}_{- 0.17}$}   &   &\multicolumn{2}{c}{10.56\,$^{+ 14.71}_{-  9.56}$}   &   \multicolumn{2}{c}{ 6.78\,$^{+  8.44}_{-  1.13}$}   &   \multicolumn{2}{c}{ 8.00\,$^{+ 15.80}_{-  1.07}$}       \\
\noalign{\smallskip}
      
\object{JW 566         }   &    356   &   K7     &   M1.5   &  CW      &   \multicolumn{2}{c}{ 0.78\,$^{+ 0.09}_{- 0.09}$}   &   \multicolumn{2}{c}{ 0.79\,$^{+ 0.16}_{- 0.16}$}   &   \multicolumn{2}{c}{ 0.74\,$^{+ 0.17}_{- 0.17}$}   &   &\multicolumn{2}{c}{ 1.00\,$^{+  0.30}_{-  0.00}$}   &   \multicolumn{2}{c}{ 1.04\,$^{+  0.38}_{-  0.09}$}   &   \multicolumn{2}{c}{ 1.31\,$^{+  0.25}_{-  0.10}$}   &   &\multicolumn{2}{c}{ 0.59\,$^{+ 0.16}_{- 0.16}$}   &   \multicolumn{2}{c}{ 0.41\,$^{+ 0.19}_{- 0.19}$}   &   \multicolumn{2}{c}{ 0.42\,$^{+ 0.15}_{- 0.15}$}   &   &\multicolumn{2}{c}{ 1.00\,$^{+  0.05}_{-  0.00}$}   &   \multicolumn{2}{c}{ 0.36\,$^{+  0.20}_{-  0.06}$}   &   \multicolumn{2}{c}{ 0.79\,$^{+  0.14}_{-  0.14}$}       \\
\noalign{\smallskip}
      
\object{JW 687         }   &    195   &   M2     &   M2.5   &  CW      &   \multicolumn{2}{c}{ 0.58\,$^{+ 0.11}_{- 0.11}$}   &   \multicolumn{2}{c}{ 0.36\,$^{+ 0.10}_{- 0.10}$}   &   \multicolumn{2}{c}{ 0.39\,$^{+ 0.07}_{- 0.07}$}   &   &\multicolumn{2}{c}{ 1.34\,$^{+  0.35}_{-  0.34}$}   &   \multicolumn{2}{c}{ 0.86\,$^{+  0.24}_{-  0.08}$}   &   \multicolumn{2}{c}{ 1.33\,$^{+  0.16}_{-  0.10}$}   &   &\multicolumn{2}{c}{ 0.57\,$^{+ 0.12}_{- 0.12}$}   &   \multicolumn{2}{c}{ 0.31\,$^{+ 0.09}_{- 0.09}$}   &   \multicolumn{2}{c}{ 0.35\,$^{+ 0.06}_{- 0.06}$}   &   &\multicolumn{2}{c}{ 1.00}   &   \multicolumn{2}{c}{ 0.10\,$^{+  0.03}_{-  0.00}$}   &   \multicolumn{2}{c}{ 0.10\,$^{+  0.09}_{-  0.00}$}       \\
\noalign{\smallskip}
      
\object{JW 648         }   &    282   &   M0     &   M3.5   &   CW     &   \multicolumn{2}{c}{ 0.69\,$^{+ 0.06}_{- 0.06}$}   &   \multicolumn{2}{c}{ 0.58\,$^{+ 0.10}_{- 0.10}$}   &   \multicolumn{2}{c}{ 0.55\,$^{+ 0.07}_{- 0.07}$}   &   &\multicolumn{2}{c}{ 1.00}   &   \multicolumn{2}{c}{ 0.46\,$^{+  0.05}_{-  0.05}$}   &   \multicolumn{2}{c}{ 0.68\,$^{+  0.05}_{-  0.03}$}   &   &\multicolumn{2}{c}{ 0.38\,$^{+ 0.21}_{- 0.21}$}   &   \multicolumn{2}{c}{ 0.22\,$^{+ 0.17}_{- 0.17}$}   &   \multicolumn{2}{c}{ 0.29\,$^{+ 0.12}_{- 0.12}$}   &   &\multicolumn{2}{c}{ 1.00\,$^{+  0.94}_{-  0.00}$}   &   \multicolumn{2}{c}{ 0.47\,$^{+  0.39}_{-  0.07}$}   &   \multicolumn{2}{c}{ 1.25\,$^{+  0.31}_{-  0.23}$}       \\
\noalign{\smallskip}
      
\object{TCC 52         }   &    215   &   M0.5   &   M2     &   CC     &   \multicolumn{2}{c}{ 0.63\,$^{+ 0.13}_{- 0.13}$}   &   \multicolumn{2}{c}{ 0.52\,$^{+ 0.20}_{- 0.20}$}   &   \multicolumn{2}{c}{ 0.59\,$^{+ 0.24}_{- 0.24}$}   &   &\multicolumn{2}{c}{ 1.00}   &   \multicolumn{2}{c}{ 0.10\,$^{+  0.00}_{-  0.00}$}   &   \multicolumn{2}{c}{ 0.10\,$^{+  0.05}_{-  0.00}$}   &   &\multicolumn{2}{c}{ 0.58\,$^{+ 0.08}_{- 0.08}$}   &   \multicolumn{2}{c}{ 0.36\,$^{+ 0.09}_{- 0.09}$}   &   \multicolumn{2}{c}{ 0.39\,$^{+ 0.07}_{- 0.07}$}   &   &\multicolumn{2}{c}{ 1.00}   &   \multicolumn{2}{c}{ 0.10\,$^{+  0.01}_{-  0.00}$}   &   \multicolumn{2}{c}{ 0.10\,$^{+  0.04}_{-  0.00}$}       \\
\noalign{\smallskip}
      
\object{[HC2000] 73    }   &    294   &   M2     &   M7     &   CW     &   \multicolumn{2}{c}{ 0.53\,$^{+ 0.11}_{- 0.11}$}   &   \multicolumn{2}{c}{ 0.36\,$^{+ 0.09}_{- 0.09}$}   &   \multicolumn{2}{c}{ 0.38\,$^{+ 0.08}_{- 0.08}$}   &   &\multicolumn{2}{c}{ 5.08\,$^{+  2.22}_{-  4.08}$}   &   \multicolumn{2}{c}{ 2.73\,$^{+  0.64}_{-  0.20}$}   &   \multicolumn{2}{c}{ 3.96\,$^{+  1.15}_{-  0.31}$}   &   &\multicolumn{2}{c}{ 0.06\,$^{+ 0.05}_{- 0.05}$}   &   \multicolumn{2}{c}{$<$ 0.1  }   &                     \multicolumn{2}{c}{ 0.10\,$^{+ 0.02}_{- 0.02}$}   &   &\multicolumn{2}{c}{ 1.00\,$^{+  3.73}_{-  0.00}$}   &   \multicolumn{2}{c}{         }   &                       \multicolumn{2}{c}{ 1.85\,$^{+  0.44}_{-  0.21}$}       \\
\noalign{\smallskip}

\hline
\end{tabular}
\end{center}
\label{Table : comparative samples}
\end{table*}

%
\begin{table*}
\scriptsize
\caption{Derived mass ratios and relative ages. The PMS tracks are those from Baraffe et al. (\cite{Baraffe_etal1998}) - B98, Palla \& Stahler (\cite{Palla_Stahler_1999}) - PS99, and Siess et al. (\cite{Siess_etal_2000}) - S00.}
\begin{center}
\renewcommand{\arraystretch}{1.4}
\setlength\tabcolsep{5pt}
\begin{tabular}{l@{\hspace{4mm}}
			r@{\,$\pm$\,}l@{\hspace{4mm}}
			r@{\,$\pm$\,}l@{\hspace{4mm}}
			r@{\,$\pm$\,}l@{\hspace{4mm}}
			c
			r@{\,$\pm$\,}l@{\hspace{4mm}}
			r@{\,$\pm$\,}l@{\hspace{4mm}}
			}

\hline\noalign{\smallskip}

	 															&  
\multicolumn{6}{c}{q=M$_{sec}$/M$_{prim}$}								&
																&
\multicolumn{4}{c}{$\Delta$\,log\,$\tau$\,=log (age$_{prim}$) - log(age$_{sec}$)} 	\\

\cline{2-7} 	\cline{9-12}	\\ 

Name 						&  
\multicolumn{2}{c}{B98} 		 	&  
\multicolumn{2}{c}{PS99} 		 	&  
\multicolumn{2}{c}{S00}   			& 
							&
\multicolumn{2}{c}{PS99} 		 	&  
\multicolumn{2}{c}{S00}   			\\

\noalign{\smallskip}
\hline
\noalign{\smallskip}
\multicolumn{12}{c}{Tau-Aur from WG01} \\
\noalign{\smallskip}
\hline
\noalign{\smallskip}
\object{FW Tau A-B     }   &   1.00   &   0.25                    &   1.00   &   0.30                    &   0.96   &   0.19                    &   &   $-0.10$   &   $ 0.79$              &   $-0.03$   &   $ 0.67$              \\
\object{332 G1         }   &   0.73   &   0.22                    &   0.59   &   0.20                    &   0.59   &   0.15                    &   &   $ 0.73$   &   $ 0.61$              &   $ 0.62$   &   $ 0.64$              \\
\object{V410 Tau A-C   }   &   0.10   &   0.16                    &   0.07   &   0.17                    &   0.08   &   0.13                    &   &   $ 0.07$   &   $ 0.50$              &   $-0.27$   &   $ 0.34$              \\
\object{UZ Tau A-Ba    }   &   0.93   &   0.12                    &   0.76   &   0.10                    &   0.83   &   0.09                    &   &   $ 0.24$   &   $ 0.46$              &   $ 0.22$   &   $ 0.45$              \\
\object{FV Tau A-B     }   &   0.90   &   0.11                    &   0.79   &   0.17                    &   0.83   &   0.16                    &   &   $-0.31$   &   $ 0.52$              &   $-0.36$   &   $ 0.50$              \\
\object{RW Aur A-B     }   &   0.68   &   0.10                    &   0.64   &   0.15                    &   0.64   &   0.17                    &   &   $-0.23$   &   $ 0.43$              &   $-0.25$   &   $ 0.35$              \\
\object{GG Tau Ba-Bb   }   &   0.30   &   0.05                    &   \multicolumn{2}{c}{$<$  0.91}      &   \multicolumn{2}{c}{$<$  0.79}      &   &   \multicolumn{2}{c}{         }      &   \multicolumn{2}{c}{         }      \\
\object{Haro 6-37 A-B  }   &   0.74   &   0.15                    &   0.61   &   0.14                    &   0.60   &   0.10                    &   &   $-0.14$   &   $ 0.48$              &   $-0.23$   &   $ 0.39$              \\
\object{UX Tau A-B     }   &   0.49   &   0.13                    &   0.35   &   0.16                    &   0.33   &   0.15                    &   &   $ 0.19$   &   $ 0.32$              &   $-0.00$   &   $ 0.24$              \\
\object{UX Tau A-C     }   &   0.17   &   0.11                    &   0.13   &   0.15                    &   0.15   &   0.14                    &   &   $ 0.57$   &   $ 0.37$              &   $ 0.09$   &   $ 0.42$              \\
\object{V710 Tau A-B   }   &   0.90   &   0.14                    &   0.67   &   0.21                    &   0.76   &   0.16                    &   &   $ 0.45$   &   $ 0.56$              &   $ 0.24$   &   $ 0.43$              \\
\noalign{\smallskip}
\hline
\noalign{\smallskip}
\multicolumn{12}{c}{Tau-Aur from HK03} \\
\noalign{\smallskip}
\hline
\noalign{\smallskip}
\object{DD Tau         }   &   0.89   &   0.22                    &   1.01   &   0.19                    &   0.98   &   0.13                    &   &   $-0.23$   &   $ 0.42$              &   $-0.18$   &   $ 0.41$              \\
\object{DF Tau         }   &   0.99   &   0.16                    &   0.87   &   0.19                    &   0.92   &   0.14                    &   &   $ 0.26$   &   $ 0.51$              &   $ 0.11$   &   $ 0.38$              \\
\object{FO Tau         }   &   1.00   &   0.23                    &   1.00   &   0.17                    &   1.00   &   0.12                    &   &   $ 0.00$   &   $ 0.38$              &   $ 0.00$   &   $ 0.44$              \\
\object{FQ Tau         }   &   0.94   &   0.22                    &   0.83   &   0.22                    &   0.93   &   0.15                    &   &   $ 0.33$   &   $ 0.45$              &   $ 0.16$   &   $ 0.32$              \\
\object{FS Tau         }   &   0.44   &   0.17                    &   0.38   &   0.15                    &   0.47   &   0.14                    &   &   $ 0.94$   &   $ 0.44$              &   $ 0.72$   &   $ 0.47$              \\
\object{FVTau c        }   &   0.67   &   0.23                    &   0.73   &   0.24                    &   0.68   &   0.21                    &   &   $-0.38$   &   $ 0.49$              &   $-0.36$   &   $ 0.46$              \\
\object{GG Tau         }   &   0.73   &   0.16                    &   0.61   &   0.18                    &   0.65   &   0.16                    &   &   $ 0.11$   &   $ 0.47$              &   $ 0.02$   &   $ 0.37$              \\
\object{GH Tau         }   &   1.00   &   0.15                    &   1.00   &   0.14                    &   1.00   &   0.12                    &   &   $-0.02$   &   $ 0.40$              &   $-0.00$   &   $ 0.37$              \\
\object{HBC 358        }   &   0.86   &   0.25                    &   0.83   &   0.22                    &   0.89   &   0.18                    &   &   $ 0.10$   &   $ 0.47$              &   $ 0.05$   &   $ 0.36$              \\
\object{IS Tau         }   &   0.48   &   0.17                    &   0.39   &   0.17                    &   0.50   &   0.14                    &   &   $ 0.16$   &   $ 0.49$              &   $-0.17$   &   $ 0.37$              \\
\object{LkCa 7         }   &   0.45   &   0.18                    &   0.38   &   0.14                    &   0.49   &   0.12                    &   &   $ 0.28$   &   $ 0.43$              &   $ 0.02$   &   $ 0.37$              \\
\object{UY Aur         }   &   0.61   &   0.16                    &   0.53   &   0.18                    &   0.58   &   0.15                    &   &   $ 0.21$   &   $ 0.42$              &   $ 0.08$   &   $ 0.38$              \\
\object{UZ Tau W       }   &   0.85   &   0.24                    &   0.73   &   0.23                    &   0.84   &   0.15                    &   &   $ 0.45$   &   $ 0.48$              &   $ 0.26$   &   $ 0.49$              \\
\object{V807 Tau       }   &   0.58   &   0.13                    &   0.31   &   0.14                    &   0.34   &   0.11                    &   &   $ 0.62$   &   $ 0.55$              &   $ 0.21$   &   $ 0.41$              \\
\object{V927 Tau       }   &   0.71   &   0.25                    &   0.85   &   0.23                    &   0.81   &   0.17                    &   &   $-0.42$   &   $ 0.48$              &   $-0.41$   &   $ 0.42$              \\
\object{V955 Tau       }   &   0.54   &   0.14                    &   0.50   &   0.14                    &   0.44   &   0.12                    &   &   $ 0.50$   &   $ 0.56$              &   $ 0.42$   &   $ 0.49$              \\
\object{V999 Tau       }   &   0.66   &   0.18                    &   0.59   &   0.19                    &   0.65   &   0.19                    &   &   $ 0.21$   &   $ 0.46$              &   $ 0.06$   &   $ 0.42$              \\
\object{V1026 Tau      }   &   0.50   &   0.23                    &   0.65   &   0.22                    &   0.60   &   0.21                    &   &   $-0.30$   &   $ 0.61$              &   $-0.31$   &   $ 0.45$              \\
\object{XZ Tau         }   &   0.71   &   0.24                    &   0.63   &   0.24                    &   0.78   &   0.17                    &   &   $ 0.62$   &   $ 0.50$              &   $ 0.49$   &   $ 0.39$              \\
\noalign{\smallskip}
\hline
\noalign{\smallskip}
\multicolumn{12}{c}{Oph from P03} \\
\noalign{\smallskip}
\hline
\noalign{\smallskip}
\object{AS 205         }   &   0.48   &   0.17                    &   0.23   &   0.10                    &   0.26   &   0.10                    &   &   $ 0.00$   &   $ 0.18$              &   $ 0.61$   &   $ 0.29$              \\
\object{WSB 4          }   &   1.00   &   0.19                    &   0.99   &   0.19                    &   1.04   &   0.14                    &   &   $ 0.13$   &   $ 0.41$              &   $ 0.18$   &   $ 0.28$              \\
\object{WSB 19         }   &   0.37   &   0.23                    &   0.48   &   0.26                    &   0.61   &   0.19                    &   &   $ 0.29$   &   $ 0.36$              &   $-0.04$   &   $ 0.41$              \\
\object{WSB 28         }   &   0.12   &   0.09                    &   \multicolumn{2}{c}{$<$  0.37}      &   \multicolumn{2}{c}{$<$  0.31}      &   &   \multicolumn{2}{c}{         }      &   \multicolumn{2}{c}{         }      \\
\object{DoAr 26        }   &   0.34   &   0.29                    &   0.39   &   0.34                    &   0.47   &   0.18                    &   &   $-0.25$   &   $ 0.60$              &   $ 0.37$   &   $ 0.54$              \\
\object{ROX 15         }   &   1.00   &   0.20                    &   1.00   &   0.17                    &   0.96   &   0.12                    &   &   $ 0.00$   &   $ 0.20$              &   $ 0.00$   &   $ 0.34$              \\
\object{SR 21          }   &   \multicolumn{2}{c}{$<$  0.20}      &   0.07   &   0.03                    &   0.10   &   0.02                    &   &   $ 0.49$   &   $ 0.41$              &   $ 0.81$   &   $ 0.38$              \\
\object{Ha 71          }   &   0.07   &   0.03                    &   \multicolumn{2}{c}{$<$  0.07}      &   0.11   &   0.03                    &   &   \multicolumn{2}{c}{         }      &   $ 1.32$   &   $ 0.43$              \\
\noalign{\smallskip}
\hline
\noalign{\smallskip}
\multicolumn{12}{c}{ONC from D12} \\
\noalign{\smallskip}
\hline
\noalign{\smallskip}
\object{JW 260         }   &   \multicolumn{2}{c}{}      &   0.88   &   0.10                    &   0.82   &   0.10                    &   &   $-0.24$   &   $ 0.26$              &   $-0.23$   &   $ 0.57$              \\
\object{JW 553         }   &   0.77   &   0.25                    &   0.60   &   0.28                    &   0.59   &   0.24                    &   &   $-1.83$   &   $ 0.60$              &   $-1.36$   &   $ 0.57$              \\
\object{JW 566         }   &   0.75   &   0.22                    &   0.52   &   0.25                    &   0.57   &   0.23                    &   &   $ 0.46$   &   $ 0.49$              &   $ 0.22$   &   $ 0.45$              \\
\object{JW 687         }   &   0.99   &   0.19                    &   0.86   &   0.21                    &   0.91   &   0.13                    &   &   $ 0.94$   &   $ 0.25$              &   $ 1.12$   &   $ 0.30$              \\
\object{JW 648         }   &   0.55   &   0.30                    &   0.39   &   0.27                    &   0.53   &   0.20                    &   &   $-0.02$   &   $ 0.45$              &   $-0.27$   &   $ 0.43$              \\
\object{TCC 52         }   &   0.91   &   0.15                    &   0.69   &   0.22                    &   0.65   &   0.22                    &   &   $ 0.00$   &   $ 0.03$              &   $ 0.00$   &   $ 0.18$              \\
\object{[HC2000] 73    }   &   0.11   &   0.10                    &   \multicolumn{2}{c}{$<$  0.27}      &   0.26   &   0.07                    &   &   \multicolumn{2}{c}{         }      &   $ 0.33$   &   $ 0.28$              \\

\hline
\end{tabular}
\end{center}
\label{Table : comparative samples : mass ratios and relative ages}
\end{table*}

\end{appendix}


\begin{thebibliography}{}
\bibitem[1999]{Armitage_etal_1999} Armitage, P.J., Clarke, C.J., Tout, C.A. 1999, MNRAS, 304, 425.
\bibitem[2007]{Bakos_etal_2007} Bakos, G. \'A., Noyes, R. W., Kov\'acs, G., et al. 2007, ApJ, 656, 552.
\bibitem[1998]{Baraffe_etal1998} Baraffe, I., Chabrier, G., Allard, F. et al. 1998, A\&A, 337, 403, [B98].
\bibitem[2000]{Bate_2000} Bate, M.R. 2000, MNRAS, 314, 33. 
\bibitem[2001]{Bate_2001} Bate, M.R. 2001, 
				in IAU Symp. 200, The Formation of Binary Stars, ed. H. Zinnecker and R.D. Mathieu (San Francisco: ASP), 429.
\bibitem[2009]{Bate_2009} Bate, M.R. 2009, MNRAS, 392, 590.				
\bibitem[2004]{Beck_etal_2004} Beck, T.L., Schaefer, G.H., Simon, M., et al. 2004, ApJ, 614, 235.
\bibitem[2008]{Beck_etal_2008} Beck, T.L. McGregor, P.J., Takami, M., Pyo, T.-S. 2008, ApJ, 676, 472.
\bibitem[1988]{Bessel_Brett_1988} Bessel, M.S. \& Brett, J.M. 1988, PASP, 100, 1134.
\bibitem[1997]{Brandner_Zinnecker_1997} Brandner, W. \& Zinnecker, H. 1997, A\&A, 321, 220.
\bibitem[2010]{Bressert_etal_2010} Bressert, E., Bastian, N., Gutermuth, R., et al. 2010, MNRAS, 409, L54.
\bibitem[2009]{Cieza_etal_2009} Cieza, L. A., Padgett, D. L., Allen, L. E., et al. 2009, ApJ, 696, L84.	
\bibitem[1997]{Cochran_etal_1997} Cochran, W.D., Hatzes, A.P., Butler, R.P., Marcy, G.W. 1997, ApJ, 483, 457.			
\bibitem[1981]{Cohen_etal_1981} Cohen, J. G., Persson, S. E., Elias, J. H., \& Frogel, J. A. 1981, ApJ, 249, 481.
\bibitem[2000]{Correia_Richichi_2000} Correia, S. \& Richichi, A. 2000, A\&AS, 141, 301.			
\bibitem[2006]{Correia_etal_2006} Correia, S., Zinnecker, H., Ratzka, Th., Sterzik, M. 2006, A\&A, 459, 909.
\bibitem[2012]{Daemgen_etal_2012} Daemgen, S., Correia, S., Petr-Gotzens, M. 2012, A\&A, 540, 46, [D12].
\bibitem[2010]{DaRio_etal_2010} Da Rio, N., Robberto, M., Soderblom, D.R., et al. 2010, ApJ, 722, 1092. 
\bibitem[2012]{Delfosse_etal_2012} Delfosse, X., Bonfils, X., Forveille, T. 2012, arXiv:1202.2467.
\bibitem[1999]{Duchene_1999} Duch\^{e}ne, G. 1999, A\&A, 341, 547.
\bibitem[1999]{Duchene_etal_1999} Duch\^{e}ne, G., Monin, J.-L., Bouvier, J., M\'enard, F. 1999, A\&A, 351, 954.
\bibitem[2007]{Duchene_etal_2007} Duch\^ene, G., Delgado-Donate, E., Haisch, K.E., Jr., et al. 2007, 
				in Protostars and Planets V, ed. B. Reipurth, D. Jewitt, K. Keil (Tucson: Univ. Arizona Press), 379.
\bibitem[2010]{Duchene_2010} Duch\^ene, G. 2010, ApJ, 709, L114.	
\bibitem[2013]{Duchene_Kraus_2013} Duch\^ene, G., Kraus, A. 2013, ARA\&A, in press.	
\bibitem[2012]{Dumusque_etal_2012} Dumusque, X., Pepe, F., Lovis, C. 2012, Nature, 491, 207.	
\bibitem[1991]{Duquennoy_Mayor_1991} Duquennoy, A. \& Mayor, M. 1991, A\&A, 248, 485.
\bibitem[2010]{Eggenberger_Udry_2010} Eggenberger, A. \& Udry, S. 2010, in Planets in Binary Star Systems, 1st edn., Vol. 366 (Berlin: Springer), 19.
\bibitem[2005]{Getman_etal_2005} Getman, K.V., Feigelson, E.D., Grosso, N., et al. 2005, ApJS, 160, 353.
\bibitem[2008]{Ghez_etal_2008} Ghez, A.M., Salim, S., Weinberg, N. N., et al. 2008, ApJ, 689, 1044.
\bibitem[1995]{Greene_Meyer_1995} Greene, T.P. \& Meyer, M.R. 1995, ApJ, 450, 233.
\bibitem[1998]{Gullbring_etal_1998} Gullbring, E., Hartmann, L., Brice\~no, C., Calvet, N. 1998, ApJ, 492, 323.
\bibitem[1994]{Hartigan_etal1994} Hartigan, P., Strom, K.M., Strom, S.E. 1994, ApJ, 427, 961.
\bibitem[2003]{Hartigan_Kenyon2003} Hartigan, P. \& Kenyon, S.J. 2003, ApJ, 583, 334, [HK03].
\bibitem[2005]{Hartmann_etal_2005} Hartmann, L., Megeath, S.T., Allen, L., et al. 2005, ApJ, 629, 881.
\bibitem[2007]{Hernandez_etal_2007} Hern\'andez, J., Hartmann, L., Megeath, T., et al. 2007, ApJ, 662, 1067.
\bibitem[1997]{Hillenbrand_1997} Hillenbrand, L.A. 1997, AJ, 113, 1733.
\bibitem[2004]{Hillenbrand_White_2004} Hillenbrand, L.A. \& White, R.J. 2004, ApJ, 604, 741.
\bibitem[1995]{Hinkle_Wallace_Livingston_1995} Hinkle, K., Wallace, L. \& Livingston, W. 1995, PASP, 107, 1042.
\bibitem[2007]{Hirota_etal_2007} Hirota, T., Bushimata, T., Choi, Y. K., et al. 2007, PASJ, 59, 897.
\bibitem[2003]{Hodapp_etal_2003} Hodapp, K.-W, Irwin, E.M., Yamada, H., et al. 2003, PASP, 115, 1388.
\bibitem[1988]{Jones_Walker_1988} Jones, B.F. \& Walker, M.F. 1988, AJ, 95, 1755.
\bibitem[1995]{Kenyon_hartmann_1995} Kenyon, S.J. \& Hartmann, L. 1995, ApJS, 101, 117.
\bibitem[2006]{Koehler_etal_2006} K\"ohler, R., Petr-Gotzens, M. G., McCaughrean, M. J., et al. 2006, A\&A, 458, 461. 
\bibitem[2009a]{Kraus_Hillenbrand_2009a} Kraus, A.L. \& Hillenbrand, L.A. 2009, ApJ, 703, 1511.
\bibitem[2009b]{Kraus_Hillenbrand_2009b} Kraus, A.L. \& Hillenbrand, L.A. 2009, ApJ, 704, 531.
\bibitem[2011]{Kraus_etal_2011} Kraus, A.L., Ireland, M.J., Martinache, F., Hillenbrand, L. 2011, ApJ, 731, 8.
\bibitem[2012]{Kraus_etal_2012} Kraus, A.L., Ireland, M.J., Hillenbrand, L.A., Martinache, F. 2012, ApJ, 745, 19.
\bibitem[2000]{Lada_etal_2000} Lada, C.J., Muench, A.A., Haisch, K.E., Jr., et al. 2000, AJ, 120, 3162.
\bibitem[2003]{Lada_Lada_2003} Lada C.J. \& Lada, E.A. 2003, A\&ARA, 41, 57.
\bibitem[1988]{Lasker_etal_1988} Lasker, B.M., Sturch, C.R., Lopez, C. 1988, ApJS, 68, 1.
\bibitem[1992]{Leggett_1992} Leggett, S.K. 1992, ApJS, 82, 351.
\bibitem[2002]{Leggett_etal_2002} Leggett, S.K., Golimowski, D.A., FAN, X. 2002, ApJ, 564, 452.
\bibitem[2003]{Leggett_etal_2003} Leggett, S.K., Hawarden, T.G., Currie, M.J., et al. 2003, MNRAS, 345, 144.
\bibitem[2006]{Leggett_etal_2006} Leggett, S.K., Currie, M.J., Warricatt, W.P., et al. 2006, MNRAS, 373, 781.
\bibitem[2006]{Levine_etal_2006} Levine, J.L., Steinhauer, A., Elston, R.J., Lada, E.A. 2006, ApJ, 646, 1215.
\bibitem[2003]{Liu_etal_2003} Liu, M.C., Najita, J., Tokunaga, A.T. 2003, ApJ, 585, 372.
\bibitem[2003]{Luhman_etal_2003} Luhman, K.L., Stauffer, J.R., Muench, A.A. et al. 2003, ApJ, 593, 1093.
\bibitem[2010]{Luhman_etal_2010} Luhman, K.L., Allen, P.R., Espaillat, C., et al. 2010, ApJS, 186, 111.
\bibitem[2012]{Manara_etal_2012} Manara, C.F., Robberto, M., Da Rio, N. et al., 2012, ApJ, 755, 154.
\bibitem[2006]{McCabe_etal2006} McCabe, C., Ghez, A.M., Prato, L., et al. 2006, ApJ, 636, 932.			
\bibitem[2002]{McGregor_2002} McGregor, P., Hart, J., Conroy, P., et al. 2002 SPIE 4841, 178.
\bibitem[2007]{Menten_etal_2007} Menten, K.M., Reid, M.J., Forbrich, J., Brunthaler, A. 2007, 474, 515.
\bibitem[1997]{Meyer_etal_1997} Meyer, M.R., Calvet, N., Hillenbrand, L.A. 1997, AJ, 114, 288.
\bibitem[2007]{Monin_etal_2007} Monin, J.-L., Clarke, C.J., Prato, L., et al. 2007, 
				in Protostars and Planets V, ed. B. Reipurth, D. Jewitt, K. Keil (Tucson: Univ. Arizona Press), 395.			
\bibitem[2011]{Morales-Calderon_etal_2011} Morales-Calder\'on, M., Stauffer, J.R., Hillenbrand, L.A., et al. 2011, ApJ, 733, 50.		
\bibitem[2007]{Mugrauer_etal_2007} Mugrauer, M., Neuh\"auser, R., Mazeh, T. 2007, A\&A, 469, 755.	
\bibitem[1998]{Muzerolle_etal_1998} Muzerolle, J., Hartmann, L., Calvet, N. 1998, AJ, 116, 2965.
\bibitem[1999]{Palla_Stahler_1999} Palla, F. \& Stahler, S.W. 1999, ApJ, 525, 772, [PS99].
\bibitem[2008]{Patience_etal_2008} Patience, J., Akeson, R.L., Jensen, E.L.N. 2008, ApJ, 677, 616. 
\bibitem[1998]{Petr_etal_1998} Petr, M. G., Foresto, V. C. D., Beckwith, S. V. W., et al. 1998, ApJ, 500, 825.
\bibitem[1997]{Prato_Simon_1997} Prato, L. \& Simon, M. 1997, ApJ, 474, 455.
\bibitem[2003]{Prato_etal2003} Prato, L., Greene, T.P., Simon, M. 2003, ApJ, 584, 853, [P03].
\bibitem[2010]{Raghavan_etal_2010} Raghavan, D., McAlister, H. A., Henry, T. J., et al. 2010, ApJS, 190, 1.
\bibitem[2009]{Rayner_Cushing_Vacca_2009} Rayner, J. T., Cushing, M. C., \& Vacca, W. D. 2009, ApJSS, 185, 289.
\bibitem[2011]{Reggiani_Meyer_2011} Reggiani, M.M. \& Meyer, M.R. 2011, ApJ, 738, 60.
\bibitem[1993]{Reipurth_Zinnecker1993} Reipurth, B. \& Zinnecker, H. 1993, A\&A, 278, 81.
\bibitem[2007]{Reipurth_etal_2007} Reipurth, B., Guimar\~aes, M.M., Connelley, M.S., Bally, J. 2007, AJ, 134, 2272, [R07].
\bibitem[2000]{Siess_etal_2000} Siess, L., Dufour, E., Forestini, M. 2000, A\&A, 358, 593, [S00].
\bibitem[2007]{Sandstrom_etal_2007} Sandstrom, K. M., Peek, J. E. G., Bower, et al. 2007, ApJ, 667, 1161.
\bibitem[2002]{Tohline_2002} Tohline, J.E. 2002, ARA\&A, 40, 349.
\bibitem[2001]{White_Ghez_2001} White, R.J.  \& Ghez, A.M. 2001, ApJ, 556, 265, [WG01].
\bibitem[1996]{Wallace_Hinkle_1996} Wallace, L. \& Hinkle, K. 1996, ApJS, 107, 312.
\bibitem[1997]{Wallace_Hinkle_1997} Wallace, L. \& Hinkle, K. 1997, ApJSS, 111, 445.
\bibitem[2008]{Wilking_etal_2008} Wilking, B.A., Gagn\'e, M., Allen, L.E. 2008, in Handbook of Star Forming Regions, Volume II, ed. B. Reipurth, 351.
\bibitem[2001]{Woitas_etal_2001} Woitas, J., Leinert, Ch., K\"ohler, R. 2001, A\&A, 376, 982.
\end{thebibliography}
\end{document}